\documentclass[longauth]{aa}  
\usepackage[dvipsnames]{xcolor}
\usepackage{graphicx}
\usepackage{txfonts}
\usepackage{url}
\usepackage{hyperref}
\usepackage{rotating}
\usepackage{amsmath,amssymb}
\usepackage{multirow}
\usepackage{makecell}
\usepackage{ulem}
\usepackage{orcidlink}
\hypersetup{
    colorlinks=true,
    linkcolor=blue,
    urlcolor=magenta,
}
\usepackage{colortbl}
\usepackage{listings}
\definecolor{dkgreen}{rgb}{0,0.6,0}
\definecolor{gray}{rgb}{0.5,0.5,0.5}
\definecolor{mauve}{rgb}{0.58,0,0.82}

\newcommand{\Gaia}{{\it Gaia }}
\newcommand{\gspspec}{{\it GSP-Spec}}

\newcommand{\T}{$T_{\rm eff}$}
\newcommand{\g}{log($g$)}

\newcommand{\meta}{[M/H]}
\newcommand{\alphaFe}{[$\alpha$/Fe]}
\newcommand{\Vrad}{$V_{\rm Rad}$}
\newcommand{\Vsini}{{\it Vsin}i}

\newcommand{\FeI}{Fe~{\sc i}}
\newcommand{\FeII}{Fe~{\sc ii}}
\newcommand{\CaII}{Ca~{\sc ii}}

\newcommand{\CaFe}{[Ca/Fe]}

\newcommand{\SFe}{[S/Fe]}
\newcommand{\NFe}{[N/Fe]}
\newcommand{\MgFe}{[Mg/Fe]}
\newcommand{\SiFe}{[Si/Fe]}
\newcommand{\TiFe}{[Ti/Fe]}
\newcommand{\CrFe}{[Cr/Fe]}
\newcommand{\NiFe}{[Ni/Fe]}
\newcommand{\ZrFe}{[Zr/Fe]}
\newcommand{\CeFe}{[Ce/Fe]}
\newcommand{\NdFe}{[Nd/Fe]}
\newcommand{\FeIM}{[\FeI/M]}
\newcommand{\FeIIM}{[\FeII/M]}
\newcommand{\FeIH}{[\FeI/H]}
\newcommand{\FeIIH}{[\FeII/H]}

\newcommand{\tabref}[1]{Table~\ref{#1}}
\newcommand{\figref}[1]{Figure~\ref{#1}}
\newcommand{\secref}[1]{Section~\ref{#1}}

\newcommand{\orcit}[1]{\protect\href{https://orcid.org/#1}{\protect\includegraphics[width=8pt]{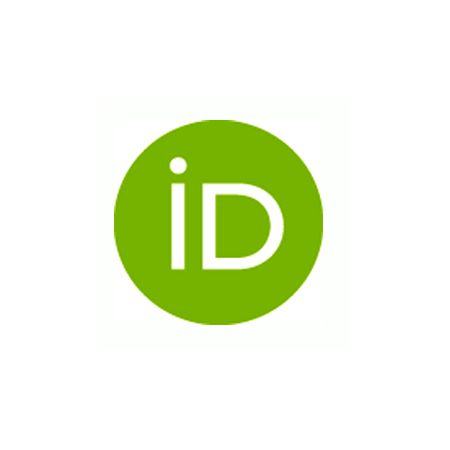}}}

\begin{document}
\titlerunning{\gspspec \ RVS analysis in Gaia DR3}

   \title{\Gaia\ Data Release 3: Analysis of RVS spectra using the {\it  General Stellar Parametriser from spectroscopy}}

   \author{
          Alejandra~ Recio-Blanco\inst{\ref{inst1}}\thanks{Corresponding author: Alejandra.Recio-Blanco@oca.eu}\orcit{0000-0002-6550-7377}
          \and
          P.~ de Laverny\inst{\ref{inst1}}\orcit{0000-0002-2817-4104}
          \and
          P. A.~ Palicio\orcit{0000-0002-7432-8709}\inst{\ref{inst1}}
          \and 
          G.~ Kordopatis\orcit{0000-0002-9035-3920}\inst{\ref{inst1}}
          \and
          M. A.~ Álvarez\orcit{0000-0002-6786-2620}\inst{\ref{inst2}}
          \and 
          M.~ Schultheis\orcit{0000-0002-6590-1657}\inst{\ref{inst1}}
          \and 
          G.~ Contursi\orcit{0000-0001-5370-1511}\inst{\ref{inst1}}
          \and 
          H.~ Zhao\orcit{0000-0003-2645-6869}\inst{\ref{inst1}}
          \and
          G.~ Torralba Elipe\inst{\ref{inst2}}
          \and
          C.~ Ordenovic\inst{\ref{inst1}}
          \and
          M.~ Manteiga\orcit{0000-0002-7711-5581}\inst{\ref{inst:0066}}
          \and
          C.~ Dafonte\orcit{0000-0003-4693-7555}\inst{\ref{inst2}}
          \and
          I.~ Oreshina-Slezak\inst{\ref{inst1}}
          \and 
          A.~ Bijaoui\inst{\ref{inst1}}
          \and
          Y.~ Fr\'emat\orcit{0000-0002-4645-6017}\inst{\ref{inst3}}
          \and        
          G.~ Seabroke\inst{\ref{inst4}}
          \and 
          F.~ Pailler\orcit{0000-0002-4834-481X}\inst{\ref{inst5}}
          \and 
          E.~ Spitoni\orcit{0000-0001-9715-5727}\inst{\ref{inst1}}
          \and
          E.~ Poggio\orcit{0000-0003-3793-8505}\inst{\ref{inst1}}
          \and
O.L.~                       Creevey\orcit{0000-0003-1853-6631}\inst{\ref{inst1}}
\and         A.~                Abreu Aramburu\inst{\ref{inst:0027}}
\and         S.~                Accart \inst{\ref{inst5}} 
\and         R.~                        Andrae\orcit{0000-0001-8006-6365}\inst{\ref{inst:0007}}
\and     C.A.L.~                  Bailer-Jones\inst{\ref{inst:0007}}
\and         I.~                Bellas-Velidis\inst{\ref{inst:0012}}
\and         N.~                     Brouillet\orcit{0000-0002-3274-7024}\inst{\ref{inst:0030}}
\and         E.~                    Brugaletta\orcit{0000-0003-2598-6737}\inst{\ref{inst:0013}}
\and         A.~                       Burlacu\inst{\ref{inst:0031}}
\and         R.~                      Carballo\orcit{0000-0001-7412-2498}\inst{\ref{inst:0032}}
\and         L.~                   Casamiquela\orcit{0000-0001-5238-8674}\inst{\ref{inst:0030},\ref{inst:0034}}
\and         A.~                     Chiavassa\orcit{0000-0003-3891-7554}\inst{\ref{inst1}}
\and       W.J.~                        Cooper\orcit{0000-0003-3501-8967}\inst{\ref{inst:0038},\ref{inst:0019}}
\and         A.~                    Dapergolas\inst{\ref{inst:0012}}
\and         L.~                    Delchambre\orcit{0000-0003-2559-408X}\inst{\ref{inst:0018}}
\and       T.E.~                 Dharmawardena\orcit{0000-0002-9583-5216}\inst{\ref{inst:0007}}
\and         R.~                       Drimmel\orcit{0000-0002-1777-5502}\inst{\ref{inst:0019}}
\and         B.~                    Edvardsson\inst{\ref{inst:0045}}
\and         M.~                     Fouesneau\orcit{0000-0001-9256-5516}\inst{\ref{inst:0007}}
\and         D.~                      Garabato\orcit{0000-0002-7133-6623}\inst{\ref{inst2}}
\and         P.~              Garc\'{i}a-Lario\orcit{0000-0003-4039-8212}\inst{\ref{inst:0046}}
\and         M.~             Garc\'{i}a-Torres\orcit{0000-0002-6867-7080}\inst{\ref{inst:0047}}
\and         A.~                         Gavel\orcit{0000-0002-2963-722X}\inst{\ref{inst:0045}}
\and         A.~                         Gomez\orcit{0000-0002-3796-3690}\inst{\ref{inst2}}
\and         I.~   Gonz\'{a}lez-Santamar\'{i}a\orcit{0000-0002-8537-9384}\inst{\ref{inst2}}
\and         D.~                Hatzidimitriou\orcit{0000-0002-5415-0464}\inst{\ref{inst:0051},\ref{inst:0012}}
\and         U.~                        Heiter\orcit{0000-0001-6825-1066}\inst{\ref{inst:0045}}
\and         A.~          Jean-Antoine Piccolo\orcit{0000-0001-8622-212X}\inst{\ref{inst5}}
\and         M.~                      Kontizas\orcit{0000-0001-7177-0158}\inst{\ref{inst:0051}}
\and       A.J.~                          Korn\orcit{0000-0002-3881-6756}\inst{\ref{inst:0045}}
\and       A.C.~                     Lanzafame\orcit{0000-0002-2697-3607}\inst{\ref{inst:0013},\ref{inst:0057}}
\and         Y.~                      Lebreton\orcit{0000-0002-4834-2144}\inst{\ref{inst:0058},\ref{inst:0059}}
\and         Y. Le Fustec \inst{\ref{inst:0031}} 
\and       E.L.~                        Licata\orcit{0000-0002-5203-0135}\inst{\ref{inst:0019}}
\and     H.E.P.~                  Lindstr{\o}m\inst{\ref{inst:0019},\ref{inst:0062},\ref{inst:0063}}
\and         E.~                       Livanou\orcit{0000-0003-0628-2347}\inst{\ref{inst:0051}}
\and         A.~                         Lobel\orcit{0000-0001-5030-019X}\inst{\ref{inst3}}
\and         A.~                         Lorca\inst{\ref{inst:0014}}
\and         A.~               Magdaleno Romeo\inst{\ref{inst:0031}}
\and         F.~                       Marocco\orcit{0000-0001-7519-1700}\inst{\ref{inst:0067}}
\and       D.J.~                      Marshall\orcit{0000-0003-3956-3524}\inst{\ref{inst:0068}}
\and         N.~                          Mary\inst{\ref{inst:0069}}
\and         C.~                       Nicolas\inst{\ref{inst5}}
\and         L.~               Pallas-Quintela\orcit{0000-0001-9296-3100}\inst{\ref{inst2}}
\and         C.~                         Panem\inst{\ref{inst5}}
\and         B.~                        Pichon\orcit{0000 0000 0062 1449}\inst{\ref{inst1}}
\and         F.~                        Riclet\inst{\ref{inst5}}
\and         C.~                         Robin\inst{\ref{inst:0069}}
\and         J.~                       Rybizki\orcit{0000-0002-0993-6089}\inst{\ref{inst:0007}}
\and         R.~                 Santove\~{n}a\orcit{0000-0002-9257-2131}\inst{\ref{inst2}}
\and         A.~                       Silvelo\orcit{0000-0002-5126-6365}\inst{\ref{inst2}}
\and       R.L.~                         Smart\orcit{0000-0002-4424-4766}\inst{\ref{inst:0019}}
\and       L.M.~                         Sarro\orcit{0000-0002-5622-5191}\inst{\ref{inst:0017}}
\and         R.~                         Sordo\orcit{0000-0003-4979-0659}\inst{\ref{inst:0002}}
\and         C.~                      Soubiran\orcit{0000-0003-3304-8134}\inst{\ref{inst:0030}}
\and         M.~                  S\"{ u}veges\orcit{0000-0003-3017-5322}\inst{\ref{inst:0084}}
\and         A.~                          Ulla\orcit{0000-0001-6424-5005}\inst{\ref{inst:0085}}
\and         A.~                     Vallenari\orcit{0000-0003-0014-519X}\inst{\ref{inst:0002}}
\and         J.~                         Zorec\inst{\ref{inst:0089}}
\and         E.~                         Utrilla\inst{\ref{inst:0014}}
\and J.~
Bakker\inst{\ref{inst:0027}}
}

   \institute{
   Université Côte d'Azur, Observatoire de la Côte d'Azur, CNRS, Laboratoire Lagrange, Nice, France.   \label{inst1}
\and CIGUS CITIC - Department of Computer Science and Information Technologies, University of A Coru\~{n}a, Campus de Elvi\~{n}a s/n, A Coru\~{n}a, 15071, Spain\relax                                                                                                                                                                                             \label{inst2}\vfill  
 \and CIGUS CITIC, Department of Nautical Sciences and Marine Engineering, University of A Coru\~{n}a, Paseo de Ronda 51, 15071, A Coru\~{n}a, Spain\relax                                                                                                                                                                                                          \label{inst:0066}\vfill
       \and
 Royal Observatory of Belgium, 3 avenue circulaire, 1180 Brussels, Belgium    \label{inst3}
       \and 
  Mullard Space Science Laboratory, University College London, Holmbury St Mary, Dorking, Surrey, RH5 6NT, United Kingdom     \label{inst4}
    \and
    CNES Centre Spatial de Toulouse, 18 avenue Edouard Belin, 31401
Toulouse Cedex 9, France    \label{inst5}
\and 
ATG Europe for European Space Agency (ESA), Camino bajo del Castillo, s/n, Urbanizacion Villafranca del Castillo, Villanueva de la Ca\~{n}ada, 28692 Madrid, Spain\relax \label{inst:0027}\vfill
\and 
Max Planck Institute for Astronomy, K\"{ o}nigstuhl 17, 69117 Heidelberg, Germany\relax \label{inst:0007}\vfill
\and 
National Observatory of Athens, I. Metaxa and Vas. Pavlou, Palaia Penteli, 15236 Athens, Greece\relax                                                          \label{inst:0012}\vfill
\and Laboratoire d'astrophysique de Bordeaux, Univ. Bordeaux, CNRS, B18N, all{\'e}e Geoffroy Saint-Hilaire, 33615 Pessac, France\relax   \label{inst:0030}\vfill
\and INAF - Osservatorio Astrofisico di Catania, via S. Sofia 78, 95123 Catania, Italy\relax                                                                                                                                                                                                                                                                       \label{inst:0013}\vfill
\and Telespazio for CNES Centre Spatial de Toulouse, 18 avenue Edouard Belin, 31401 Toulouse Cedex 9, France\relax                                                                                                                                                                                                                                                 \label{inst:0031}\vfill
\and Dpto. de Matem\'{a}tica Aplicada y Ciencias de la Computaci\'{o}n, Univ. de Cantabria, ETS Ingenieros de Caminos, Canales y Puertos, Avda. de los Castros s/n, 39005 Santander, Spain\relax                                                                                                                                                                   \label{inst:0032}\vfill
\and GEPI, Observatoire de Paris, Universit\'{e} PSL, CNRS, 5 Place Jules Janssen, 92190 Meudon, France\relax                                                                                                          \label{inst:0034}\vfill
\and Centre for Astrophysics Research, University of Hertfordshire, College Lane, AL10 9AB, Hatfield, United Kingdom\relax                                                                                                                                                                                                                                         \label{inst:0038}\vfill
\and INAF - Osservatorio Astrofisico di Torino, via Osservatorio 20, 10025 Pino Torinese (TO), Italy\relax                                                                                                                                                                                                                        \label{inst:0019}\vfill
\and Institut d'Astrophysique et de G\'{e}ophysique, Universit\'{e} de Li\`{e}ge, 19c, All\'{e}e du 6 Ao\^{u}t, B-4000 Li\`{e}ge, Belgium\relax                                                                                                                                                                                                                    \label{inst:0018}\vfill
\and 
Theoretical Astrophysics, Division of Astronomy and Space Physics, Department of Physics and Astronomy, Uppsala University, Box 516, 751 20 Uppsala, Sweden\relax                                                                                                                                                                                           \label{inst:0045}\vfill
\and European Space Agency (ESA), European Space Astronomy Centre (ESAC), Camino bajo del Castillo, s/n, Urbanizacion Villafranca del Castillo, Villanueva de la Ca\~{n}ada, 28692 Madrid, Spain\relax                                                                                                                                                             \label{inst:0046}\vfill
\and Data Science and Big Data Lab, Pablo de Olavide University, 41013, Seville, Spain\relax                                                                                                                                                                                                                                                                       \label{inst:0047}\vfill
\and Department of Astrophysics, Astronomy and Mechanics, National and Kapodistrian University of Athens, Panepistimiopolis, Zografos, 15783 Athens, Greece\relax                                                                                                                                                                                                  \label{inst:0051}\vfill
\and Dipartimento di Fisica e Astronomia ""Ettore Majorana"", Universit\`{a} di Catania, Via S. Sofia 64, 95123 Catania, Italy\relax                                                                                                                                                                                                                               \label{inst:0057}\vfill
\and LESIA, Observatoire de Paris, Universit\'{e} PSL, CNRS, Sorbonne Universit\'{e}, Universit\'{e} de Paris, 5 Place Jules Janssen, 92190 Meudon, France\relax                                                                                                                                                                                                   \label{inst:0058}\vfill
\and Universit\'{e} Rennes, CNRS, IPR (Institut de Physique de Rennes) - UMR 6251, 35000 Rennes, France\relax                                                                                                                                                                                                                                                      \label{inst:0059}\vfill
\and Niels Bohr Institute, University of Copenhagen, Juliane Maries Vej 30, 2100 Copenhagen {\O}, Denmark\relax                                                                                                                                                                                                                                                    \label{inst:0062}\vfill
\and DXC Technology, Retortvej 8, 2500 Valby, Denmark\relax                                                                                                                                                                                                                                                                                                        \label{inst:0063}\vfill
\and Aurora Technology for European Space Agency (ESA), Camino bajo del Castillo, s/n, Urbanizacion Villafranca del Castillo, Villanueva de la Ca\~{n}ada, 28692 Madrid, Spain\relax                                                                                                                                                                               \label{inst:0014}\vfill
\and IPAC, Mail Code 100-22, California Institute of Technology, 1200 E. California Blvd., Pasadena, CA 91125, USA\relax                                                                                                                                                                                                                                           \label{inst:0067}\vfill
\and IRAP, Universit\'{e} de Toulouse, CNRS, UPS, CNES, 9 Av. colonel Roche, BP 44346, 31028 Toulouse Cedex 4, France\relax                                                                                                                                                                                                                                        \label{inst:0068}\vfill
\and Thales Services for CNES Centre Spatial de Toulouse, 18 avenue Edouard Belin, 31401 Toulouse Cedex 9, France\relax                                                                                                                                                                                                                                            \label{inst:0069}\vfill
\and Dpto. de Inteligencia Artificial, UNED, c/ Juan del Rosal 16, 28040 Madrid, Spain\relax                                                                                                                                                                                                                                                                       \label{inst:0017}\vfill
\and INAF - Osservatorio astronomico di Padova, Vicolo Osservatorio 5, 35122 Padova, Italy\relax  \label{inst:0002}\vfill
\and Institute of Global Health, University of Geneva\relax                                                                                                                                                                                                                                                                                                        \label{inst:0084}\vfill
\and Applied Physics Department, Universidade de Vigo, 36310 Vigo, Spain\relax                                                                                                                                                                                                                                                                                     \label{inst:0085}\vfill
\and Sorbonne Universit\'{e}, CNRS, UMR7095, Institut d'Astrophysique de Paris, 98bis bd. Arago, 75014 Paris, France\relax         \label{inst:0089}\vfill
}
   \date{Received March ??, 2022; accepted May, 17, 2022}

 
  \abstract
   {The chemo-physical parametrisation of stellar spectra is essential for understanding the nature and evolution of stars and of Galactic stellar populations. A worldwide observational effort from the ground has provided, in one century, an extremely heterogeneous collection of chemical abundances for about  two million stars in total, with fragmentary sky coverage. }
   {This situation is revolutionised by the \Gaia third data release (DR3), which contains the parametrisation of Radial Velocity Spectrometer (RVS) data performed by the General Stellar Parametriser-spectroscopy,
   \gspspec, module. Here we describe the parametrisation of the first 34 months of \Gaia RVS observations.}
   {\gspspec\ estimates the chemo-physical parameters from  combined RVS spectra of single stars, without additional inputs from   astrometric, photometric, or spectro-photometric BP/RP data. The main analysis workflow described here,  MatisseGauguin, is based on projection and optimisation methods and provides the stellar atmospheric parameters;  the individual chemical abundances of N, Mg, Si, S, Ca, Ti, Cr, \FeI, \FeII, Ni, Zr, Ce and Nd; the differential equivalent width of a cyanogen line; and the parameters of a diffuse interstellar band (DIB) feature. Another workflow, based on an artificial neural network (ANN) and referred to with the same acronym, provides a second set of atmospheric  parameters that are useful for classification control. For both workflows, we implement a detailed quality flag chain considering different error sources.}
   {With about 5.6 million stars, the \Gaia DR3 \gspspec\ all-sky catalogue is the largest compilation of stellar chemo-physical parameters ever published and the first one from space data. Internal and external biases have been studied taking into account the implemented flags. In some cases, simple calibrations with low degree polynomials are suggested.
   The homogeneity and quality of the estimated parameters enables chemo-dynamical studies of Galactic stellar populations, interstellar extinction studies from individual spectra, and clear constraints on stellar evolution models. We highly recommend that users adopt the provided quality flags for scientific exploitation.}
   {The \Gaia DR3 \gspspec\ catalogue is a major step in the scientific exploration of Milky Way stellar populations. It will be followed by increasingly large and  higher quality catalogues in future data releases, confirming the \Gaia promise of a new Galactic vision. }

   \keywords{Stars: fundamental parameters; Stars: abundances; ISM: lines and bands; Galaxy: stellar content; Galaxy: abundances; Methods: data analysis
               }

   \maketitle
%

\section{Introduction}

 The chemo-physical characterisation of stars is at the core of stellar physics and Galactic studies, but also, through the analysis of unresolved stellar populations, of extragalactic physics.  Stellar spectra encode a wealth of information that we have now learned to decrypt. The light emitted by a star is absorbed by the atoms and molecules present in its own atmosphere. This creates spectral absorption lines whose profiles depend on the physical properties of the star and the abundances of the different absorbing chemical species. 
 Our understanding of stellar spectra, used to decode the enclosed information on the nature of stars, relies on a complex and extensive theoretical framework, including (among others) nuclear, atomic, and molecular physics, stellar atmosphere physics, element nucleosynthesis, and radiative transfer theory. 
 
 Before development of the necessary background  theoretical knowledge, stellar spectra motivated the definition of stellar types and luminosity classes. These were the fruit of a classification effort categorising stars based on the identification and strength of their spectral features. Therefore, the chemo-physical parametrisation of stellar spectra has its roots in the large observational campaigns of the beginning of the 20th century \citep[cf.][]{Cannon1928} and the seminal works leading to the Morgan-Keenan classification \citep[cf.][]{MK1943}.

The development of CCD detectors and, more recently, of multiobject facilities has resulted in the ability of even small telescopes to acquire large numbers of stellar spectra. Pioneering projects such as the Geneva Copenhaguen Survey \citep{GCS}, RAVE \citep{RAVE}, and SEGUE \citep{Segue}, followed by archival parametrisation projects like AMBRE \citep{AMBRE12} and a worldwide observational effort illustrated by the {\it Gaia}-ESO Survey \citep{GES}, LAMOST \citep{Lamost}, APOGEE \citep{Apogee}, and GALAH \citep{Galah}  characterise era of Galactic spectroscopic surveys. In parallel to the above-mentioned ground-based efforts, the design and preparation of the \Gaia space mission, including the Radial Velocity Spectrometer  \citep[RVS; for a historical overview see][and references therein]{Katz04, Cropper18}, opened new horizons in the observation of  Milky Way stellar populations, and delivered on the promise of an unprecedentedly extensive spectroscopic survey \citep[][]{Wilkinson2005}.
 
 This rapid evolution of observational capabilities  brought to the fore the need for automated parametrisation tools, enabling fast and homogeneous processing of extensive data sets. Once again, pioneering efforts followed the trail of the first spectroscopic surveys   \citep[e.g.][among others]{Ferre} and the \Gaia space project \citep{2006MNRAS.370..141R}. A variety of mathematical approaches have been developed and applied since then. These include different optimisation, projection, and classification methods used as part of model-driven or data-driven approaches \citep[see e.g.][and references therein]{LijangReview,Carlos16,Jofre19}. 
 
 \Gaia observations started on 25 July 2014. The wavelength range covered by the RVS is $[846 - 870]$~nm, and its medium resolving power is $R=\lambda / \Delta \lambda \sim 11\,500$ \citep{Cropper18}. 
 The present work describes the parametrisation of the first 34 months of \Gaia RVS observations by the General Stellar Parametriser from spectroscopy   (\gspspec) module of the Astrophysical parameters inference system \citep[Apsis,][]{Apsis}.  
 Apsis is the heritage of the previously described scientific pathway and the outcome of a long-term effort: from the development of innovative methodologies \citep[][]{2006MNRAS.370..141R,Albert10,2012StatMethod.9.55B,GeorgesCaII} to their integration into the \Gaia Data Processing and Analysis Consortium (DPAC) framework \citep{Apsis2013}, their tailoring to {\it Gaia}/RVS prelaunch characteristics \citep{2016A&A...585A..93R, Dafonte16} and their first publication as part of the \Gaia Data Release 3 \citep[DR3;][]{GDR3}. This effort results in the largest catalogue of stellar chemo-physical parameters ever published, which is simultaneously the first of its kind from a space spectroscopic survey and with all-sky coverage. 

Section~\ref{Sec:Goals} presents the \gspspec\ goals and output parameters. This is followed by a description of the input \Gaia RVS data (Sect.~\ref{Sect:InputData}) and reference synthetic spectra grids (Sect.~\ref{Sec:SyntheticGrids}). The spectral line selection used for individual abundance analysis is explained in Sect.~\ref{Sec:Lines}. The two \gspspec\ analysis workflows, MatisseGauguin and artificial neural network (ANN), are described in detail in Sects.~\ref{Sec:MATISSEGAUGUIN} and~\ref{Sec:ANN}, respectively. Section~\ref{sec:val} presents the performed validation of \gspspec\ outputs as part of \Gaia DR3 operations, and defines the implemented quality flags. Section~\ref{sec:biases} is devoted to the comparison of \gspspec\ results to literature data and suggested calibrations. Finally, in Sects.~\ref{Sec:Results} and~\ref{sec:Conclusions} we present the \gspspec\ catalogue and our conclusions.


\section{Goals and outputs of \gspspec} \label{Sec:Goals}
The \gspspec\ module implements a purely spectroscopic treatment. It estimates the stellar  chemo-physical parameters from combined RVS spectra of single stars. No additional information is considered from astrometric, photometric, or spectro-photometric BP/RP data.\footnote{A separate Apsis module, the GSP from photometry is in charge of the stellar parametrisation from BP/RP data, using constraints from astrometric and stellar isochrones \citep{GSPphot}.}

In particular, \gspspec\ estimates (i) the stellar effective temperature \T, reported as \verb|teff_gspspec|; (ii) the stellar surface gravity expressed in logarithm \g\footnote{$g$ being in cm.s$^{-2}$}, reported as \verb|logg_gspspec|; (iii) the stellar mean metallicity \meta\footnote{In the following, we adopt the standard abundance notation for a given element $X$: [X/H] $= \log{({\rm X/H})_\star} - \log{({\rm X/H})_\odot}$, where (X/H) is the abundance by number, and $\log \epsilon(X) \equiv \log{({\rm X/H})} + 12$.} defined as the solar-scaled abundances of all elements
heavier than He and reported as \verb|mh_gspspec|; (iv)  the enrichment of $\alpha$-elements\footnote{O, Ne, Mg, Si, S, Ar, Ca, and Ti are considered as $\alpha$-elements and  vary in lockstep.} with respect to iron (\alphaFe), reported as \verb|alphafe_gspspec|; (v) the individual abundances of 13 chemical species ([N/Fe], [Mg/Fe], [Si/Fe], [S/Fe], [Ca/Fe], [Ti/Fe], [Cr/Fe], \FeIM, \FeIIM\footnote{\FeI\ and \FeII abundance enhancements with respect to the mean metallicity are estimated and respectively called {\it fem$\_$gspspec} and {\it feIIm$\_$gspspec} in the {\it AstrophysicalParameters} table.}, [Ni/Fe], [Zr/Fe], [Ce/Fe], [Nd/Fe], (reported as \verb|Xfe_gspspec| or \verb|Xm_gspspec|, with X being the chemical species), including the number of used spectral lines (\verb|Xfe_gspspec_nlines|) and the line-to-line scatter ( \verb|Xfe_gspspec_linescatter|); (vi) a CN differential abundance proxy reported as \verb|cnew_gspspec|; (vii) the equivalent width (EW) and fitting parameters of the diffuse interstellar band (DIB) at 862\,nm, reported as \verb|dibew_gspspec| \verb|dibp1_gspspec| and \verb|dibp2_gspspec|; (viii) a goodness-of-fit ($gof$) over the entire spectral range reported as \verb|logchisq_gspspec|; and (ix) a quality flag chain (the use of which is highly recommended) considering different error sources affecting the output parameters and reported as \verb|flags_gspspec|.  

Two different procedures, MatisseGauguin and ANN, described in  \cite{2016A&A...585A..93R}, are applied to estimate the stellar atmospheric parameters (\T, \g, \meta,\ and \alphaFe). Individual chemical abundances and DIB parameters are estimated by the GAUGUIN algorithm \citep{2016A&A...585A..93R} and Gaussian fitting methods \citep{2021A&A...645A..14Z}, respectively, and are only produced by the MatisseGauguin analysis workflow\footnote{It is worth mentioning that MatisseGauguin algorithms have been conceived  assuming a white Gaussian noise framework}. The goodness-of-fit and the quality flag chain are provided for both the MatisseGauguin and ANN parametrisation.
It is worth noting that, for each star, parameter uncertainties are estimated from 50 Monte-Carlo  realisations\footnote{This number of realisations has been optimised through simulations to ensure a good sampling of the associated parameter distributions, and taking into account the computation time allocated to \gspspec. We note that this Monte-Carlo procedure does not take into account uncertainties in the radial velocity correction, which have been considered through analysis flags (c.f. \ref{subsec:vrad}).}  of its RVS spectrum, considering flux uncertainties. For each realisation, a new complete analysis is implemented, including atmospheric parameters, individual chemical abundances, and CN and DIB parameters. From this analysis, upper and lower confidence values are respectively provided from the 84th and 16th quantiles of the resulting parameter and abundance distributions  and reported with the suffix \verb|_upper| and \verb|_lower|, respectively (cf. Sect.~\ref{sec:mg_mcmc}).

\begin{figure}[t]
\includegraphics[width=0.5\textwidth]{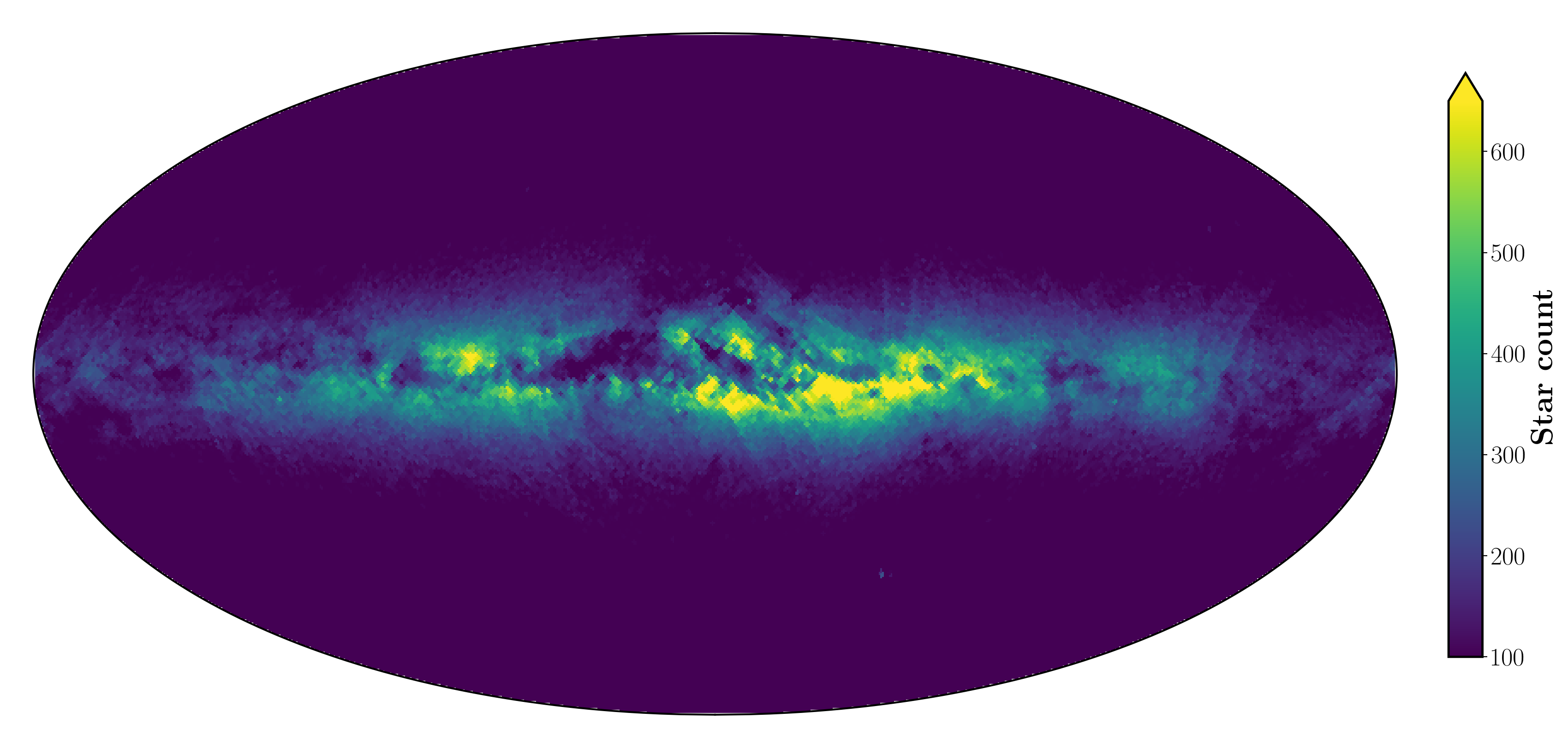}
\caption{Global all-sky spatial density  distribution of all the \gspspec\ parametrised stars. This HEALPix map in Galactic coordinates has a spatial resolution of 0.46$^\circ$ and at least 100 stars are contained in each resolution element. }
\label{Fig:AllSky}
\end{figure}

\begin{figure}[t]
\includegraphics[width=0.5\textwidth]{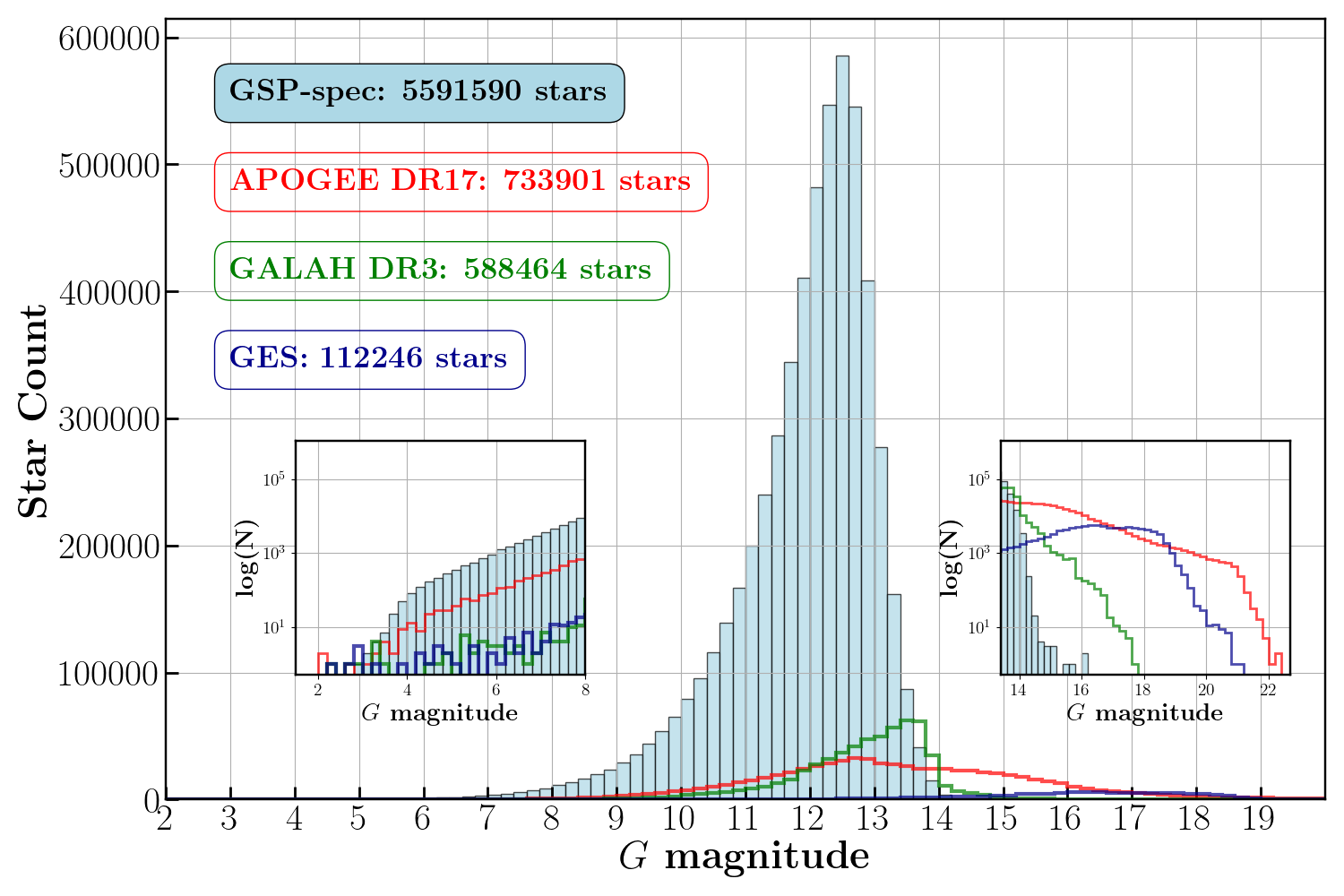}
\caption{{\it Gaia}-magnitude distribution of all the \gspspec\ parametrised stars. The APOGEE, GALAH, and GES magnitude distributions are shown for comparison in red, green, and blue, respectively.}
\label{Fig:HistoMag}
\end{figure}

The DR3 \gspspec\ analysis is available through two archive tables:  
the MatisseGauguin workflow provides 101 fields for 5\,594\,205  stars in the {\it AstrophysicalParameters} table, and the ANN workflow  provides 13 fields for 5\,524\,387 stars in the {\it AstrophysicalParametersSupp} table with an added \verb|_ann| suffix in the parameter names.  Figure~\ref{Fig:AllSky} shows the spatial distribution in ($l,b$) Galactic coordinates of all the \gspspec\ parametrised stars. One can see that most stars are located close to the Galactic plane, as expected, although larger latitudes are still very well sampled with at least 100 stars per resolution element. The small-scale structures close to the Galactic plane are caused by  interstellar absorption. Figure~\ref{Fig:HistoMag} illustrates the magnitude distribution of all the \gspspec\ parametrised stars in the $G$-band. The parametrised stars can be seen to cover a large range of magnitudes, starting from the brightest objects (about 4\,000 of them have $G$<6, i.e. about two-thirds of the sky visible to the naked eye) to the faintest ones up to $G\sim$16 (more than half a million and $\sim$100\,000 have $G$>13 and  $G$>13.5, respectively). 
This very high number statistics can also be appreciated for the magnitude bins with the highest number of stars. For instance,  the bin 12.4$\le G$-mag<12.6 contains as many stars as published by the large ground-based spectroscopic survey GALAH.
For comparison,  Fig.~\ref{Fig:HistoMag} also shows the magnitude distributions of the largest ground-based spectroscopic surveys whose spectral resolution is larger than the RVS one: APOGEE-DR17 \citep{APOGEEDR17}, GALAH-DR3 \citep{Buder2021}, and the  Gaia-ESO Survey (GES) \citep{GilmoreGES, RandichGES}. The highest number statistics of the Gaia \gspspec\ catalogue is achieved for G$<$13.6~mag. For magnitudes fainter than about G$\sim$14.0, APOGEE dominates with about 100 000 stars. GES also complements Gaia DR3 data at such fainter magnitudes  with several tens of thousands, while GALAH has only a few thousand stars fainter than this data release of \gspspec.
We note that the number of stars parametrised by \gspspec\  will strongly increase with the next \Gaia data releases, being about a factor ten larger in DR4 as a result of the spectra signal-to-noise ratio (S/N) increase with repeated observations (and hence with observing time). 

The \gspspec\ analysis module is coded in Java following DPAC requirements, and is executed at the Data Processing Centre C hosted by the Centre National d'Etudes Spatiales (CNES) in Toulouse, France. During DR3 operations, about 6.9 million spectra were processed by the module in $\sim$110\,000~hours,
spread over $\sim$2\,100 cores (execution time of around 130~h, all the cores not
being fully dedicated to \gspspec). The necessary RAM to run \gspspec\ is 25-30~GB. 
Therefore, the total execution time 
to derive the two sets (MatisseGauguin and ANN) of four atmospheric parameters, the 13 individual chemical abundances, the CN differencial abundance proxy, the DIB fitting parameters, and all the associated uncertainties and goodness of fit is about one second per spectrum for one Monte-Carlo realisation of the noise.

An illustration of the \gspspec\ parameterisation was published as a \href{https://www.cosmos.esa.int/web/gaia/iow_20210709}{\Gaia Image of the Week}\footnote{https://www.cosmos.esa.int/web/gaia/iow$\_$20210709}. \gspspec\ parameters are also used in the \Gaia DR3 chemical cartography analysis \citep{Recio22, Schultheis22}. \gspspec\ is the main spectroscopic parametriser module of the \Gaia Apsis pipeline, independent of other modules, and feeds  some of them executed  afterwards in the module chain. The \gspspec\ methodology was largely tested on ground-based spectroscopic observations resulting from different projects, such as RAVE \citep{RAVE}, GES \citep{Recio-Blanco14}, and AMBRE \citep{AMBRE13}, among others.

\section{Input \Gaia RVS data}
\label{Sect:InputData}
  As  input, \gspspec \ uses  combined  RVS  spectra (averaged over multiple transits) and  their flux uncertainties  per  wavelength pixel (wlp) over the 846-870~nm spectral domain. Prior to the \gspspec\ module operations, the stellar radial velocity \citep[\Vrad,][]{Katz22} is used to Doppler shift RVS CCD spectra to the rest frame  before combining them into a mean RVS spectrum \citep[][]{Seabroke22}. The actual RVS wavelength range extends into the filter wings \citep[845-872~nm, see][Fig.~16]{Cropper18}, and the cut to 846-870~nm minimises border effects. In addition, 
the  spectra  are normalised at the local pseudo-continuum and are resampled to a wavelength bin width of 0.01~nm (2400 wavelength points, $wlp$ hereafter) by the DPAC/{Coordination Unit}6 (CU6) pipelines. 

It is important to note that \gspspec\ reassesses the continuum placement during the parameterisation procedure (see Sect.~\ref{SectNorm}). Moreover,  the  spectra  are  rebinned  from  2400  to  800 $wlp$, sampled every 0.03~nm (without  reducing  the  spectral  resolution  thanks  to  the RVS  oversampling),  which increases  their  S/N.  The  RVS  spectra  analysed  by \gspspec\ during DR3  operations  were  selected  to  have  S/N>20  before  resampling. The considered S/N corresponds to the \verb|rv_expected_sig_to_noise| value provided by the CU6 analysis \citep[][]{Seabroke22}.

It is worth mentioning that, although the mean RVS spectra serve as an  input to \gspspec, subsequent filtering of \Vrad\  was not propagated to \gspspec\ outputs for DR3.  This means that there are a very small number of stars with \gspspec\ parameters, but not \Vrad\ (Appendix~\ref{Sect:rv}).  
A subset of RVS mean spectra (999\,995, all having \Vrad) are published for the first time in DR3 \citep[][]{Seabroke22}. These articles detail the overlap of the published mean RVS spectra with \gspspec\ parameters (and other \Gaia parameters).

\section{Input and training synthetic spectra grids} 
\label{Sec:SyntheticGrids}
\begin{figure*}
\sidecaption
\includegraphics[height=10.cm,width=12.cm]{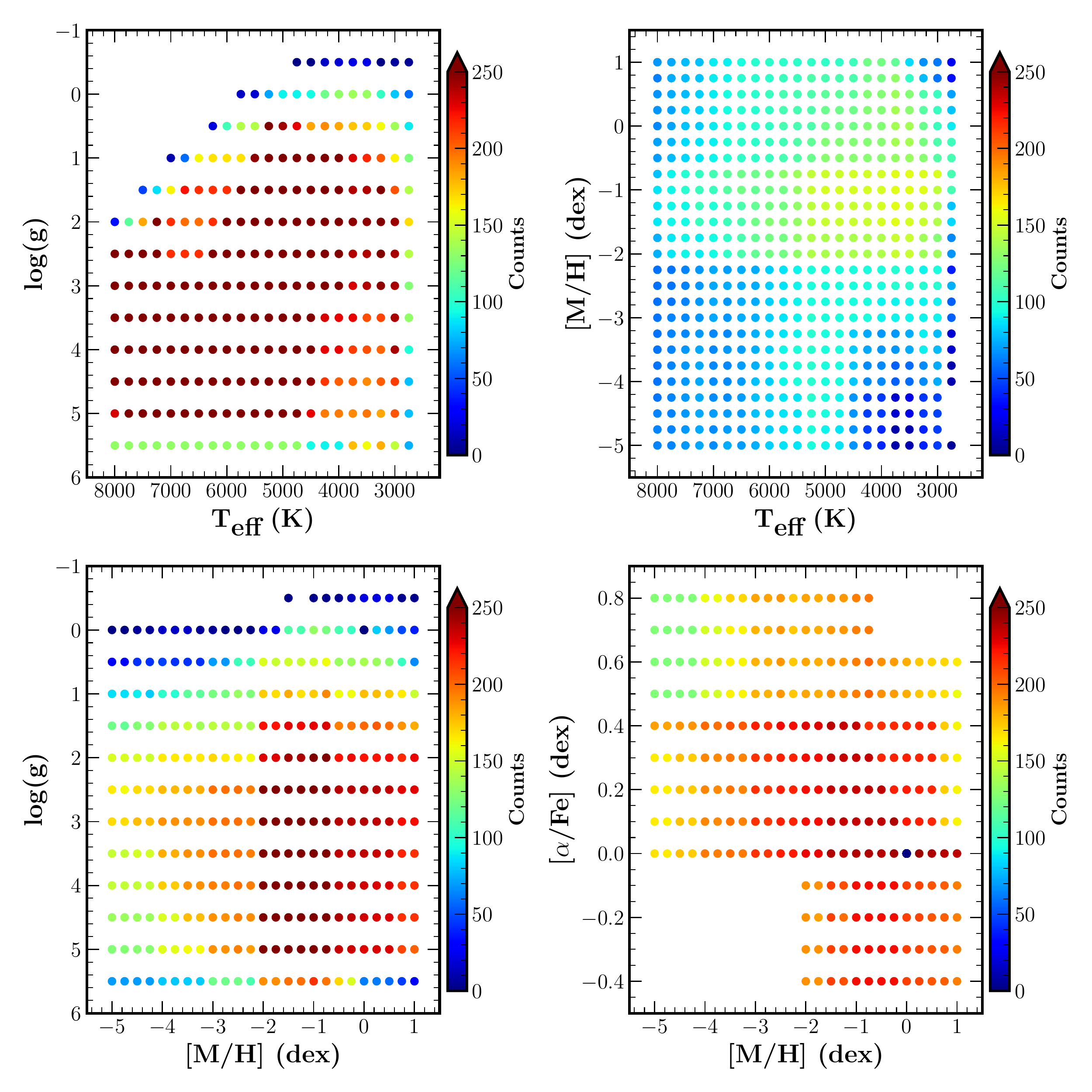}
\caption{Distribution in the 4-D parameter space of the \gspspec \ reference grid, that contains the 51,373 synthetic spectra adopted for the stellar parametrisation. The colour-code refers to the number of available spectra in each 2-D projection. For the derivation of the chemical
abundance of a given chemical element $X$ with the GAUGUIN method, 21 spectra are computed for most combinations of the four atmospheric parameters by varying the individual abundance of $X$ (12 different species
were considered: N, Mg, Si, S, Ca, Ti, Cr, Fe, Ni, Zr, Ce, Nd).}
\label{Fig.Grid}
\end{figure*}
\gspspec\ performs a model-driven parametrization for which stellar flux dependencies on atmospheric parameters and surface chemical abundances are interpreted through the comparison of the observed spectra with theoretical ({\it synthetic}) ones. For this purpose, we have computed large grids of synthetic RVS spectra with different combinations of  stellar atmospheric parameters (\T, \g, \meta\  and \alphaFe) and individual chemical abundances ([X/Fe], with $X$ being the considered element, with the exception of \FeI\ and \FeII\ for which [X/M] is used). They span the entire parameter space of Galactic stellar populations with a detailed coverage that allows to reach the required parametrization precision. The use of these grids is three-fold: i) training the \gspspec\ MATISSE (cf. Sect.~\ref{Sec:Matisse}) and ANN (cf. Sect.~\ref{Sec:ANN}) algorithms before their application; ii) acting as reference models for the algorithm performing on-the-fly regressions (GAUGUIN), and iii) anchoring the normalization and DIB analysis procedures to reference flux values.

As a consequence, a 4-dimensional grid of spectra in \T, \g, \meta\  and \alphaFe\ (cf. Sect.~\ref{sec:4Dgrid}) and 5-dimensional grids for twelve chemical elements with the fifth dimension being [X/Fe] (cf. Sect.~\ref{sec:5Dgrids}) are provided as input for \gspspec\ together with the learning functions of the parametrisation algorithms. These synthetic spectra are calculated through a procedure previously implemented for the AMBRE Project \citep{AMBRE12}. We refer to a detailed  description of the AMBRE grid to  \citet[][]{AMBRE13}. In the following, we particularly focus on several improvements  considered for the \gspspec\ module.

\subsection{Set of MARCS atmosphere models}
\label{Sec:MARCS}
The reference spectra are computed using MARCS atmosphere models \citep{MARCS}. 
We first selected 13,848 models that covered the following parameter space:
2600 to 8000~K for \T\ in steps of 200 or 250~K (below or above 4000~K, respectively), -0.5 to 5.5 for  \g\ (step of 0.5~dex), and -5.0 to 1.0~dex for the mean metallicity (step of 0.25~dex for \meta$>$-2.0~dex and 0.5~dex for lower \meta\ values). 
For each metallicity, all the available \alphaFe-enrichments
were considered. In practice, this corresponds to models with \alphaFe-values varying between at most -0.4~dex and +0.8~dex, around the classical
relation observed for Galactic populations: \alphaFe = 0.0~dex for \meta $\ge$ 0.0~dex,
\alphaFe = +0.4~dex for \meta $\le$ -1.0~dex and
\alphaFe = -0.4 $\times$ \meta\ for -1.0 $\le$ \meta $\le$ 0.0~dex. 
We point out, however, that not all values of \alphaFe\ were always available for a given set of \T, \g, and \meta.
Moreover, we only selected models for dwarfs (defined as \g$>$3.5) with plane-parallel geometry and a microturbulent-velocity parameter of 1.0~km/s whereas spherical geometry with a mass of 1~M$_{\odot}$ and $V_{\rm micro}$=2~km/s were considered for giants (\g$\le$3.5). 
Then, in order to improve the covering of the parameter space (particularly in the \alphaFe\ dimension for which
we adopted a step of 0.1~dex), we filled this first
selection of MARCS models by models interpolated linearly, 
using the tool developed by T. Masseron and available on the MARCS website\footnote{https://marcs.astro.uu.se/}.
The resulting grid of MARCS atmosphere models adopted in the present work contains 35,803 models.

\subsection{The 4-D spectra grid in  \T, \g, \meta\  and \alphaFe}
\label{sec:4Dgrid}
For each adopted MARCS atmosphere model, a synthetic spectrum has been computed with the TURBOSPECTRUM code \citep[version 19.1.2,][]{Plez2012} between 842.0~nm and 874.0~nm (i.e. a wider spectral domain than the one covered by the RVS spectra, in order not to be affected by border effects when simulating the RVS-like spectra) and adopting an initial wavelength step of 0.001~nm (i.e. corresponding to a spectral resolution larger than $\sim$300,000).
We considered the
Solar abundances of \cite{Grevesse07}, and specific atomic and molecular line lists. These line lists contain millions of lines and have been checked (and, when necessary, some atomic lines were calibrated) with observed spectra of benchmark reference stars  \citep[see][for more details]{Contursi21}. For dwarfs (defined as above for the MARCS models by \g$>$3.5), the spectra were computed assuming one-dimensional plane-parallel atmospheric model while for giants (\g$\le$3.5) a spherical geometry is considered. Both cases assume hydrostatic and local thermodynamic equilibria.
Similar stellar masses as in the MARCS models were adopted for the computation. Moreover, consistent \alphaFe-enrichments were considered in the model atmosphere and the synthetic spectrum calculation. 
Finally, we used an empirical law for the microturbulence parameter. This parametrized relation is a function of \T , \g \ and \meta \ and  has been derived from $V_{\rm micro}$ literature values for the {\it Gaia}-ESO Survey (Bergemann et al., in preparation). The spectra were computed in the air and then converted into vacuum wavelengths thanks to the relation of \citet{Birch94}. It is worth noting that no stellar rotation or macro-turbulence broadening were included in these spectra. The impact of this assumption in the derived stellar parameters has been estimated from simulations and accounted through quality flags (Sect.~\ref{tab:vsiniQF}). These flags are a function of the $vbroad$ parameter value of each star (available in the {\it gaia\_source} table) but also of \T, \g\ and \meta.

The high-resolution spectra were then convolved and resampled
in order to mimic real observed RVS spectra. For that purpose,
we adopted a broadening instrumental profile corresponding to the RVS spectral resolution, keeping only the 846-870~nm domain and adopting the sampling of 0.03~nm chosen for the parametrisation within \gspspec\ (800 $wlp$, see Sect.~\ref{Sect:InputData}). 
In practice, this convolution was performed thanks to tools developed for the DR3 version of the CU6 pipeline \citep{Sartoretti18}. It assumes a Gaussian ALong-scan line spread function and adopts the median resolving power value known at the beginning of CU8's  DR3 processing phase \citep[$R$=11,500,][]{Cropper18}. 

Finally, for the stellar atmospheric parameters estimation (see Sect.~\ref{Sec:MATISSEGAUGUIN} \& \ref{Sec:ANN}), this original grid
of RVS-like synthetic spectra has been filled adopting a cubic Catmull-Rom \citep{CATMULL1974317}, a quadratic or linear 1-D interpolation, depending on the number of neighbour models available. The final 4-D grid contains 51,373 spectra with a constant step of 250~K, 0.5, 0.25~dex and 0.1~dex in \T, \g, \meta \ and \alphaFe, respectively. Figure~\ref{Fig.Grid} illustrates the covered parameter space.

\subsection{5-D spectra grids for individual chemical abundance estimations} \label{sec:5Dgrids}
For the derivation of individual chemical abundances with the GAUGUIN method (Sect.~\ref{Sec:GAUGUIN}), we have computed sets of 5-D grids
for which the first four dimensions are the ones of the 4-D grid described above while the fifth dimension corresponds to the abundance values of a specific chemical species [X/Fe] (with the exception of \FeI\ and \FeII\ for which [X/M] is used). The considered chemical elements, $X$, are N, Mg, Si, S, Ca, Ti, Cr, Fe, Ni, Zr, Ce, Nd. These species have been chosen due to the availability of at least one of their atomic lines in the RVS spectral domain, following a careful line quality selection (see Sec.~\ref{Sec:Lines}). 

For these 5-D grids, we
considered a subsample of the MARCS models selected in Sect.~\ref{Sec:MARCS}: \T $>$3500~K, \meta 
$>$-3.0~dex and any values of \g \ and \alphaFe, except for Ca, Fe and Ti.
Some atomic lines of these three atoms can indeed be detected at the very metal-poor regime and we therefore computed their 5-D grids for any \meta\ values, i.e. down to \meta=$-5.0$~dex. The adopted variations in the chemical element dimension are from -2.0 to +2.0~dex around 
$\epsilon$(X) = $\epsilon$(X)$_\odot$+\meta +$K_\alpha$, with a step of 0.2~dex (i.e. 21 different abundance values). $K_\alpha$ is assumed to be equal to zero for all elements except the $\alpha$-species for which it follows a similar variation with the metallicity as \alphaFe: $K_\alpha$=0.0 for \meta $\ge$ 0.0~dex,
$K_\alpha$=+0.4 for \meta $\le$ -1.0~dex and $K_\alpha$=-0.4 $\times$ \meta\ for -1.0 $\le$ \meta $\le$ 0.0~dex. 

In total, we have computed twelve 5-D grids of $\sim$478,400 spectra each, except for Ca, Fe and Ti whose grids contain $\sim$590,750 spectra since they cover the entire metallicity regime of the atmosphere model grids.

\section{Line and wavelength interval selection for individual abundance analysis}
\label{Sec:Lines}

As mentioned above, the reference synthetic spectra grids contain all the atomic and molecular lines collected by \cite{Contursi21}. Most of these lines are too weak and/or blended and can therefore not easily be used to derive reliable chemical diagnostics. To choose the adequate spectral intervals for individual abundance estimation, we implemented a careful line selection procedure and a thorough definition of the wavelength intervals for abundance estimation and local normalisation described below.

\medskip
\noindent{\it Selection of unblended lines}

\smallskip
\noindent First, we looked for unblended lines through visual inspection of synthetic spectra at high-resolution ($R\sim$100\,000) and at the resolution of RVS ($R\sim$11\,500). The atmospheric parameters of four well-known reference stars were adopted: two cool giants (Arcturus and $\mu$~Leo),  one cool dwarf (the Sun), and one hot dwarf (Procyon)\footnote{ The adopted parameters for these stars can be found in \citet{Contursi21}.}. In particular, we looked at (i) the flux contribution of each chemical species (including the 12 atomic elements and the most abundant molecules) by computing specific spectra with highly enhanced abundances, and (ii) the existing blends assuming super-solar metallicities and high enhancements in $\alpha$-elements. This led to an initial selection of about 130 isolated atomic lines belonging to a dozen different atoms and five CN lines\footnote{In our tests, CN was the sole identified molecule with rather unblended lines but this work has to be extended towards cooler stars (\T < 4,000~K) for future \Gaia releases.} that could be useful for chemical diagnostics.
In particular, we identified interesting lines of some heavy elements (Zr, Ce, and Nd) and one line of singly ionised iron at $\lambda=$~858.794~nm, as suggested by \citet{Contursi21}, to complement iron abundance based on Fe~{\sc i} lines (see Sect.\ref{SecFe2}). The correct simulation of these lines was verified through the comparison of synthetic spectra to high-resolution observed spectra for the four mentioned benchmarks.

Second, the previous selection was confirmed by examining the observed RVS spectra of a few  stars with atmospheric parameters close to those of the reference ones. By visual inspection, we kept only the lines showing the highest sensitivity to abundance variations in at least one of the inspected spectra, excluding those for which  blends were still suspected from lines of different chemical elements within $\sim$0.3~nm. 

\medskip
\noindent{\it Selection of abundance and local normalization windows.}

\smallskip
\noindent
To further optimise the line selection and the chemical analysis procedure, we carefully defined two wavelength windows around the selected lines used in the abundance estimation and the local normalisation, respectively. These were defined after visual inspection of the Arcturus, solar, and Procyon spectra at the RVS resolution, maximising the wavelength domains (and therefore the information on the abundance and the continuum placement) and avoiding nearby lines. To ensure the reliability of the finally selected windows, chemical abundances were derived using GAUGUIN for a set of about 10\,000 RVS spectra and slight variations of the window interval. This allowed us to exclude window definitions producing very discrepant results ($\gtrsim$~0.5 dex) with respect to other lines of the same element and their average value.

For line doublets and triplets, the merger of each line within a single abundance determination and normalisation window was adopted whenever possible. These are referred to as $merged$ $multiplets$ hereafter. In the  particular case of the Ca~{\sc ii} IR triplet, to minimise NLTE effects, two abundance windows were defined at the line wings, avoiding the cores (i.e. up to six independent abundance estimates can be provided for the three Ca~{\sc ii} lines).


\medskip
\noindent{\it Line selection based on line-to-line abundance scatter}

\smallskip
\noindent
Finally, from the above set of unblended lines, we performed an additional selection to optimise the line-to-line scatter. For some species (N, Si, S, Ca, Ti, Cr, Fe~{\sc i}), the final element abundance was computed by combining the results of the different single line (or merged multiplets) abundances of the same element (cf. Sect.~\ref{subsec:combinedAbun})\footnote{ Other derived abundances (Mg, Ni, Fe~{\sc ii}, Zr, Ce, Nd) rely on only a single line or a single abundance determination in the case of merged multiplets.}. For that purpose, we computed 50 Monte-Carlo flux  realisations of the above-mentioned set of RVS spectra, considering their corresponding flux covariances. For each spectral line, the median and the inter-quartile range (IQR) were estimated from the derived abundance distribution. We then explored all the possible line combinations to evaluate the contribution of each line to the mean abundance, as well as its effect on the total number of estimates. A mean abundance is derived for each line combination, weighted by the inverse of the individual line abundance IQR. These weights were set to zero if they had IQR>0.5~dex to avoid low-quality estimations. For each chemical element, the combination of lines minimising the line-to-line scatter and maximising the correlation\footnote{We first performed a linear fit and then obtained the slope, the intercept, the median, and the MAD of the distance |data-fit|. } of the mean abundance with \meta\ (for iron-peak elements) or \alphaFe\ (for $\alpha$-elements) was selected. 

The final list of 33 lines (some being merged multiplets) selected for the individual abundance analysis is provided in Appendix~\ref{Appendix:lines} and \tabref{Table_linelist}, together with their associated windows for chemical analysis and normalisation. We refer also to the two figures of Sect.~\ref{Sec:MATISSEGAUGUIN} for some examples of observed and model spectra that help to identify most of these lines.  We note that Zr, Ce, and Nd lines passed all the above tests and are therefore adopted for abundance estimation. Similarly, the Fe~{\sc ii} line is also conserved. As a consequence, the \gspspec\  module is able to estimate abundances of neutral and singly ionised iron (see Sect.\ref{SecFe2}).

\section{\gspspec \ MatisseGauguin analysis workflow}
\label{Sec:MATISSEGAUGUIN}


This section describes the MatisseGauguin  analysis workflow in sequential order. We reiterate that MatisseGauguin produces the \gspspec\ fields published in the $AstrophysicalParameters$ table, including stellar atmospheric parameters, individual chemical abundances, and DIB parameters.

\begin{figure}[t]
    \begin{center}
    \includegraphics[width=0.5\textwidth]{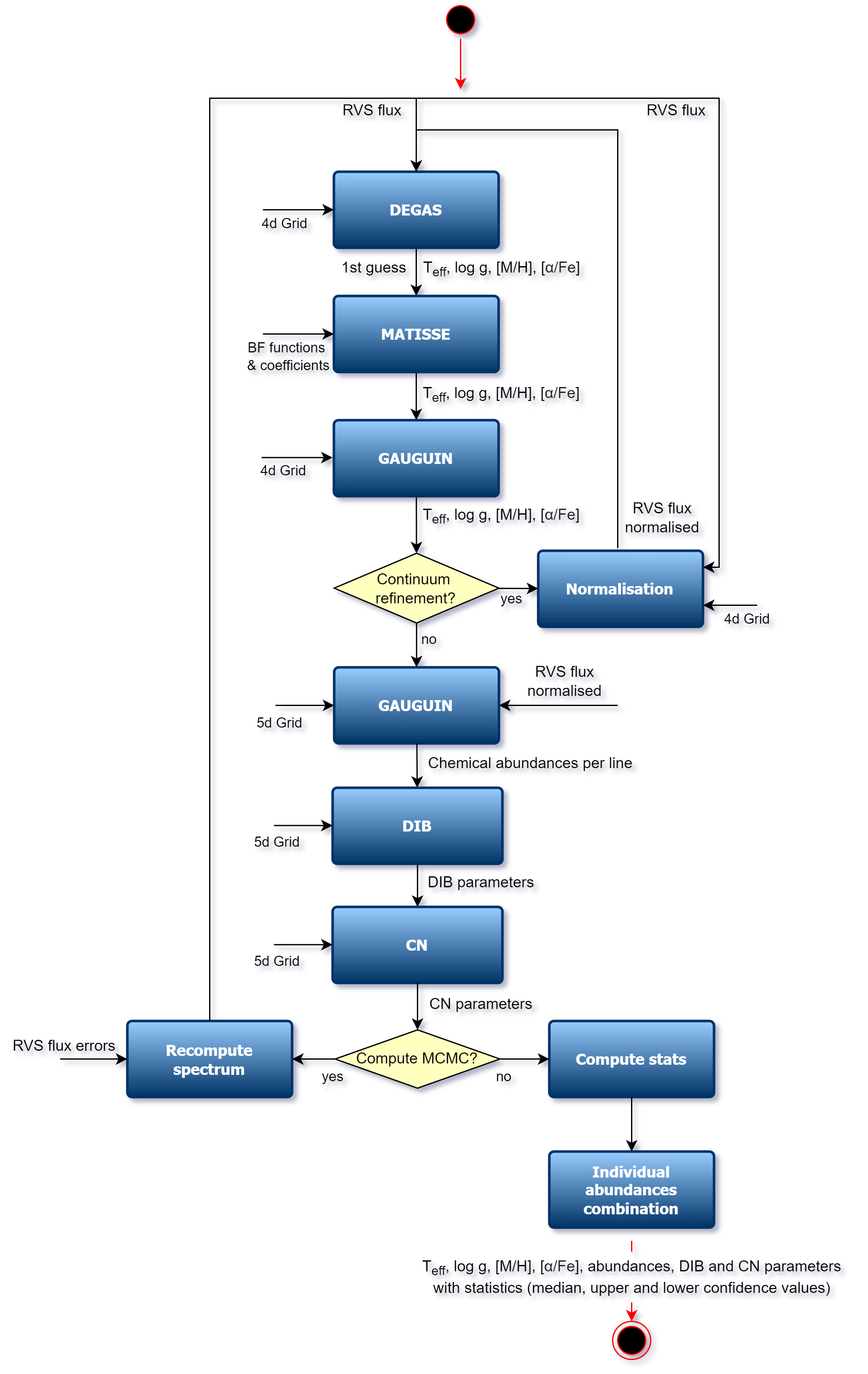}
    \caption{Complete MatisseGauguin workflow that estimated stellar atmospheric parameters (\T, \g, \meta, and \alphaFe), individual chemical abundances of 12 species, CN, and DIB parameters (see Sect.\ref{Sec:MATISSEGAUGUIN} for detailed description). }
    \label{fig:mg_workflow}
    \end{center}
\end{figure}

\begin{figure*}
\includegraphics[width=1.1\textwidth, trim=4cm 0 0 0 ]{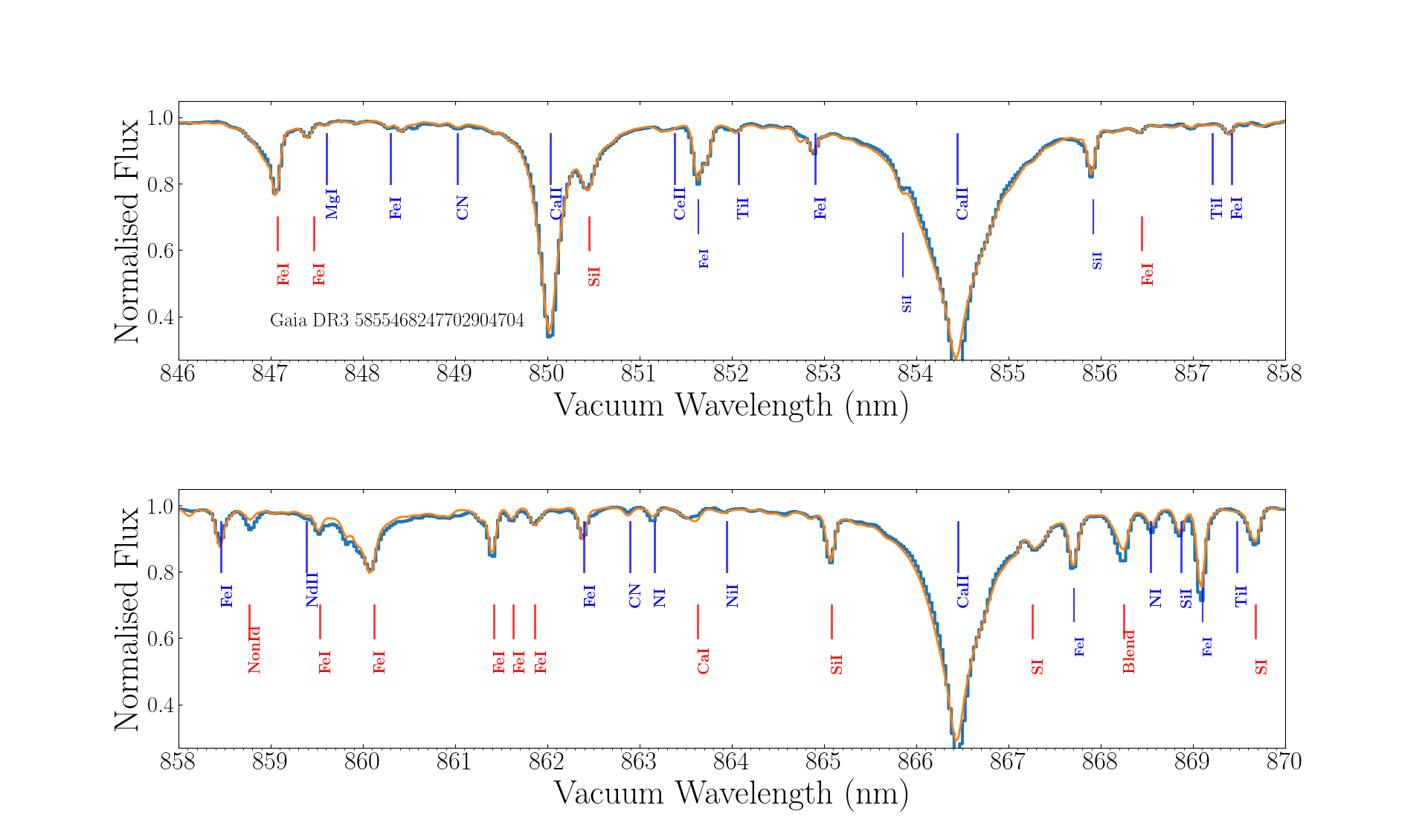}
\caption{Observed (blue histogram) and synthetic (orange line) spectra of the Cepheid variable star \Gaia DR3 5855468247702904704. The observed spectrum has a very high S/N (equal to 884) and its histogram bin size corresponds to the wavelength sampling adopted for the analysis (0.03~nm, 800~$wlp$). The synthetic spectrum was computed from the \gspspec\ MatisseGauguin atmospheric parameters (\T=5477~K, \g=1.44, \meta=0.07~dex, \alphaFe=0.11~dex) and individual chemical abundances, was then convolved by a rotational profile to reproduce the CU6 estimated broadening velocity (15.6~km.s$^{-1}$) and, finally, was degraded to the RVS spectral resolution and sampling. The atomic lines identified in blue belong to the chemical species whose abundances were derived by the GAUGUIN method (the local normalisation performed for the chemical analysis of these selected lines was not considered in the figure for clarity reasons). The lines in red were not analysed in the shown spectrum because of suspected blends in the present case.
The feature around 868.3~nm is a blend of S{\sc i}+Fe{\sc i}+Si{\sc i} plus probably other potential unidentified lines. The NonId feature at $\sim$858.8~nm is a blend of the \FeII\ line described in Sect.~\ref{SecFe2} (seen in orange) and of unidentified lines that cannot be reproduced with the present line list.
}
\label{Fig:FitEx1}
\end{figure*}

\begin{figure*}
\includegraphics[width=1.1\textwidth, trim=4cm 0 0 0]{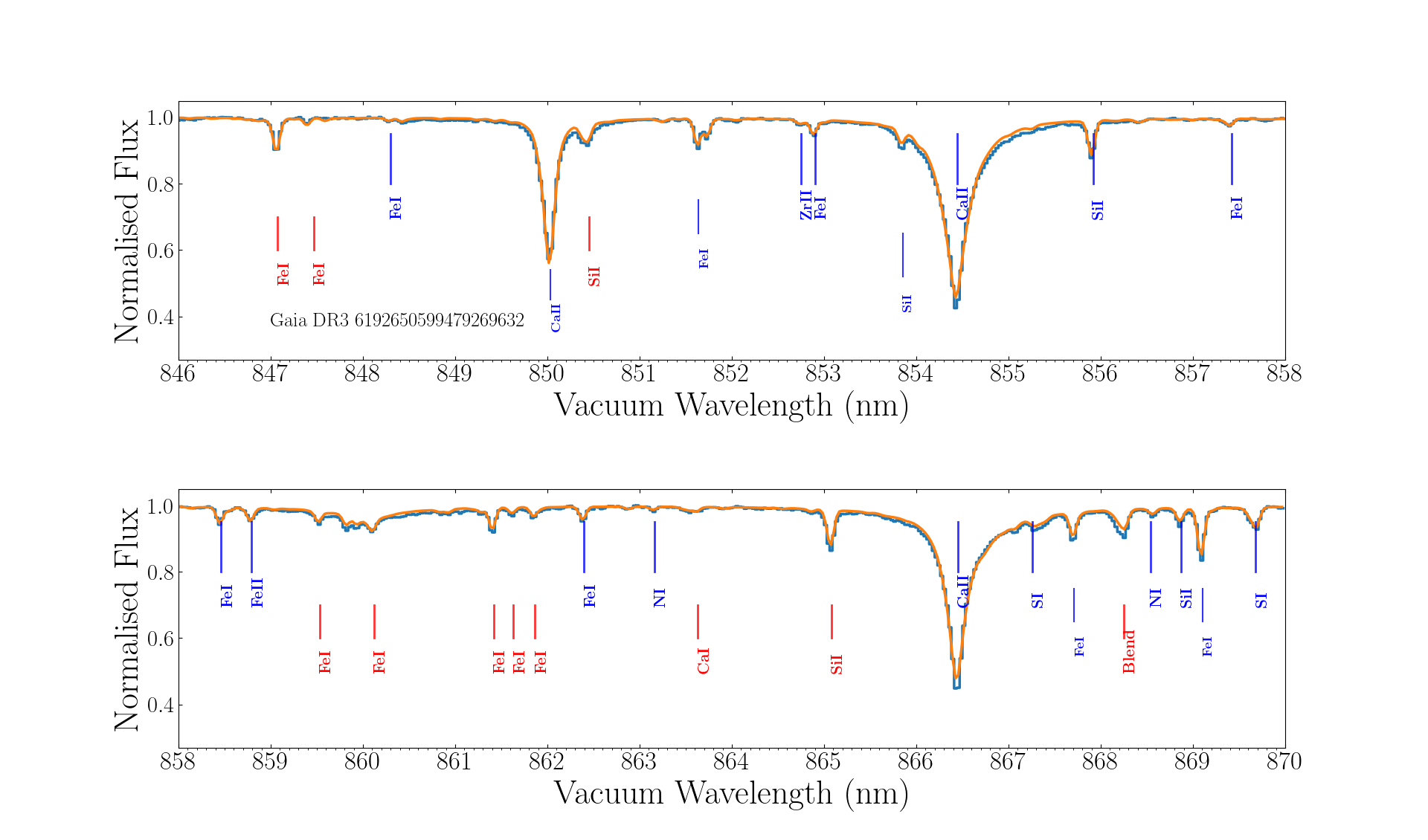}
\caption{Same as Fig.~\ref{Fig:FitEx1} but for the hot dwarf \Gaia DR3 6192650599479269632 whose MatisseGauguin atmospheric parameters are \T=6754~K, \g=4.38, \meta=-0.03~dex, and \alphaFe=0.15~dex (S/N=408). No rotational profile was applied as no broadening velocity was estimated (suspected low-rotating star).}
\label{Fig:FitEx2}
\end{figure*}

The complete workflow of  MatisseGauguin is summarised in Fig.~\ref{fig:mg_workflow}. In addition, to illustrate the MatisseGauguin  parametrisation performance, the challenging automated fit of two observed high-S/N spectra\footnote{These two spectra are not part of the set of RVS spectra published in the \Gaia DR3.} is presented in Figs.~\ref{Fig:FitEx1} and \ref{Fig:FitEx2}. The presented synthetic spectra are computed using the atmospheric parameters and chemical abundances estimated by \gspspec\ for those two stars. The identification of several spectral lines is included in the figures.  It is worth noting that the combination of the automated MatisseGauguin parametrisation with our reference spectra is able to find an excellent match with the observations that confirms the quality and the high precision of the observed RVS spectra and the input reference spectra grids.

\subsection{MATISSE stellar atmospheric parameters} 
\label{Sec:Matisse}
To initialise the whole MatisseGauguin procedure, a first guess of \T, \g, \meta, and \alphaFe\ is derived using the DEGAS decision-tree method \citep{Albert10}, which considers the entire parameter space of the 4D grid \citep[see also][for first applications to observed spectra]{2011A&A...535A.106K, Kordopatis13}. 

Subsequently, the MATISSE algorithm \citep{2006MNRAS.370..141R} is applied following an iterative procedure in the parameter estimation. This allows the user to overcome problems caused by a non-linear variation of the spectral flux with the stellar parameters. MATISSE is a local multi-linear regression method, resulting from the projection of the full input spectrum onto a set of vectors (called BF functions in Fig.~\ref{fig:mg_workflow}). These vectors (and the associated coefficients) account for the sensitivity, at each wavelength, of the stellar flux  to variations of a given parameter  ($\Delta$\T, $\Delta$\g, $\Delta$\meta\ or $\Delta$\alphaFe); they are derived during a training phase based on the noise-free 4D reference grids, and correspond to regions of the entire parameter space, spanning $\pm$500~K in \T, $\pm$0.5~dex in \g, $\pm$0.25~dex in \meta, and $\pm$0.20~dex in \alphaFe. The noise optimisation is taken into account by employing a Landweber algorithm during the covariance matrix inversion and which is adapted to each scientific application \citep[see][for more details]{2006MNRAS.370..141R}.   The MATISSE projection is first applied at the DEGAS solution in a local environment of $\pm$500~K in \T, $\pm$0.5~dex in \g, $\pm$0.25~dex in \meta,\ and $\pm$0.20~dex in \alphaFe\ (corresponding to the parameter space region of each training function). This produces a second solution around which MATISSE is applied again. This iterative procedure is repeated until convergence (i.e. the solution stays within the local environment), within a maximum of ten iterations.

\subsection{GAUGUIN refinement of the atmospheric parameters} \label{Sec:GAUGUINParams}
The GAUGUIN algorithm is then applied around the final MATISSE solution of the previous step, considering a local environment of $\pm$250~K in \T, $\pm$0.5~dex in \g, $\pm$0.25~dex in \meta, and $\pm$0.20~dex in \alphaFe. GAUGUIN \citep{2012StatMethod.9.55B, 2016A&A...585A..93R} is a classical, local optimisation method implementing a Gauss-Newton algorithm. It is based on a local linearisation around a given set of parameters that are associated with a reference synthetic spectrum (via linear interpolation of the derivatives). 
It is designed to find the direction in the parameter space that has the highest negative gradient as a function of  distance (defined as the flux difference between the observed and synthetic spectra). Once this direction is found, the method proceeds in an iterative way, by modifying the initial guess of the studied parameter and re-calculating the gradient again, until convergence of the parameter solution. 
A few iterations are carried out through linearisation around the new solution until the algorithm converges towards the minimum distance. In practice, and to avoid trapping into secondary minima, we recall that GAUGUIN is initialised by parameters independently determined by MATISSE. At the end of this process, the final MatisseGauguin solution in \T, \g, \meta, and \alphaFe\ is provided as input to the spectrum normalisation procedure. 

\subsection{Spectra re-normalisation and iterations on atmospheric parameters}
\label{SectNorm}
The parameter solution of the previous step is used to re-estimate the continuum placement. This step is particularly important in the case of cool stars, which have pseudo-continuum flux values that can be much lower than one. The continuum placement and normalisation procedure is described in detail in \cite{2020A&A...639A.140S}. In this step, the spectrum flux is normalised over the entire RVS wavelength domain. For this purpose, the observed spectrum ($O$) is compared to an interpolated synthetic one from the 4D reference grid ($S$) with the same atmospheric parameters. First, the most appropriate wavelength points of the residuals ($Res$ = $S/O$) are selected using an iterative procedure implementing a linear fit to $Res$ followed by a $\sigma-$clipping. The residual trend is then fitted with a third-degree polynomial. Finally, the refined normalised spectrum is obtained after dividing the observed spectrum by a linear function resulting from the fit of the residuals.

This renormalised spectrum is then fed back to the first step described in Sect.~\ref{Sec:Matisse} to re-estimate the atmospheric parameters using the new spectra normalisation. This loop is performed five times (a sufficient number to reach convergence), iterating on the parameters and the continuum placement. 

The parameters of the converged solution in \T,  \g, \meta, and \alphaFe\ is then saved, as well as the final normalised spectrum. A goodness-of-fit ($gof$) between the observed and a synthetic spectrum interpolated to the atmospheric parameters is computed. The logarithm of this $gof$ is reported in the $AstrophysicalParameters$ table under the $logchisq\_gspspec$ field. The provided $gof$ value reports the goodness of fit with respect to the observed spectrum, not including Monte-Carlo variations of the flux (see Sect.~\ref{sec:mg_mcmc}). 

\subsection{GAUGUIN chemical abundances per spectral line}
\label{Sec:GAUGUIN}
Considering the final atmospheric parameters solution and normalised spectrum, each of the 33 selected atomic lines (see Table~\ref{Table_linelist}) is then analysed with GAUGUIN to estimate the chemical abundance of the related chemical element causing the line absorption.

First, for each line $l$ associated with the chemical element $X$, a specific 1D grid in the $[X/Fe]$ abundance space is generated. To this purpose, the corresponding 5D grid presented in Sect.~\ref{sec:5Dgrids} is interpolated at the stellar \T, \g, \meta, and \alphaFe\ values of the adopted MatisseGauguin solution (cf. Sect.~\ref{SectNorm}). This 1D reference spectra grid covers the entire normalisation wavelength range.  It includes a large range of abundance variations in $\epsilon(X$). Second, a local normalisation around the line is performed \citep{2020A&A...639A.140S}. A minimum quadratic distance is then calculated between the reference grid and the observed spectrum, providing a first guess of the abundance estimate [X/Fe]$_0^l$. This initial guess is then optimised using the GAUGUIN algorithm, which iterates through linearisation around the successive new solutions. The algorithm stops when the relative difference between two consecutive iterations is less than a given value (one-hundredth of the grid abundance step) and provides the final abundance estimation of each line ([X/Fe]$^l$).

\subsection {Diffuse interstellar band parameters}
\label{DIBparam}

\begin{figure}[t]
\includegraphics[width=0.52\textwidth, trim=3cm 0 1cm 2cm]{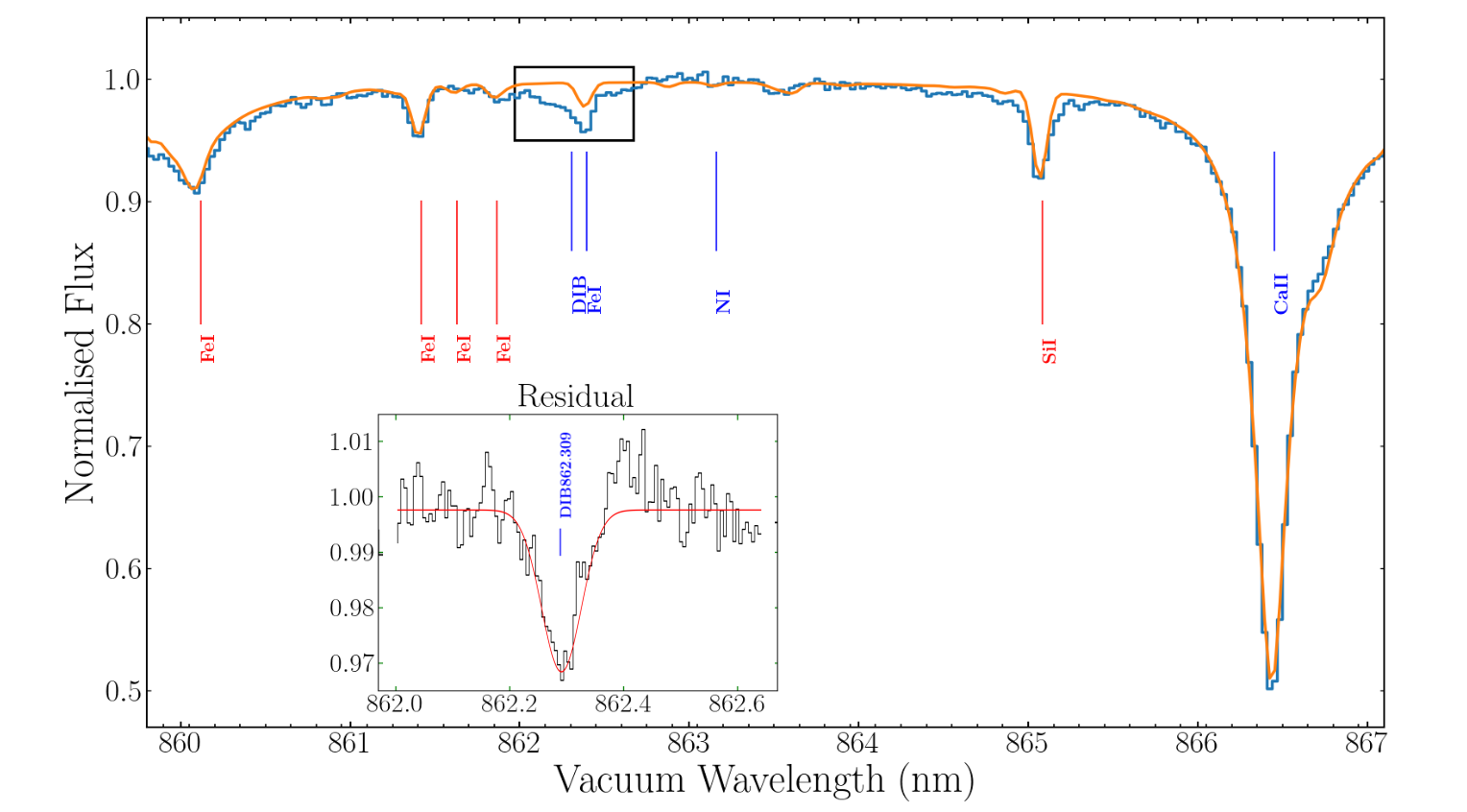}
\caption{Similar to Fig.~\ref{Fig:FitEx1} but for the metal-poor hot subgiant \Gaia DR3 4378933739135936000 around its DIB feature. The insert is a zoom onto the  flux residual between observed and model spectra around the DIB. It has been renormalised and the DIB characteristics are measured thanks to the Gaussian fit shown in red (EW=0.0244~nm and central wavelength $p_1$=862.309~nm). The MatisseGauguin atmospheric parameters of this star are \T=6414~K, \g=3.75, \meta=-0.61~dex, and \alphaFe=+0.42~dex (S/N=293 and CU6 broadening velocity equal to 17.1~km.s$^{-1}$).}
\label{Fig:DIB}
\end{figure}

Once the atmospheric parameters and the individual abundances have been derived, the next step of the MatisseGauguin workflow  is to evaluate the presence of any DIB signature around $\sim$862~nm.
 For each RVS spectrum, we first  perform a  local renormalisation on the spectrum  around the DIB feature (over 35\,{\AA} around 862~nm). We then pass  a preliminary detection
 of the DIB profile  by fitting a Gaussian profile to produce initial guesses for the fitting and eliminate cases where noise is at the same level as or exceeds the depth of the possible detection of the DIB. Only detections above the 3\,$\rm \sigma$-level are considered as true detections. In order to perform the main fitting process of the DIB, we then separate our sample into  cool (3\,500<\T$\leq$ 7\,000\,K) and hot star samples (\T$\geq$ 7\,000\,K). For cool stars, the observed spectrum is divided by a synthetic spectrum whose atmospheric parameters are provided by MatisseGauguin. The residual, assumed to correspond to the DIB profile, is then renormalised and fitted by a Gaussian function (see Fig.~\ref{Fig:DIB}). For hot stars for which no lines are found close to the DIB feature, a Gaussian process similar to \citet{Kos2017} is applied where the DIB profile is fitted by a Gaussian process regression (\citealt{GB12}). 
 
 For each detected DIB feature,  we determine its equivalent width (EW), the central wavelength of the fitted Gaussian ($p_1$), its depth ($p_0$), the width of the Gaussian profile ($p_2$), and their uncertainties which are estimated based on Monte-Carlo Markov Chain realisations (see Sect.~\ref{sec:mg_mcmc}). We remind the reader that, for a Gaussian, $p_2$~=~$FWHM$/(2*sqrt(2*ln 2)) with $FWHM$ being the full width at half maximum. The EW is computed assuming a Gaussian profile: EW~=~$\sqrt 2\pi \times |p_0| \times p_2 / C$ where $C$ is the continuum level.
 
 Finally, two quality flags ($DIB_q$ and $QF$) for the DIB parameters were implemented in order to allow a selection of the best determinations, depending on the science application \citep[see e.g.][]{Schultheis22}. 
 The first quality flag  $DIB_q$ is included in the \gspspec\ quality flag string chain and its value varies from zero (highest quality) to five (lowest quality) and is equal to nine when no DIBs are measured; we refer to Sect.~\ref{subsec:DIBqf} for its definition. 
 The second flag $QF$ is defined during the preliminary detection of the DIB profile and provides the reason why $DIB_q$ has been fixed to nine for a given spectrum.
 If the depth of the fitted profile is smaller than 3-$\sigma$ the noise level, we do not consider this case to be a true detection and assign it $QF$ =-1. Finally, stars with effective temperatures cooler than 3500\,K are automatically disregarded because their spectrum is crowded by molecular lines, leading to undetectable DIB ($QF$=-2).

\subsection{Cyanogen differential abundance proxy} 
In the  spectra of cool stars, a couple of cyanogen lines can be seen (their wavelength identification can be found in Fig.~\ref{Fig:FitEx1}, although the lines are weaker in the illustrated spectrum with respect to cooler stars). Five interesting CN lines were initially identified when building the line list. The tests performed in the line-selection process presented in Sect.~\ref{Sec:Lines} selected one of these CN lines as a reliable CN over- or underabundance proxy in the spectra of cool stars.

This CN line is centred at 862.884~nm and a window of 0.15~nm  has been selected around it for its analysis. As for the DIB feature of cool stars, the observed spectrum is divided by the corresponding synthetic spectrum, interpolated to the atmospheric parameters of the star derived by MatisseGauguin. This synthetic spectrum assumes the solar-scaled values of carbon and nitrogen abundances [C/Fe]=[N/Fe]=0.0~dex.
We then estimated the EW of the residual by adopting the same Gaussian fitting procedure as for the DIB parameters of cool stars.
This CN proxy is therefore an indicator of the strength of the line with respect to the standard value, and reveals a CN underabundance or overabundance (positive or negative EW, respectively). 
In addition, the central wavelength and the width of the residual feature are also derived from the above-mentioned Gaussian fit, as already implemented for the DIB. 

\subsection{Propagation of flux uncertainties}
\label{sec:mg_mcmc}
The estimation of a star's atmospheric parameters, chemical abundances from individual lines, DIB, and CN-index parameters described above is performed
from the input RVS spectrum, without considering the associated flux uncertainties
per $wlp$. To estimate parameter uncertainties induced by the spectral noise, the complete MatisseGauguin workflow is rerun 50 times to analyse the same number of different Monte-Carlo realisations of the stellar spectrum. Upon each realisation, the input stellar flux per $wlp$ $F_i$ is modified according to the corresponding flux uncertainty of that $wlp$, $\sigma_{Fi}$. In particular, each realisation is computed by adding or subtracting a $\Delta F_i$ at each $wlp$ $i$, randomly sampling a Gaussian distribution of standard deviation  equal to $\sigma_{Fi}$ and centred at zero. 

The complete Monte-Carlo implementation produces a total set of 50 values for each estimated parameter (\T, \g, \meta, \alphaFe,  individual line abundances [X/Fe]$^l$, DIB, and CN indexes). For each of the corresponding parameter distributions, we compute the median and the lower and upper confidence values, from the 50th, 16th, and 84th quantiles, respectively. The median value of each parameter is saved as the adopted parameter estimation in the \gspspec\ catalogue. Both the lower and upper confidence levels are also published. In summary, this procedure allows parameter uncertainties to be properly estimated for each star, and for them to be tailored to the quality of the associated spectrum, but also to its stellar type and chemical abundance pattern. It is important to note that, in this way, the uncertainties on individual line abundances [X/Fe]$^l$ propagate the atmospheric parameters ones, as new [X/Fe]$^l$ values are computed upon each realisation for the new \T, \g, \meta, and \alphaFe\ estimations. In addition, asymmetric uncertainties around the finally considered median value are provided thanks to the lower and upper confidence levels. This Monte Carlo treatment is made possible thanks to the extremely fast application of the \gspspec\ analysis (cf. Sect.~\ref{Sec:Goals}).

\subsection{Individual element chemical abundances} \label{subsec:combinedAbun}
As explained in Sect.\ref{Sec:GAUGUIN}, GAUGUIN provides chemical abundances for each of the 33 atomic lines of Table \ref{Table_linelist}, called [X/Fe]$^l$. 
The final chemical abundances per element [X/Fe] are derived by combining the independent abundance estimates [X/Fe]$^l$ of all the available lines of the same species. To this purpose, a mean abundance per element  is calculated, weighted by the inverse of the [X/Fe]$^l$ uncertainty of each line (defined as the upper minus lower confidence values of the [X/Fe]$^l$ abundance distribution provided by the 50 Monte-Carlo realisations). The published abundances are \NFe, \MgFe, \SiFe, \SFe, \CaFe, \TiFe, \CrFe, \FeIM, \FeIIM\footnote{We provide iron abundances with respect to the mean metallicity following the implementation of the reference grids. The classical [Fe/H] can be easily obtained by adding \meta\ to \FeIM\ or \FeIIM.}, \NiFe, \ZrFe, \CeFe, and \NdFe. Their associated lower and upper confidence values are also published and were calculated as the weighted mean of the [X/Fe]$^l$ ones.

\section{The \gspspec\ ANN workflow}
\label{Sec:ANN}
The ANN algorithm is based on supervised learning and provides a different parameterisation of the RVS spectra, independent from the MatisseGauguin workflow. ANN projects the RVS spectra onto the label
space of the astrophysical parameters. We trained the network on the same grid of reference synthetic spectra as MatisseGauguin (see Sect.~\ref{Sec:SyntheticGrids}), in this case adding noise according to the different S/N scales in the observed spectra \citep{Manteiga2010}.

The ANN architecture is feed-forward with three fully
connected neuron layers.
The input layer has as many neurons as $wlp$ in the spectrum (800) whereas the output layer has four neurons corresponding to the number of estimated parameters. The number of neurons in the hidden layer was empirically determined between 50 and 100 for nets trained with low- to high-S/N spectra, respectively.
In the same way, we determined the learning rate in the range [0.001, 0.2].
The activation function selected for input and output layers is linear, whereas the logistic function was selected for the hidden layer. 


\begin{figure}[t]
    \begin{center}
    \includegraphics[width=0.5\textwidth]{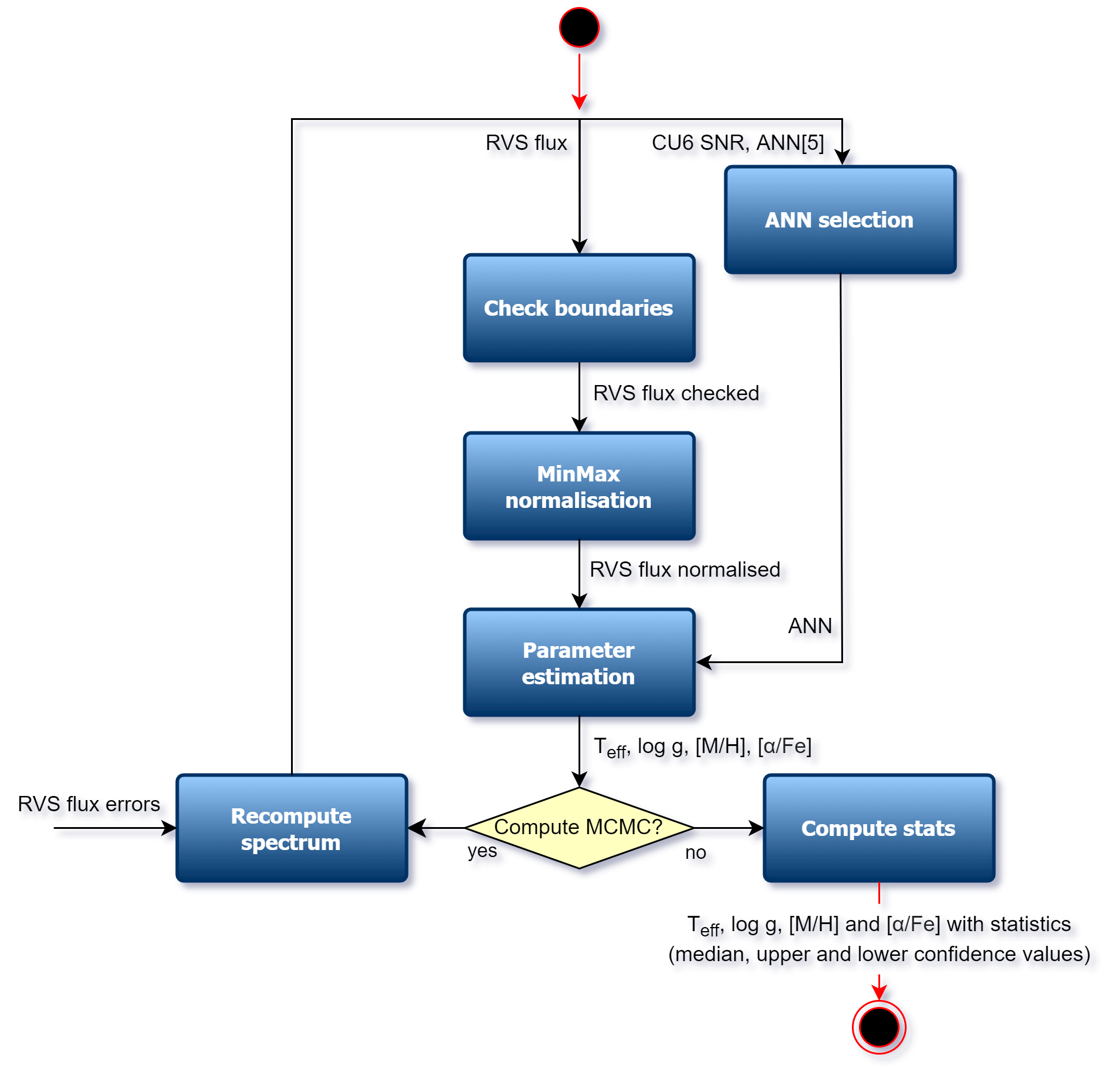}
    \caption{ANN workflow that provides the second set of the main stellar atmospheric parameters (\T, \g, \meta\ and \alphaFe).}
    \label{fig:ann_workflow}
    \end{center}
\end{figure}

\begin{table}[t]
    \centering
    \caption{Equivalent S/Ns  between ANN networks and RVS spectra.}
    \begin{tabular}{|l|c|c|c|c|c|}
        \hline
        ANN & 25 & 30 & 35 & 40 & 50 \\
        \hline
        RVS & [20, 24] & (24, 40] & (40, 68] & (68, 108] & (108, $\infty$] \\
        \hline
    \end{tabular}
    \label{tab:ann_SNReq}
\end{table}

The training procedure is performed with the backpropagation function, which can be interpreted as a problem of minimisation of the error existing between the obtained and desired outputs. 
In order to avoid overtraining, and to select the ANN that leads to the best generalisation, the early stopping procedure was used, finalising the training process when the performance starts to degrade, and obtaining the net that minimises the error.

The effectiveness of the ANN depends on the input ordering. For that reason, we perform ten trainings with different ordering, selecting the one with minimum error. For each train, weights initialise in the range [-0.2,0.2], and we established a limit of 1000 iterations because we observed that beyond that number, the training process does not improve but the computational cost increases.

The ANN parameterisation procedure that estimates the second set of \gspspec\ atmospheric parameters (\T, \g, \meta\ and \alphaFe) is published in the {\it AstrophysicalParametersSupp} table and is summarised in Fig.~\ref{fig:ann_workflow}.
Specifically, the present ANN version included in \gspspec\ proceeds as follows:

\medskip


     ANN selection: ANN behaves well in the presence of noise~\citep{Manteiga2010}, confirming that it is a robust method when estimating  astrophysical parameters for relatively low-S/N spectra. As there is no noise model for the \Gaia RVS spectra, we empirically determined the relation between the noise given by CU6 and the Gaussian noise that we need to use to train the nets. The corresponding values are shown in \tabref{tab:ann_SNReq}. 
    For each RVS input spectrum, we then used its S/N value, provided by CU6, to select which net performs the parameter estimation.
    
\smallskip

     Check boundaries: Some RVS spectra have zero flux values at the beginning or at the end of their spectral range. These are often caused by radial velocity corrections and could lead to large flux variations in the borders and cause ANN malfunctions. 
    To avoid this behaviour, we truncated these zero flux values and adopted the mean of the flux spectrum for these $wlp$.
 
 \smallskip
    
    Normalisation: A minimum--maximum scaling procedure is applied to the RVS spectra, equalising it to avoid geometric biases during the training stage in order to guarantee that all the inputs are in a comparable range.

\smallskip

    Parameter estimation: Once the net has been selected, it is fed with the normalised spectrum to estimate \T, \g, \meta\ and \alphaFe. The net returns these estimations normalised, and so a denormalisation procedure is applied to return the values in the expected range.

\smallskip
    
     Monte-Carlo iterations using flux uncertainties: The same procedure as for MatisseGauguin (see Sect.~\ref{sec:mg_mcmc}) is also applied for ANN to estimate the parameter uncertainties caused by flux errors. We therefore obtain the median and the lower and upper confidence values of each AP again.

\section{Validation and {\it flags$\_$gspspec} quality flag chain} \label{sec:val}
\begin{table*}[h]
    \caption{Definition of each character in the \gspspec\ quality flag string chain ({\it flags$\_$gspspec}), including the possible values (col.3) and the related subsection and tables providing further information (col.4). Flag names are split into three categories: Parameter flags (green), individual abundance flags (blue), and EW flags (orange). All flags concern the MatisseGauguin parameters, while only the parameter flags except KMgiantPar are applied to ANN results.}
    \centering
    \begin{tabular}{|l|l|l|c|}
\hline
\hline
{\bf Chain character} & {\bf ~~~~~~~~~~~~~~~~~~~~~~~~~~~~Considered } & {\bf ~~~~~~Possible }& {\bf Related  }\\
{\bf number - name} & {\bf ~~~~~~~~~~~~~~~~~~~~~~~~~ quality aspect} & {\bf adopted values} & {\bf subsection and table}\\
\hline
\hline
1 \color{teal} \bf vbroadT & $vbroad$ induced bias in \T & 0,1,2,9 & \ref{subsec:rotation} \& \ref{tab:vsiniQF}\\
\hline
2 \color{teal} \bf vbroadG & $vbroad$ induced bias in \g & 0,1,2,9 &\ref{subsec:rotation}  \& \ref{tab:vsiniQF}\\
\hline
3 \color{teal} \bf vbroadM & $vbroad$ induced bias in \meta & 0,1,2,9 & \ref{subsec:rotation} \& \ref{tab:vsiniQF}\\
\hline
4 \color{teal} \bf vradT & \Vrad\ induced bias in \T & 0,1,2,9  & \ref{subsec:vrad} \& \ref{tab:vradQF}\\
\hline
5 \color{teal} \bf vradG & \Vrad\ induced bias in \g & 0,1,2,9  & \ref{subsec:vrad} \& \ref{tab:vradQF}\\
\hline
6 \color{teal} \bf vradM & \Vrad\ induced bias in \meta & 0,1,2,9 & \ref{subsec:vrad} \& \ref{tab:vradQF}\\
\hline
7 \color{teal} \bf fluxNoise & Flux noise induced uncertainties & 0,1,2,3,4,5,9  & \ref{subsec:fluxnoise} \& \ref{tab:noiseQF}, \ref{tab:noiseQFANN}\\
\hline
8 \color{teal} \bf  extrapol & Extrapolation level of the parametrisation& 0,1,2,3,4,9   &  \ref{subsec:extrapolation} \& \ref{tab:extrapol},  \ref{tab:extrapolANN}\\
\hline
9 \color{teal} \bf  negFlux & Negative flux $wlp$ & 0,1,9  & \ref{subsec:fluxproblems} \& \ref{tab:FluxProb}\\
\hline
10 \color{teal} \bf  nanFlux & NaN flux $wlp$ & 0,9  & \ref{subsec:fluxproblems} \& \ref{tab:FluxProb} \\
\hline
11 \color{teal} \bf  emission & Emission line detected by CU6& 0,9  & \ref{subsec:fluxproblems} \& \ref{tab:FluxProb} \\
\hline
12 \color{teal} \bf  nullFluxErr & Null uncertainties $wlp$ & 0,9  &\ref{subsec:fluxproblems} \& \ref{tab:FluxProb}\\   
\hline
13 \color{teal} \bf  KMgiantPar  & KM-type giant stars & 0,1,2  & \ref{subsec:KM} \& \ref{tab:KMflag}\\
\hline
\hline
14 \color{blue} \bf NUpLim  & Nitrogen abundance upper limit & 0,1,2,9  & \ref{subsec:abundqf} \& \ref{tab:XUpLim}\\
\hline
15 \color{blue} \bf NUncer & Nitrogen abundance uncertainty quality & 0,1,2,9  & \ref{subsec:abundqf} \& \ref{tab:XUncer}\\
\hline
16 \color{blue} \bf MgUpLim & Magnesium abundance upper limit & 0,1,2,9  & \ref{subsec:abundqf} \& \ref{tab:XUpLim}\\
\hline
17 \color{blue} \bf MgUncer & Magnesium abundance uncertainty quality & 0,1,2,9  & \ref{subsec:abundqf} \& \ref{tab:XUncer}\\
\hline
18  \color{blue} \bf SiUpLim & Silicon abundance upper limit & 0,1,2,9  & \ref{subsec:abundqf} \& \ref{tab:XUpLim}\\
\hline
19 \color{blue} \bf SiUncer & Silicon abundance uncertainty quality & 0,1,2,9  & \ref{subsec:abundqf} \& \ref{tab:XUncer}\\
\hline
20 \color{blue} \bf SUpLim & Sulphur abundance upper limit & 0,1,2,9  & \ref{subsec:abundqf} \& \ref{tab:XUpLim}\\
\hline
21 \color{blue} \bf SUncer & Sulphur abundance uncertainty quality & 0,1,2,9  & \ref{subsec:abundqf} \& \ref{tab:XUncer}\\
\hline
22 \color{blue} \bf CaUpLim & Calcium abundance upper limit & 0,1,2,9  & \ref{subsec:abundqf} \& \ref{tab:XUpLim}\\
\hline
23 \color{blue} \bf CaUncer & Calcium abundance uncertainty quality & 0,1,2,9  & \ref{subsec:abundqf} \& \ref{tab:XUncer}\\
\hline
24 \color{blue} \bf TiUpLim & Titanium abundance upper limit & 0,1,2,9  & \ref{subsec:abundqf} \& \ref{tab:XUpLim}\\
\hline
25 \color{blue} \bf TiUncer & Titanium abundance uncertainty quality & 0,1,2,9  & \ref{subsec:abundqf} \& \ref{tab:XUncer}\\
\hline
26  \color{blue} \bf CrUpLim & Chromium abundance upper limit & 0,1,2,9  & \ref{subsec:abundqf} \& \ref{tab:XUpLim}\\
\hline
27 \color{blue} \bf CrUncer & Chromium abundance uncertainty quality & 0,1,2,9  & \ref{subsec:abundqf} \& \ref{tab:XUncer}\\
\hline
28 \color{blue} \bf FeUpLim & Neutral iron abundance upper limit & 0,1,2,9  &\ref{subsec:abundqf}  \& \ref{tab:XUpLim}\\
\hline
29 \color{blue} \bf FeUncer & Neutral iron abundance uncertainty quality & 0,1,2,9  &\ref{subsec:abundqf} \& \ref{tab:XUncer}\\
\hline
30 \color{blue} \bf FeIIUpLim & Ionised iron abundance upper limit & 0,1,2,9  &\ref{subsec:abundqf}  \& \ref{tab:XUpLim}\\
\hline
31 \color{blue} \bf FeIIUncer & Ionised iron abundance uncertainty quality & 0,1,2,9  &\ref{subsec:abundqf} \& \ref{tab:XUncer}\\
\hline
32 \color{blue} \bf NiUpLim & Nickel abundance upper limit & 0,1,2,9  & \ref{subsec:abundqf} \& \ref{tab:XUpLim}\\
\hline
33 \color{blue} \bf NiUncer & Nickel abundance uncertainty quality & 0,1,2,9   &\ref{subsec:abundqf} \& \ref{tab:XUncer}\\
\hline
34  \color{blue} \bf ZrUpLim & Zirconium abundance upper limit & 0,1,2,9  & \ref{subsec:abundqf} \& \ref{tab:XUpLim}\\
\hline
35 \color{blue} \bf ZrUncer & Zirconium abundance uncertainty quality & 0,1,2,9   &\ref{subsec:abundqf} \& \ref{tab:XUncer}\\
\hline
36 \color{blue} \bf CeUpLim & Cerium abundance upper limit & 0,1,2,9  & \ref{subsec:abundqf} \& \ref{tab:XUpLim}\\
\hline
37  \color{blue} \bf CeUncer & Cerium abundance uncertainty quality & 0,1,2,9  &\ref{subsec:abundqf}  \& \ref{tab:XUncer}\\
\hline
38 \color{blue} \bf NdUpLim & Neodymium abundance upper limit & 0,1,2,9  & \ref{subsec:abundqf} \& \ref{tab:XUpLim}\\
\hline
39 \color{blue} \bf NdUncer & Neodymium abundance uncertainty quality & 0,1,2,9  & \ref{subsec:abundqf} \& \ref{tab:XUncer}\\
\hline
\hline
40 \color{orange} 
\bf DeltaCNq & Cyanogen differential equivalent width quality &0,9  &\ref{subsec:CNqf} \& \ref{tab:CNqflag}\\
\hline
41 \color{orange} 
\bf DIBq & DIB quality flag & 0,1,2,3,4,5,9   & \ref{subsec:DIBqf} \& \ref{tab:DIBflag}\\
\hline
\hline
    \end{tabular}
    \label{tab:MG_QFchain}
\end{table*}

The \gspspec\ output after operations has been carefully checked and validated, considering different potential error sources. Following this validation procedure, a quality flag chain ({\it flags$\_$gspspec})  is implemented (cf. Table~\ref{tab:MG_QFchain})\footnote{We note that the flags associated with the ANN results correspond to the first 12 flags of this table.}. In this chain, a value of 0 is the best, and 9 is the worst, generally implying the parameter masking. This allows the user to publish all kinds of quality results, satisfying the more or less restrictive needs of different science applications.
Nevertheless, this implies that considering these quality flags is mandatory for correct use of the \gspspec\ parameters and abundances. If not applied, results of low quality for a given application could be unconsciously included in the analysis, severely affecting its conclusions. 

The following subsections review the different reasons for  failure, potential bias, and the uncertainty sources considered in the \gspspec\ validation, and following the characters ordering in the quality flag chain. Several associated figures and tables can be found in Appendix~\ref{Appendix:flags}.

\subsection{Parameterisation biases induced by rotational and macroturbulence line broadening ($vbroad$ flags)}
\label{subsec:rotation}

\gspspec\ is trained with reference spectra assuming no rotation (see Sect.~\ref{Sec:SyntheticGrids}). At the RVS spectral resolution, the parameterisation tolerance to broadened spectra through rotational (\Vsini) and/or macroturbulence broadening has to be flagged according to tests with synthetic data. 

Potential biases in \T, \g,\  and \meta\  induced by stellar rotation were therefore modelled using a dedicated set of synthetic RVS spectra, which were broadened with different \Vsini\ values from 0 to 70~km.s$^{-1}$. For simplicity, we assumed in what follows that the line broadening factor produced by CU6 ($vbroad$) is well reproduced by only mimicking a stellar rotation. 
The estimated biases ($\Delta$\T, $\Delta$\g, $\Delta$\meta) induced by rotational broadening are a function of \T\ and \g. Metallicity dependencies are also observed, with metal-poor objects being more affected than metal-rich ones (a consequence of their smaller number of lines that can be used for the parametrisation). 

First, using this data set, we identified the limiting \Vsini\ values inducing a bias larger than $\Delta$\T=2\,000~K.
We then modelled the parameter dependence of the \Vsini\ values leading to that maximum admitted bias by fitting a third-order polynomial with variable \T, \g,\ and \meta, as shown in Fig.~\ref{Fig:FlagVsini}. 
To avoid extrapolation issues, upper and lower limits were imposed on the third-order interpolation polynomial during the post-processing. 
This function was finally adopted during post-processing to mask the corresponding \gspspec\ results (Flag $vbroad$T=9 in Table~\ref{tab:MG_QFchain}).
Similarly, we applied this procedure to define three other values for this quality flag, depending on the amplitude of the predicted induced bias in \T: Flag $vbroad$T=0, 1, or 2 for stars with a possible bias $\Delta$\T$\leq$250~K, 250<$\Delta$\T$\leq$500~K, and 500<$\Delta$\T<2\,000~K, respectively. 

Exactly the same procedure was adopted for defining the flags associated with a bias in \g\ and \meta\ induced by the rotational and macroturbulence line broadening. Their detailed definition is given in Table~\ref{tab:vsiniQF}.

\subsection{Parameterisation biases induced by radial velocity uncertainty ($vrad$ flags)}
\label{subsec:vrad}

In a very similar way, we investigated the possible bias induced by radial velocity uncertainties, because the \gspspec\ parametrisation is performed whilst assuming that the observed spectra are perfectly at rest-frame. The examination of \gspspec\ unfiltered results reveals that large \Vrad\ errors (provided by CU6) are preferentially found in specific regions of the output atmospheric parameter space (combinations of \T, \g,\ and \meta\  where no stars are expected, or at extremely high or low \alphaFe). This is an important illustration of the expected parametrisation sensitivity to \Vrad\ uncertainties.

We therefore investigated the amplitude of possible biases in \T, \g,\ and \meta\ caused by \Vrad\ errors varying between 0 and 10~km.s$^{-1}$ using specific synthetic spectra.
Again, metal-poor stars (with a lower number of lines available for the parametrisation) were found to be more affected than metal-rich ones. As described above for the $vbroad$ flags, specific third-order polynomials with variable \T, \g,\ and \meta\ were then fitted to define the values associated with three $vrad$ flags. Their precise definition is given in Table~\ref{tab:vradQF}.
        
\subsection{Parameter uncertainties due to flux noise ($fluxNoise$ flag)}
\label{subsec:fluxnoise}

The parametrisation is affected by uncertainties in the observed fluxes, that is, the noise at each wavelength leading to a mean S/N over the entire wavelength domain. To quantify this effect and as already explained in Sect.\ref{sec:mg_mcmc}, flux uncertainties are taken into account by \gspspec\  through 50 Monte-Carlo realisations of the spectral flux for each star. The \gspspec\ parameterisation is then performed for those 50 spectrum realisations and parameter uncertainties (noted $\sigma$, hereafter) are defined from the 16th and 84th quantiles of the obtained distributions. 
To enable a rapid selection of results in the \gspspec\ catalogue from the estimated parameter uncertainties, we defined a specific quality
flag ($fluxNoise$). This flag simultaneously considers uncertainties in \T, \g, \meta,\ and \alphaFe, labelling results of progressively higher precision from $fluxNoise$=5 to  $fluxNoise$=0. The exact conditions imposed during the post-processing for the noise uncertainty quality flags are indicated in Tables~\ref{tab:noiseQF} and \ref{tab:noiseQFANN} for MatisseGauguin and ANN, respectively. It is worth noting that stars with extremely poor quality parameters, such as, for instance, those without any distinction between giants and dwarfs ($\sigma$\g>2~dex) or between F, G, and K stellar types ($\sigma$\T>2\,000~K) are filtered out during the post-processing ($fluxNoise$=9) and do not appear in the finally published catalogue.

\subsection{Extrapolation level ($extrapol$ flag)}
\label{subsec:extrapolation}

Due to extrapolation, the \gspspec\ parameter  solution could be located outside the parameter space of the training grid (cf. Fig.~\ref{Fig.Grid}) for either one or several parameters. In addition, censored training occurs near the grid borders. In order to flag those extrapolated results for which the parametrisation is less reliable, we have implemented a specific flag ($extrapol$) that is indicative of the extrapolation level. 

The definition of this flag is reported in Tables~\ref{tab:extrapol} and \ref{tab:extrapolANN} for MatisseGauguin and ANN, respectively, depending on the availability (or not) of  a $gof$  and the  distance between the parameter solutions and the grid borders. The flag value  depends on the level of extrapolation: from results near the grid limits ($extrapol$=4) to no extrapolation at all (and therefore a more reliable solution, $extrapol$=0). 
Again, sources without a $gof$ and with \T\ values outside the 2\,500 to 9\,000~K interval or \g\ values outside the -1 to 6~dex range were filtered out ($extrapol$=9) during the post-processing and do not appear in the final catalogue. 

\subsection{RVS flux issues or emission line flags}
\label{subsec:fluxproblems}
MATISSE and GAUGUIN being model-driven methods that essentially aim to maximise the goodness of fit between an observation and a set of templates, any significant and/or systematic difference between the RVS spectra and the reference grid can introduce biases in the results. These differences can be associated with the RVS spectra processing, or be inherent to the stellar physics assumptions adopted when computing the reference grid (stellar activity being one example). When such issues randomly affect a $wlp$, then it can be very difficult  (if not impossible) to properly take them into account during the analysis. We have implemented four specific flags to identify such cases. Their definition is presented below and is summarised in Table~\ref{tab:FluxProb}.

RVS spectral anomalies can manifest as $wlp$ that have a negative flux (flag  $negFlux$), or a flux (or associated variance) that is not a number ($nanFlux$ and $nullFluxErr$ flags, respectively). Whereas such caveats do not necessarily alter the RV determination, they can hamper  parameterisation estimates relying  specifically on the affected $wlp$.  
For instance, some tens of stars have a couple of $wlp$ with negative flux. They are predominantly found in the cores of the strongest \CaII\ lines and result from an oversubtraction of the straylight during the spectrum production. This leads to a modified line profile and could indeed affect the parametrisation.
Similarly, NaN flux values can appear in the spectra. As explained in \cite{Seabroke22}, $wlp$ are masked in the CCD sample. When these are averaged, a chance alignment of these masks when there are few CCD spectra pixels contributing to a particular wavelength bin in the combined spectrum could lead to a NaN flux value, which happens more often near the edges. The \gspspec\ treatment partly
overcomes this problem thanks to the rebinning (from 2400 to 800~$wlp$) of the oversampled input spectra. For this rebinning, a median flux is computed every three $wlp$, excluding NaN values. As a consequence, NaN flux values in the rebinned spectra only remain if the three averaged $wlp$ are equal to NaN. To filter out those rare cases, we have implemented the specific $nanFlux$ flag.  Finally, if no flux variance is associated with a $wlp$, then the derived parameter uncertainty is unreliable or impossible to estimate. This is reported by the $nullFluxErr$ flag.

As presented in Table~\ref{tab:extrapol} , while the $nanFlux$ and the $nullFluxErr$ flags lead to a systematic exclusion of the source from the final catalogue (only values equal to 9 have been implemented), the $negFlux$ flag can also be equal to 1 (one or two $wlp$ with negative flux values) or 0 (no negative $wlp$ at all). However, for the reasons described above, we recommend preferentially selecting stars with $negFlux$=0.  

On the other hand, emission lines due to stellar activity are inherent to the stellar properties and carry important information about the observed star. However, the physical conditions that lead to the emission lines are not considered in our grid of synthetic spectra. Therefore,  if a star shows signs of activity, its \gspspec\ parameters should also be discarded and considered unreliable.  We used the $CU6\_is\_emission$ flag provided by the CU6 to detect such stars, and forced them to have a \gspspec\ flag $emission$=9 to reject them.

\subsection{Parametrisation quality of K and M type giants}
\label{subsec:KM}
The parametrisation of cool stars with effective temperatures below 4\,000~K is known to be complex due to their crowded spectra, which results from the increasing presence of atomic and, especially, molecular lines. This aggravates normalisation issues and parameter degeneracies, in particular for metal-rich stars.
During the \gspspec\ validation process, a correlation was found between the minimum flux value ($F_{\rm min}$) of the  spectra of giant stars with \T$\la$4\,000~K and their estimated \g. In particular, in this cool temperature regime, objects with higher \g\ values present larger $F_{\rm min}$ values than expected when compared to those of slightly hotter giants with similar \g\ and S/N values. This reveals a parametrisation problem, as the pseudo-continuum should present lower values for cooler stars for which the line-crowding increases, and not vice versa. 
We have therefore implemented a specific flag ({\it KM-typestars}) that takes this issue
into account. This flag depends on the $F_{\rm min}$ value and the $gof$ in order to take account of the influence of the S/N on $F_{\rm min}$. As reported in Table~\ref{tab:KMflag}, stars with {\it KM-typestars} equal to 1 and 2 have corrected \T\ and \g\ with uncertainties reflecting the \gspspec\ parameterisation problems encountered for these stars: \T=4250$\pm$500~K and \g=1.5 $\pm$1.
        
\subsection{Quality of individual chemical abundances}
\label{subsec:abundqf}

We checked the reliability of all the abundance estimates, including their uncertainties across the Kiel diagram (\g\ vs. \T\ plot) and taking into account the S/N. As a result of this process, we defined two flags for each individual abundance. Their definitions are given in Table \ref{tab:XUpLim} and \ref{tab:XUncer} (with associated coefficients in Table ~\ref{tab:XFeCoefs}).

On one hand, as expected, the estimation quality depends on the strength of the spectral lines of the studied element, which varies with \T, \g, \meta, and the abundance of the element itself. To help the user to deal with this effect, we implemented the individual abundance upper limit flag ($XUpLim$), which is an indicator of the line depth with respect to the noise level. This flag is based on an estimate of the detectability limit ({\it upper-limit}) that depends on the line atomic data, the stellar parameters, the line broadening, and the S/N. We note that, for the definition of this $XUpLim$ flag, we adopted a \gspspec\ internal estimate of the S/N that could slightly differ from the published \verb|rv_expected_sig_to_noise|. The closer is the derived abundance to this upper limit, the higher the flag value and the abundances should therefore be used more cautiously.

On the other hand, for low-S/N spectra (with a limiting S/N depending on the analysed lines), the reliability of the associated abundance uncertainties can be underestimated. This is due to the fact that the maximum allowed abundance value in the reference grids is [X/Fe]=2.0~dex, preventing higher values in the abundance distribution associated with the flux noise Monte-Carlo realisations. This effect depends on the line detectability and the S/N. As a consequence, we defined a second individual abundance flag ($XUncer$) labelling the reliability of the associated abundance uncertainty taking into account its dependence on the stellar type (\T, \g,\ and \meta), the S/N estimate, and/or the $gof$. Moreover, it is worth noting that the distance between the [X/Fe] upper confidence level and the grid upper border is also a good indicator of the estimate reliability. 

\subsubsection{Validation of heavy element abundances}
\begin{figure}[t]
    \begin{center}
    \includegraphics[width=0.55\textwidth, trim=4.5cm 0 0 0, clip]{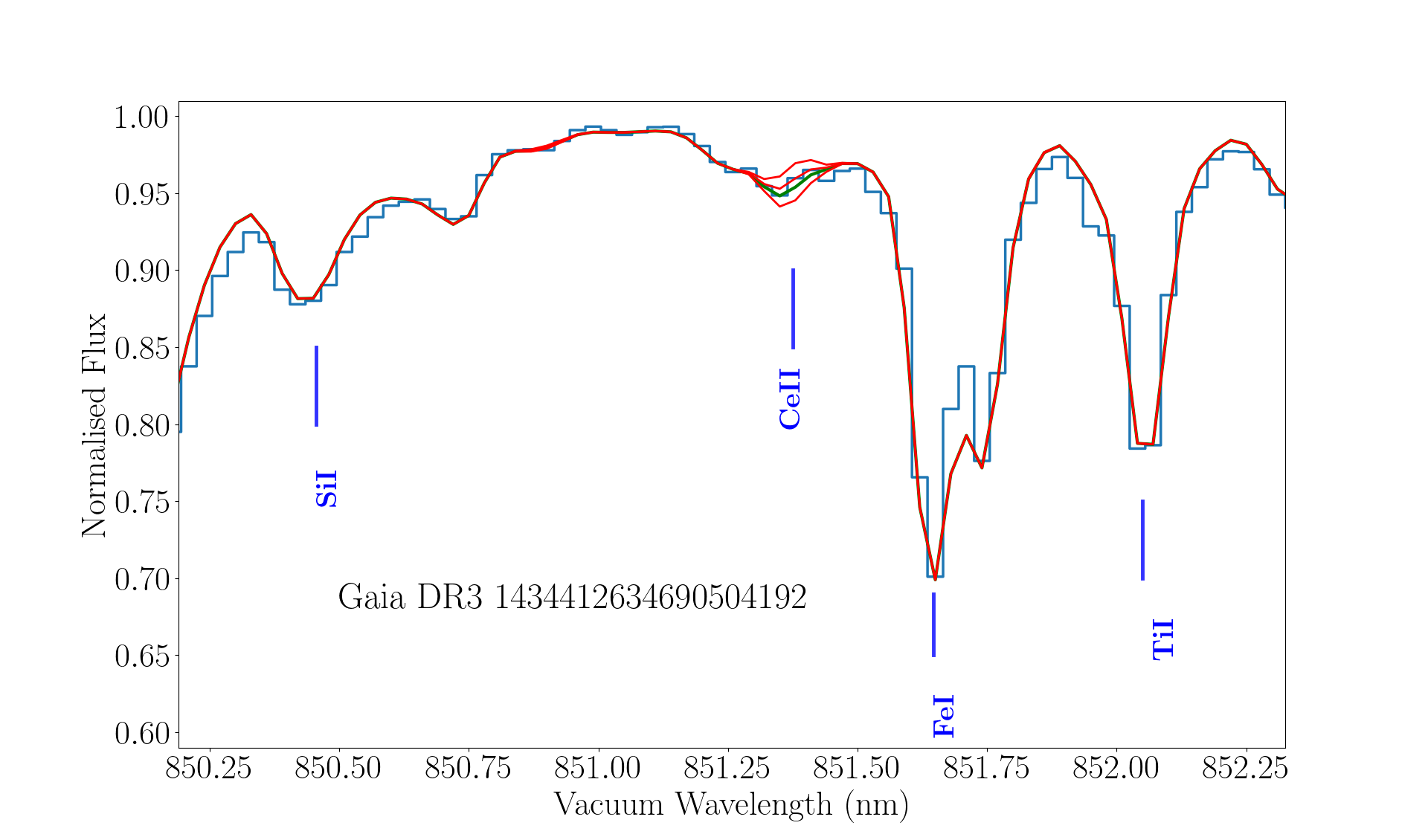}
    \caption{Fit of the RVS spectrum (blue histogram) of the RGB star \Gaia DR3 1434412634690504192 around its cerium line. 
    The model in green corresponds to the GAUGUIN solution \CeFe=0.26~dex (in excellent agreement with the literature value) whereas those in orange have \CeFe=-2.0~dex (almost no cerium) and $\pm$0.2~dex around the GAUGUIN abundance, respectively. The S/N is 907 and the broadening velocity is equal to 10.4~km.s$^{-1}$. See text for more details.} 
    \label{FigCe}
    \end{center}
\end{figure}

An important illustration of the quality of the \gspspec\ abundance analysis with GAUGUIN is provided by the derivation of heavy element abundances, the estimation of which seemed too challenging for the RVS resolution.  As an example, Fig.~\ref{FigCe} shows the RVS spectrum of a red giant branch (RGB) star around its cerium line. The Ce abundance (\CeFe=0.26~dex with lower and upper confidence levels being 0.18 and 0.38~dex, respectively) was derived from the MatisseGauguin parameters: \T=4157~K, \g=1.09, \meta=-0.4~dex, and \alphaFe=0.12~dex. The  \gspspec\ abundance flags
are $CeUpLim$=$CeUncer$=0.
This star was previously analysed by \citet[][]{Forsberg19} who derived a very consistent \CeFe=0.22~dex, adopting very similar atmospheric parameters. 
It is important to note that \gspspec\ cerium abundances are on the same scale as those found by \citet{Forsberg19} with a null median difference for the overlapping sample.
This confirms the high quality of the \gspspec\ chemical analysis and of the Gaia/RVS spectra.

\subsubsection{Validation of the singly ionised iron abundance} 
\label{SecFe2}
\begin{figure}[t]
    \begin{center}
    \includegraphics[width=0.5\textwidth]{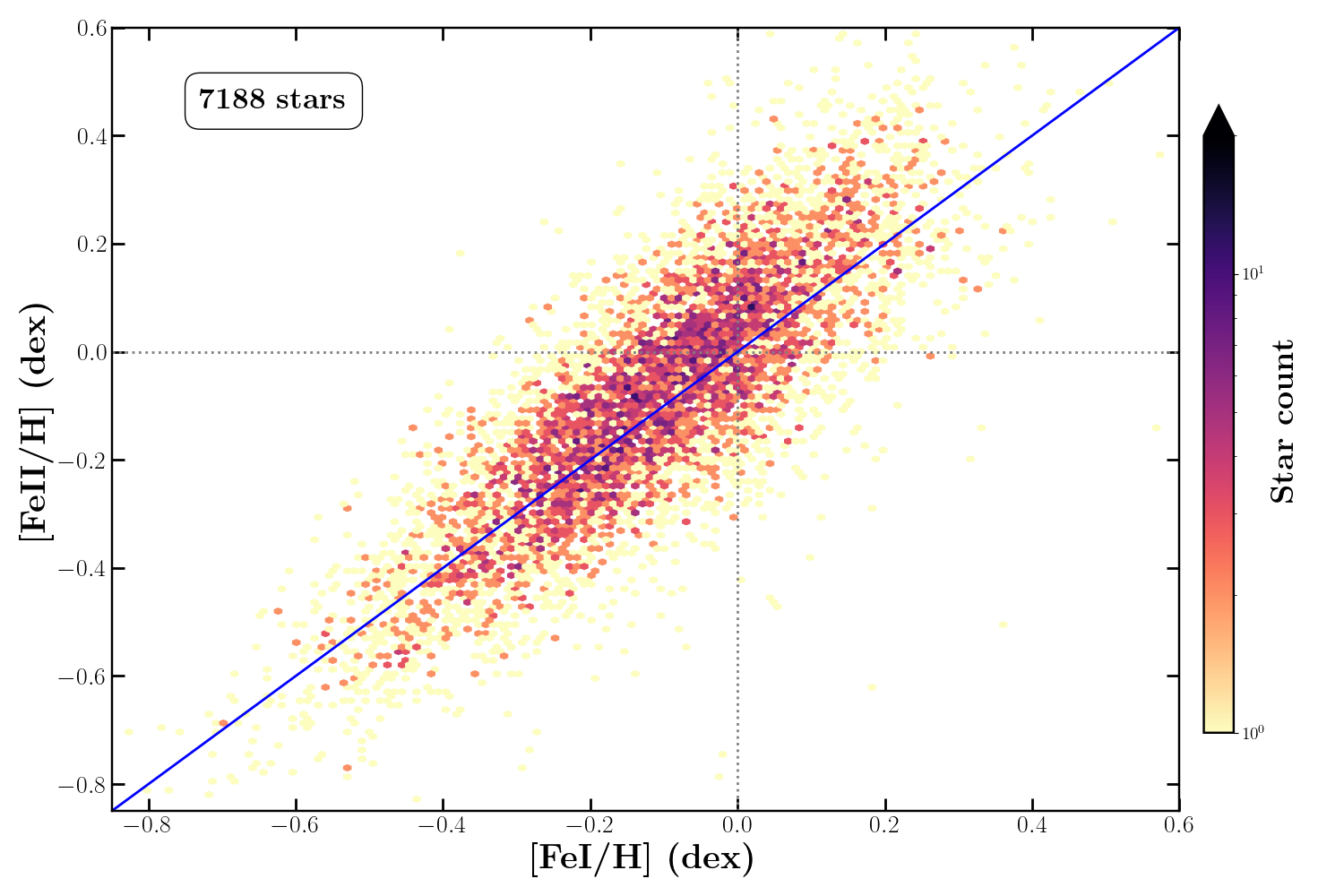}
    \caption{Comparison between iron abundances measured from the proposed \FeII~line at 858.79~nm and from all the other \FeI~lines. The Spearman correlation coefficient is equal to 0.82. See text for more details.} 
    \label{FigFe2}
    \end{center}
\end{figure}

The specific case of \FeII\ abundances merits discussion. When building the \gspspec\ line list, \citet{Contursi21}  identified an unknown line at 858.79~nm (in the vacuum) and proposed that it is actually an \FeII\ feature. Because of its unblended nature in the RVS spectra of hot stars (see Fig.~\ref{Fig:FitEx2}), it has been included in the line list used by GAUGUIN for the individual abundance analysis (Table ~\ref{Table_linelist}).

In Fig.~\ref{FigFe2}, \FeII\ abundances are compared to \FeI\ ones in the atmospheric parameters regime where both estimates are possible at the same time.  Both iron abundances were calibrated as suggested in Sect.~\ref{sec:biases}. We selected stars with all 13 atmospheric parameter flags equal to zero together with a rather strict quality selection using the two abundance flags: $XUpLim\le$1 and  $XUncer$=0. We also selected only stars in which the \FeII~line is easily detected (6\,000<\T<7\,200~K). The agreement between both iron abundances is excellent. The Spearman correlation coefficient is equal to 0.82 and increases up to 0.89 when selecting the $\sim$2\,500 stars with S/N>300. 

We can therefore safely conclude that this 858.79~nm line is indeed a very good metallicity proxy and probably corresponds to an absorption produced by an iron-peak element, \FeII\ being the best candidate as suggested by \citet{Contursi21}.

\subsection{Quality of cyanogen differential equivalent width ($DeltaCNq$)}
\label{subsec:CNqf}

To validate CN parameters, literature data were used to identify cool RGB and AGB stars for which CN lines are expected to be present. The flag associated with the EW of this CN abundance proxy ($DeltaCNq$) is defined in Table~\ref{tab:CNqflag};
it depends on the three line-broadening flags ($vbroad$), the S/N, the $gof,$ and the measured line position ($p_1$).

\subsection{DIB quality flag ($DIBq$)}
\label{subsec:DIBqf}
To quantify the quality of the DIB analysis, we defined a specific flag, ranging from $DIBq$=0 (highest 
quality) to 5 (lowest quality). When no DIB is measured ($DIBq$=9), another flag $QF$, not included in the flag chain, details the reasons as to why no measurements were performed (see Sect.~\ref{DIBparam}). Its definition depends on the $p_0$ and $p_2$ parameters but also on the global noise level ($Ra$) defined by the standard deviation of the ($data$ -- $model$) residual between 860.5 and 864~nm as well as on the local noise level $Rb$, that is, the ($data$--$model$) residual within the DIB profile. Table~\ref{tab:DIBflag} explains the definition of the $DIBq$ flag and Fig.~\ref{QFdib}  shows its flow chart. As discussed in \citet{Schultheis22}, we recommend the adoption of the most reliable DIB parameters ($DIBq$=0,1,2) as well as (i) a good central wavelength measurement $\rm 862.0 < p1 < 862.6$\,nm, (ii) a rather small uncertainty on the EW measurement (err($EW$)/$EW$ < 0.35), and (iii) a good stellar parametrisation (first 13 \gspspec\ flag being smaller than 2).

\section{Known parameter and abundance biases}\label{sec:biases}
After the previous evaluation of the parameter quality through a flagging system, internal and external biases were studied, taking into account the implemented flags. The result of this analysis is presented in this section for MatisseGauguin atmospheric parameters and abundances (Sect.~\ref{MGBiases} includes a summary of the proposed solutions at the end) and  for ANN atmospheric parameters (Sect.~\ref{ANNBiases}). Several figures and tables associated with this section can be found in Appendix~\ref{AppendixBiases}.
In some cases, simple calibrations with low-degree polynomials are suggested. It is worth noting that published DR3 \gspspec\ data are deliberately uncalibrated, and so {users are able to (i) use the raw data that come from the} \gspspec\ processing, (ii) apply, whenever suggested, the calibrations presented in this paper,  and (iii) perform a new calibration tailored to their scientific analysis.

On one hand, although specific work has been done on the optimisation of the reference synthetic spectra grids, the observed biases can be partially due to  mismatches between observations and reference synthetic spectra if some physical aspects not considered in the modelling (e.g. stellar rotation, macroturbulence, departures from local thermodynamic and hydrostatic equilibria) become non-negligible for some parameters of certain types of stars. This has been partially taken into account with parameter flags (e.g. $vbroadT$, $vbroadG$, $vbroadM$ flags). We recall that the parametrisation  of cool stars is often challenging \citep[see e.g. Sect.~\ref{subsec:KM} and][]{Soubiran21}, and even higher resolution surveys in the literature can exhibit biases.

On the other hand, it is worth noting that the observed biases  with respect to the literature can also have their origin in methodological and theoretical assumption differences with respect to those adopted in this work, such as different atmosphere models, atomic data, or  reference solar abundances. In addition, several ground-based spectroscopic surveys have applied $ad hoc$ offset corrections as a result of their calibration procedures. Finally, the presented global biases with respect to the literature depend on the relative proportion of stars in the various reference catalogues as a function of the S/N and the analysed parameter space. 

Finally, it is important to mention that reference catalogues have their own biases. Although literature references are generally calibrated (while Gaia archive data are not), this does not remove all the existent trends, as shown by some recent works \citep[e.g.][]{Soubiran21}. As a consequence, it cannot be excluded that the observed trends in the comparison with external catalogues are partly due to biases that are still present in the literature data.

As a consequence of all the above mentioned points, the results of the bias analysis presented in the following have to be cautiously and thoroughly considered. We recommend that the user adapt any bias correction to the targeted scientific goal and selected sample.

\begin{figure}[t]
\begin{center}
\includegraphics[width=0.35\textwidth, angle=0]{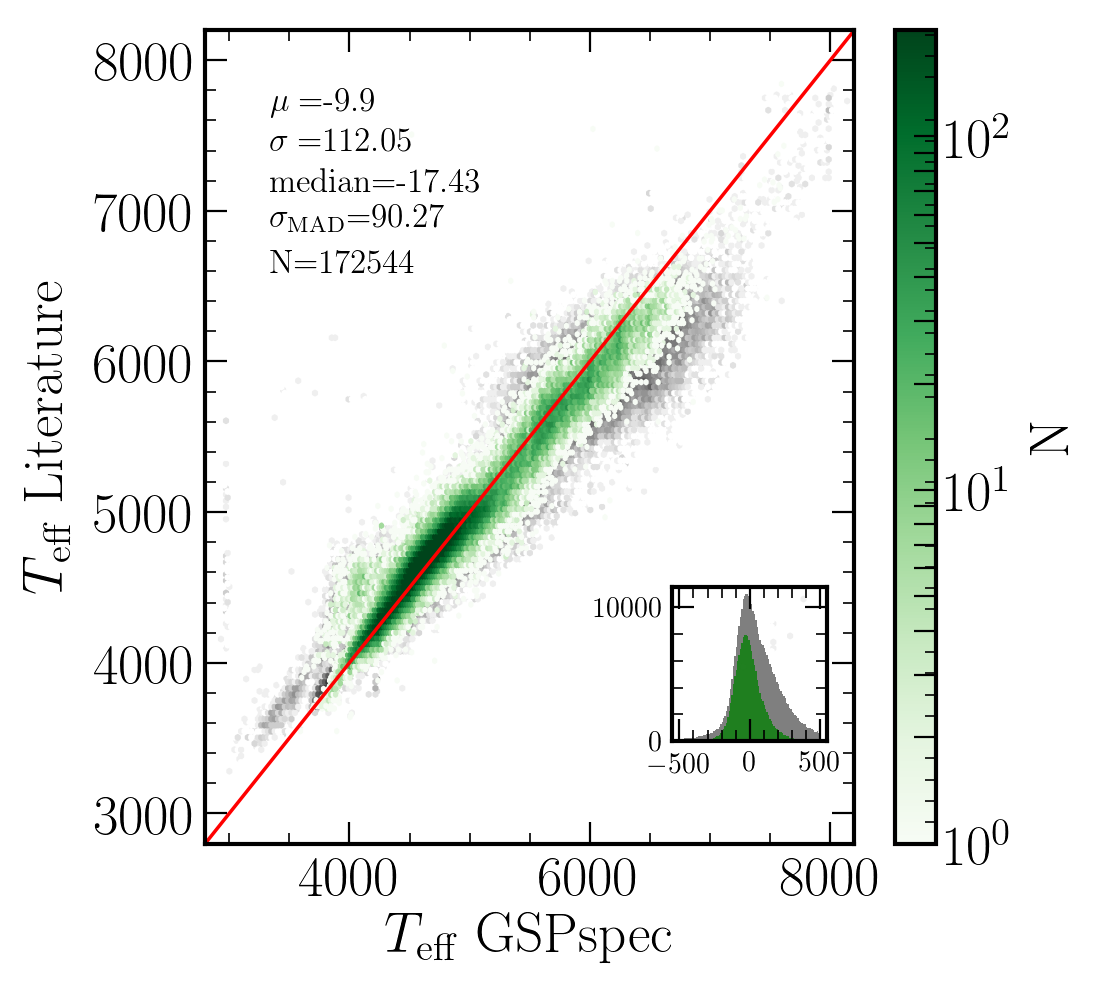}

\includegraphics[width=0.32\textwidth, angle=0]{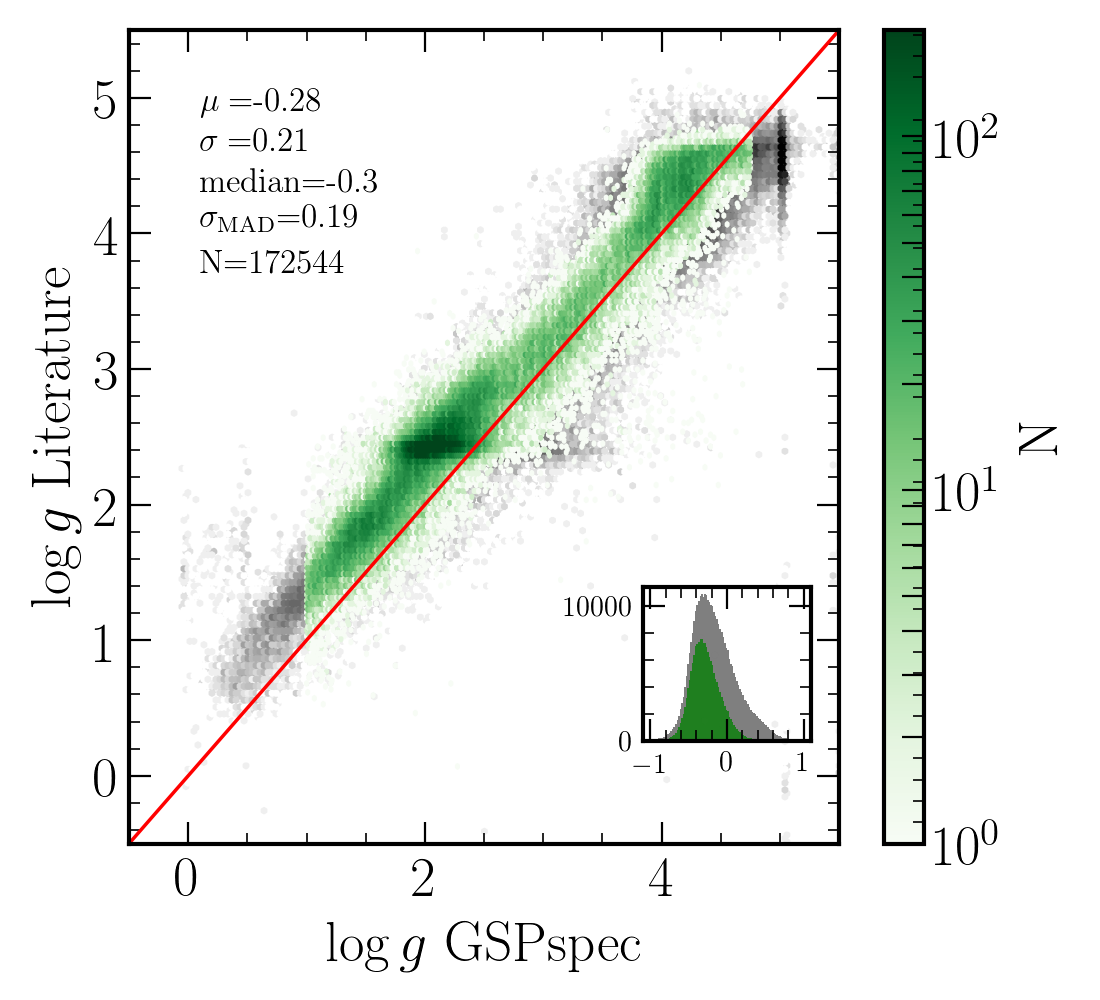}

\includegraphics[width=0.35\textwidth, angle=0]{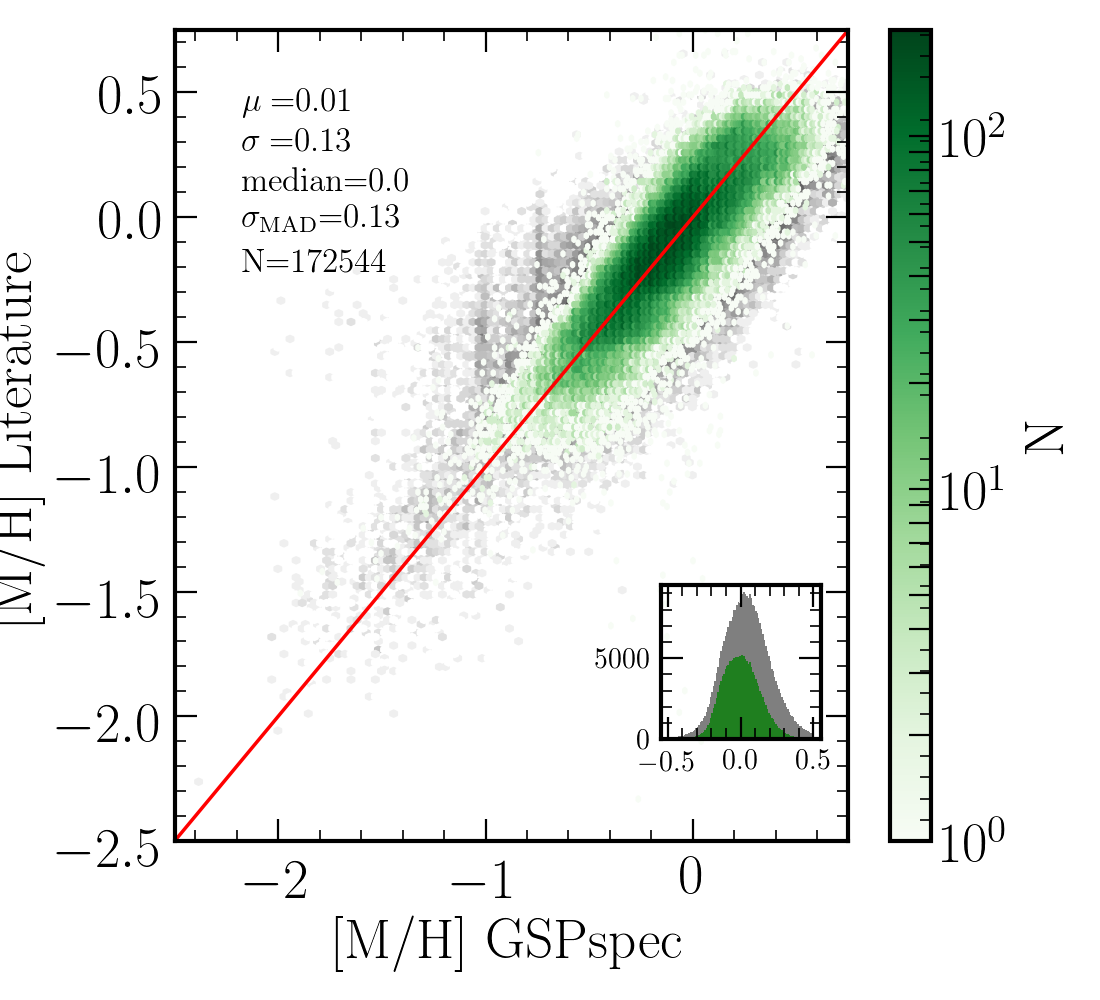}
\caption{Density plots comparing \gspspec\ MatisseGauguin parameters with literature data (APOGEE-DR17, GALAH-DR3, RAVE-DR6). 
Green and grey show the best- and medium-quality subsamples, respectively (see text for details about these samples). The histograms inside each plot show the difference between the literature and the \gspspec\ parameters. Mean ($\mu$), standard deviation ($\sigma$), median, robust standard deviation (derived from the MAD), and the number of stars ($N$) of the offsets for the best-quality subset are annotated inside each box.  }
\label{fig:VSTs_atmospheric}
\end{center}
\end{figure}

\subsection{\gspspec\ MatisseGauguin biases }
\label{MGBiases}
The MatisseGauguin workflow produces both atmospheric parameters and individual chemical abundances. Estimation biases have been evaluated for each case and are presented in the two following subsections. We have chosen to present the \alphaFe\ biases together with those of individual abundances, as the underlying spectral indicators are dominated by the \CaII\ IR triplet lines and, as a consequence, the behaviour of  \alphaFe\  is very similar to that of the [Ca/Fe] abundance. Subsection \ref{MGBiasSummary} summarises the observed biases and the proposed solutions.

%
%
%
%
    
    
    

\begin{figure}[t]
\begin{center}
\includegraphics[width=0.4\textwidth, angle=0]{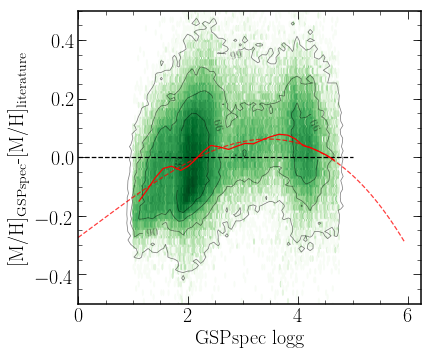}

\includegraphics[width=0.4\textwidth, angle=0]{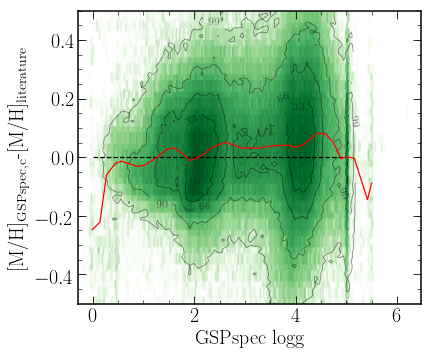}
\caption{Comparison of \gspspec\ and literature metallicities. Top: 2D histogram of the differences between the \gspspec\ metallicities and the literature values as a function of uncalibrated \g\  for our best-quality sample. The red full line is the running mean of the difference, and the dashed line is the fit to the running mean, defining the correction to apply. Bottom: Medium-quality sample showing the differences between the calibrated metallicities and the literature values. }
\label{fig:mh_calibration}
\end{center}
\end{figure}

\begin{figure*}[h]
\begin{center}
\includegraphics[width=\linewidth, angle=0]{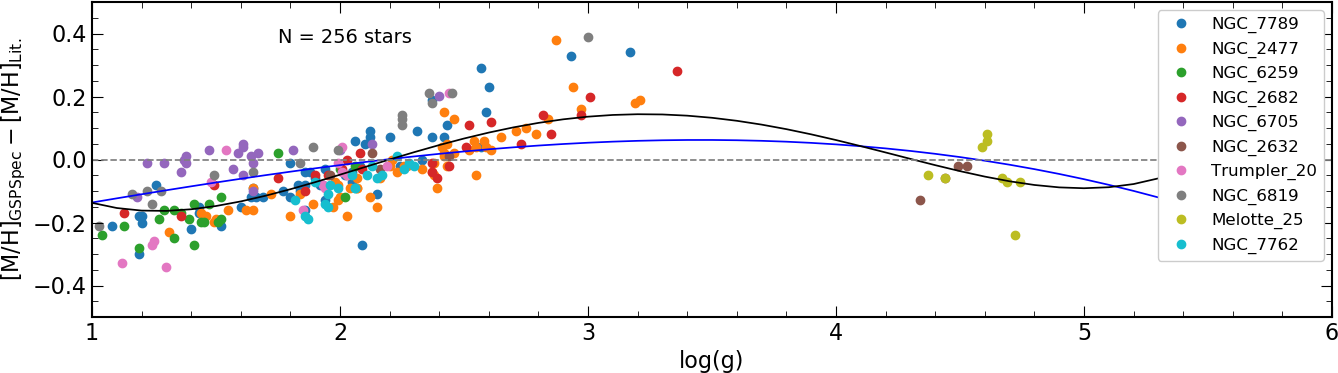}
\caption{Metallicity bias with respect to the literature as a function of \g\ for the open cluster stars, excluding dwarfs with S/N lower than 50. The colour code used for each cluster is indicated in the legend. The solid blue line corresponds to the general metallicity correction while the black line refers to that specifically obtained from the open clusters.}
\label{fig:calib_metOpenClusters}
\end{center}
\end{figure*}
   
\subsubsection{Analysis of \T, \g, and \meta}
\label{Sec:VST_parameters}

In this section, we compare the \gspspec\ MatisseGauguin \T, \g, and \meta\ with the latest data releases of three major ground-based spectroscopic surveys, namely APOGEE-DR17 \citep[][]{APOGEEDR17}, GALAH-DR3 \citep[][]{Buder2021}, and RAVE-DR6 \citep[][]{Steinmetz2020}. We filtered the literature samples based on both the associated  uncertainties of the published parameters ($\le$ 500\,K, 0.5, 0.3~dex, for \T, \g\ and metallicity/iron abundance, respectively)  and the reliability flags (following the suggestions of each of the respective surveys). In total, a sample of $\sim 8\cdot 10^5$ stars (among which $\sim 7.5\cdot 10^5$ unique targets) were selected in such a way.   The three panels in \figref{fig:VSTs_atmospheric} show how the main atmospheric parameters compare when all of the first 13 \gspspec\ flags are equal to zero (best quality sample, $\sim 1.7\cdot10^5$ stars plotted in green) and when we allow them to be smaller than or equal to one, except for the $KMgiantPar,$ which we insist must be equal to zero and the $fluxnoise$ flag that we relax to smaller than or equal to three (medium quality sample, plotted in grey, $\sim 3.7\cdot 10^5$ stars).

%
\begin{table}[t]
\centering
    \caption{Polynomial coefficients for the calibration of the MatisseGauguin gravities and metallicities.}
    \label{tab:MG_calibrations}
    \begin{tabular}{l|ccccc}
        Parameter & $p_0$ & $p_1$ & $p_2$ & $p_3$  & $p_4$\\ 
        \hline
         \g & 0.4496 & -0.0036 & -0.0224 & & \\
         \meta & 0.274 & -0.1373 & -0.0050 & 0.0048 & \\
         \meta$_{\rm OC}$ & -0.7541& 1.8108 &-1.1779  & 0.2809 & -0.0222\\
        \hline
    \end{tabular}
\end{table}

\begin{table*}[h]
    \centering
     \caption{Polynomial coefficients, recommended parameter intervals, and {\it extrapol} flag values for  Matisse-Gauguin \alphaFe\ and individual abundance calibrations (Eq.~\ref{Eq:calib_abund}). The uncertainties associated with these coefficients are provided in Table~\ref{tab:calibrations_errors}.  }
    \begin{tabular}{l|ccccc|cc|c}
    \hline
    \hline
    \cellcolor{gray!20} Element   &  \cellcolor{gray!20} $p_0$   & \cellcolor{gray!20}  $p_1$   & \cellcolor{gray!20}  $p_2$   &  \cellcolor{gray!20} $p_3$   &  \cellcolor{gray!20} $p_4$   & \multicolumn{2}{c}{\cellcolor{gray!20} Recommended interval} & \cellcolor{gray!20} {\it extrapol} flag\\ 
    \hline
    \hline
    & \multicolumn{5}{c}{\cellcolor{gray!20} As a function of \g} &  \cellcolor{gray!20} Min \g   &  \cellcolor{gray!20} Max \g & \cellcolor{gray!20} \\
    \arrayrulecolor{gray!60}\hline 
 \alphaFe   &   -0.5809    &    0.7018   &  -0.2402   &  0.0239  & 0.0000    &   1.01   &   4.85 &0\\
    \CaFe   &   -0.6250    &    0.7558   &   -0.2581   &   0.0256   &  0.0000    &   1.01   &   4.85 & 0\\
    \MgFe   &   -0.7244    &    0.3779   &   -0.0421   &  -0.0038   &  0.0000    &   1.30   &   4.38 &0\\
     \SFe   &  -17.6080    &   12.3239   &   -2.8595   &   0.2192   &  0.0000    &   3.38   &   4.81 &0\\
    \SiFe   &   -0.3491    &    0.3757   &   -0.1051   &   0.0092   &  0.0000    &   1.28   &   4.85 &0\\
    \TiFe   &   -0.2656    &    0.4551   &   -0.1901   &   0.0209   &  0.0000    &   1.01   &   4.39 &0\\
    \CrFe   &   -0.0769    &   -0.1299   &    0.1009   &  -0.0200   &  0.0000    &   1.01   &   4.45 &0\\
    \FeIH   &    0.3699    &   -0.0680   &    0.0028   &  -0.0004   &  0.0000    &   1.01   &   4.85 &0\\
    \FeIIH   &   35.5994    &  -27.9179   &    7.1822   &  -0.6086   &  0.0000    &   3.53   &   4.82 &0\\
    \NiFe   &   -0.2902    &    0.4066   &   -0.1313   &   0.0105   &  0.0000    &   1.41   &   4.81 &0\\
     \NFe   &    0.0975    &   -0.0293   &    0.0238   &  -0.0071   &  0.0000    &   1.21   &   4.79 &0\\
 \hline
  \alphaFe\   &  -0.2838   &   0.3713   &  -0.1236   &  0.0106   &  0.0002   &   0.84   &  4.44 & $\leq$~1\\
    \CaFe\    &  -0.3128   &   0.3587   &  -0.0816   &    -0.0066   &  0.0020   &  0.84   &  4.98 & $\leq$~1\\
   \arrayrulecolor{black} \hline
  & \multicolumn{5}{c}{\cellcolor{gray!20} As a function of $t=$\T/5750} &  \cellcolor{gray!20} Min \T   &  \cellcolor{gray!20} Max \T & \cellcolor{gray!20} \\
    \arrayrulecolor{gray!60}\hline
  \alphaFe\  & -6.6960 & 20.8770 & -21.0976 & 6.8313 & 0.0000 & 4000 &  6830 & $\leq$~1 \\
  \CaFe\  & -7.4577 & 23.2759 & -23.6621 & 7.7657 & 0.0000 & 4000 &  6830 & $\leq$~1 \\
   \SFe\  & 0.1930 & -0.2234 & 0.0000 & 0.0000 & 0.0000 & 5700 &  6800 & $\leq$~1 \\ 
\arrayrulecolor{black} \hline
\end{tabular}
    \label{tab:calibrations}
\end{table*}

For our best quality sample, we find a median offset for 
\T, \g, \meta~ of $-17$\,K, $-0.3$~dex and $0.0$\,dex, respectively, and a robust standard deviation (i.e. $\sim$1.48 times the median absolute deviation) of $90$\,K,  $0.19$~dex, and $0.13$\,dex. These trends are globally similar when taking into account each reference catalogue separately (see Appendix~\ref{appendix:param_biases_surveys}).
Whereas \T\ and \meta\ are globally well recovered (however, see next paragraph), \g\ determination is slightly biased. \gspspec\ MatisseGauguin finds consistently lower gravities, the offset being larger for giants than for dwarfs. Based on these findings, we suggest the following calibration for \g:  
\begin{equation}
\log(g)_{\rm calibrated}=\log(g)+ \sum_{i=0}^2 p_i \cdot \g^i
.\end{equation}
The $p_i$ coefficients  were obtained by fitting the trends with respect to the above-mentioned literature compilation, and are reported in the first row of Table~\ref{tab:MG_calibrations}.

Furthermore, we note that despite finding, overall, a zero offset in metallicity, a further investigation of the trends compared to the literature shows that giants (\g$\lesssim 1.5$) have slightly underestimated metallicities, whereas dwarfs (\g$\gtrsim 4$) have slightly overestimated values (see top plot of Fig.~\ref{fig:mh_calibration}). These trends can be corrected by fitting a low-order polynomial to the residuals as a function of \emph{uncalibrated} \g, and correcting the raw metallicities by this polynomial.
The correction takes the form of:
\begin{equation}
{\rm [M/H]}_{\rm calibrated}={\rm [M/H]}+ \sum_{i=0}^{deg} p_i \cdot \g^i
.\end{equation}
The $p_i$ coefficients are provided in Table~\ref{tab:MG_calibrations}. Two different corrections are proposed. The first one, a third-order polynomial, was obtained by fitting the trends with respect to the above-mentioned literature compilation. The result of this calibration is illustrated in the bottom plot of Fig.~\ref{fig:mh_calibration}. The second proposed correction, a fourth-order polynomial, is based on a set of open cluster stars with known metallicity from the literature and high membership probability \citep{CantatGaudin20,CastroGinard21,Tarricq21}. The advantage of open cluster data is that they ensure a constant metallicity at all \g\ values for the same object. However, as open clusters are thin-disc objects, the considered \meta\ range is restricted to the metal-rich regime. This alternative correction is illustrated in Fig.~\ref{fig:calib_metOpenClusters} and reported in the last row of Table~\ref{tab:MG_calibrations}.

\subsubsection{Analysis of \alphaFe\ and individual chemical abundances }
\label{sec:Elemental_calibrations}
\begin{figure*}[h]
\begin{center}
\includegraphics[width=\linewidth, angle=0]{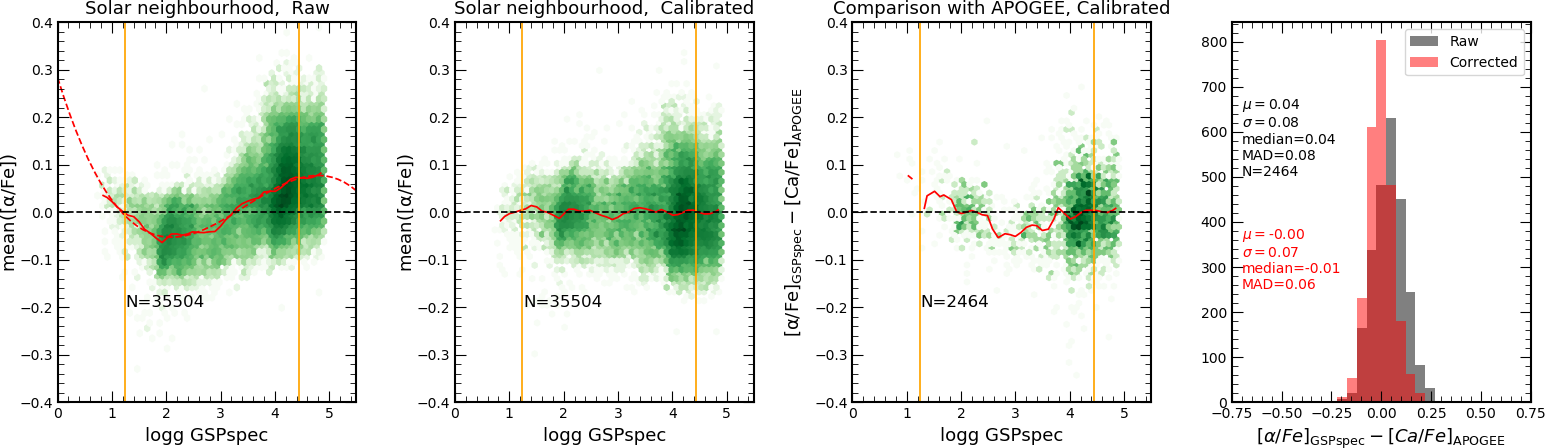}
\caption{Correction of \alphaFe\ trends as a function of \g. The left panel shows the 2D histogram of the stars with $3750$~K$\leq$\T$<$5750~K, \g$<4.9$  in green, with all of their quality flags equal to zero, located at the solar neighbourhood, with velocities close to the LSR and metallicities close to solar values in the raw (i.e. uncalibrated) \alphaFe-\g~space, colour-coded by $\log(N)$. The running mean is plotted as a full red line, and its fit is the red dashed line. The dashed black line is included as a visual reference for the y-axis. Vertical orange lines indicate the \g\  range over which the calibration is assumed to be reliable (differences between the fit and the running mean smaller than 0.05 dex). The second panel is similar to the left one, but the calibration has now been applied. The third panel shows the difference between the calibrated \alphaFe\ and the calcium values from APOGEE DR17 as a function of MatisseGauguin \g, where we have relaxed the $extrapol$ flag to be less than or equal to one. Finally, the right panel shows the histograms of the differences compared to the literature data before (in grey) and after (in red) the calibration. Quantifications of the mean, median, standard deviation, and robust standard deviation ($1.4826\cdot$\,MAD) are shown in the top left corner for the uncalibrated values (in grey) and in the bottom right corner for the calibrated values (in red).}
\label{fig:calib_alphaFe_only}
\end{center}
\end{figure*}

\begin{figure*}[h]
\begin{center}
\includegraphics[width=\linewidth, angle=0]{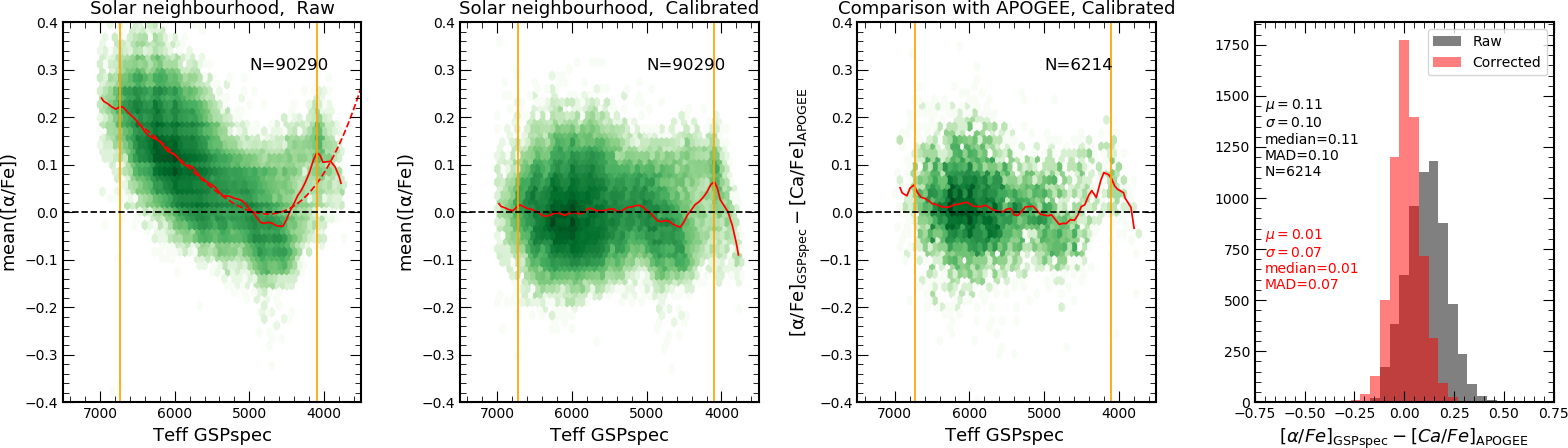}
\caption{Same as Fig.~\ref{fig:calib_alphaFe_only} but using the effective temperature as a reference parameter, instead of \g. The associated polynomial coefficients and applicability intervals are provided in Table~\ref{tab:MG_calibrations}}.
\label{fig:calib_alphaFe_Teff}
\end{center}
\end{figure*}


To evaluate, calibrate, and remove possible gravity dependencies on the measured \alphaFe,  \FeIH, \FeIIH,\ and [X/Fe] abundance values (with X being an arbitrary element), we follow the strategy described below. 
It assumes that the abundance distribution (expressed relative to the solar values) should be close to zero in the solar neighbourhood for stars with metallicities close to solar and velocities close to the Local Standard of Rest (to avoid stars with large eccentricities). This strategy furthermore has the advantage of avoiding any calibration based on external catalogues. The procedure that we carry out is the following. 
\medskip

    We first select only stars that have their first 13 quality flags (see \tabref{tab:MG_QFchain}) less or equal to one, except for their $KMgiantPar$ flag and $extrapol$ flag which we set to be equal to zero. In addition, we also impose that the abundance flag associated with the upper limit ($XUpLim$) is equal to zero, whereas the one associated with the uncertainties ($XUncer$) is set to less than or equal to one. Finally, we set an upper limit for their uncertainty (defined as the difference between the upper value and the lower value divided by two) and the line scatter to be less than 0.2~dex for both. 

    \smallskip Amongst the selected stars, we further select the ones that are located within $0.25$\,kpc of the Sun;  have a global metallicity \meta$=0.0\pm0.25$ dex (to avoid possible effects due to metallicty zero-point offsets); and have an azimuthal velocity $V_\phi$ close to the Local Standard of Rest ($V_{\rm LSR}\pm25$\,km\,s$^{-1}$).\footnote{Velocities are computed as in \citet[][]{Recio22}.} By choosing such a sample, we ensure that we select stars with a high probability of having, on average, similar chemical properties to the Sun. Therefore, their $\rm [X_1/X_2]$ abundance distributions (with $X_1$ and $X_2$ associated with two different elements or families of elements) are expected to be centred on zero.   
    
    \smallskip We then compute the running mean of $\rm [X_1/X_2]$ as a function of \g, in bins of $\delta$\g$=0.2$~dex (red full line on the first row of plots in Figs.~\ref{fig:calib_alphaFe_only}, \ref{fig:calibs_alphas}, and \ref{fig:calibs_irons}). This trend, for an unbiased abundance estimation, should be centred on zero, regardless of the dispersion of the underlying distribution (which is a manifestation of either a true Galactic dispersion, or of the precision of \gspspec, or both).  
    
    \smallskip Finally, we fit the trend defined by the running mean with a third- or fourth-order polynomial (choosing the correct compromise, depending on the data behaviour, to avoid overfittings), where each point has a weight inversely proportional to the dispersion of $[X_1/X_2]$ within the considered \g-bin. This fit defines the correction that could be  applied to our data (red dashed line on the leftmost panels in Figs.~\ref{fig:calib_alphaFe_only}, \ref{fig:calibs_alphas}, and \ref{fig:calibs_irons}). 
    The correction takes the form of: 
    \begin{equation}
    \label{Eq:calib_abund}
        [X_1/X_2]_{\rm calibrated}=[X_1/X_2] + \sum_{i=0}^{deg} p_i \cdot \g^i
    ,\end{equation}
where $deg$=3 or 4 and X$_2$ is either Fe or H, depending on the chemical species (see Table~\ref{tab:calibrations}).

    \smallskip
    We also define the \g\  range over which the calibration is expected to be valid (vertical orange lines in the figures). The latter is evaluated by estimating the difference between the running mean and the fit, $\Delta_{\rm fit}$, and excluding the points at \g$\pm0.4$ from the boundaries for which $\Delta_{\rm fit}$ is larger than 0.05\,dex (chosen arbitrarily). We note that the application of the calibration outside this \g\ confidence range should be used with caution, if not avoided. To increase the validity \g\ range for the abundances with high number statistics (\alphaFe\ and \CaFe), we propose another calibration by relaxing the \gspspec\ quality flag associated 
    to the extrapolation ($\leq1)$. This leads to an alternative fourth-order polynomial fitting (second and third last rows of Table.~\ref{tab:calibrations}) and allows a qualitative view of how the correction behaves outside the \g-confidence range of the third-order polynomial. 

    
    \smallskip We then verify on the same sample that the correction improves the trends (second column of plots of Figs.~\ref{fig:calib_alphaFe_only}, \ref{fig:calibs_alphas}, and \ref{fig:calibs_irons}). 
    
    \smallskip Finally, we use literature data, which contain a wider variety of metallicities, to verify that the calibration is indeed  improving the offsets (third and fourth columns of  Figs.~\ref{fig:calib_alphaFe_only}, \ref{fig:calibs_alphas}, and \ref{fig:calibs_irons}). The literature data we use in this case are composed of APOGEE-DR17 and GALAH-DR3 for all of the elements except sulphur, and AMBRE for this latter abundance \citep[][]{Jeremy21}. In the case of \alphaFe,\, the comparison is made with respect to literature \CaFe\ values, as in the RVS domain the \alphaFe\ indicators are dominated by the CaT lines. We note that, for these abundance comparisons, no agreement was required between \gspspec\ and the literature in the related stellar atmospheric parameters or the assumed solar abundances. 
    

It can be seen from  Fig.~\ref{fig:calib_alphaFe_only},  Fig.~\ref{fig:calibs_alphas}, and Fig.~\ref{fig:calibs_irons} that the provided calibrations for the \alphaFe\ and individual chemical abundance offsets significantly  decrease their gravity dependence (and even remove it completely for several species), and that they set them close to the solar values. Moreover, the comparison with literature data is also improved, reducing the offset and/or the dispersion. 
The values of the polynomial coefficients of Eq.~\ref{Eq:calib_abund} ($p_i$) together with their domain of validity in \g, to avoid extrapolations, are listed in Table~\ref{tab:calibrations}. We also provide the uncertainties on the polynomial coefficients  (derived from the fit) in Table~\ref{tab:calibrations_errors}, as a possible criterion to evaluate the robustness of the calibration. 
 We note that, for some elements, a lower order polynomial might be sufficient to fit the data, but the verification made on the datasets suggests that we nevertheless correct without overfitting with a third-order polynomial. \\
 
 Interestingly, the methodology described above does not allow calibration of the Zr and Nd abundances, as as insufficient number of stars are selected with the criteria previously described.  Furthermore, we note that the distribution of the literature \NdFe\  values \citep[GALAH-DR3,][]{Buder2021} found for our solar neighborhood sample does not peak at 0 dex; therefore, an offset correction of \NdFe\ would be meaningless. For the same reason, we do not apply any correction to the \gspspec\ \CeFe\  abundances since our cross-match with literature \citep{Hinkel14, APOGEEDR17, Forsberg19, Buder2021} reveals a similar offset in the cerium distribution with respect to the solar value. It is also important to note that the \g\ domain covered by the abundances of these three heavy elements is not very large, minimising gravity trends in the results.\\
 
Finally, it could be convenient for some scientific purposes to calibrate the abundance trends as a function of the effective temperature instead of the gravity. This is particularly the case when hot dwarf stars are included in the used sample. For this reason, we provide an example of this alternative calibration applied to \alphaFe\ and illustrated in Fig.\ref{fig:calib_alphaFe_Teff}. The derived third-order polynomial is provided at the end of Table~\ref{tab:calibrations}. In those cases, the \g\ variable in Eq.~\ref{Eq:calib_abund} should be replaced by the effective temperature. The use of similar calibrations as a function of \T\ for other chemical abundances or to correct gravity trends has to be evaluated by the user, depending on the target sample and scientific goals. Such calibrations are not provided here for clarity.

\subsubsection{Summary of \gspspec\ MatisseGauguin biases and proposed solutions}
\label{MGBiasSummary}
In this section, we summarise the observed MatisseGauguin parameter and abundance biases, as well as the recommended solutions.
\smallskip

\noindent \emph{Effective temperature}: 

\smallskip
No significant biases are observed in \T. For hot stars, rotational mismatches could nevertheless affect the results.

\smallskip

{\it Proposed solution}: The {\it vbroadT} flag (or the vbroad value) has to be checked and/or used to clean the samples.

\medskip
\noindent \emph{Surface gravity}: 

\smallskip
A bias in \g\ is present. The median value is 0.3~dex on the entire parameter space. It shows a slight trend with \T, getting worse as \T\ decreases. This could be related to the progressive dominance of the CaT lines as \g\ indicators (they become stronger along the giant branch as \T\ decreases), and to the absence of Paschen lines for \T$\la$5500~K. No clear relation with line broadening mismatches seems to exist.

\smallskip
{\it Proposed solution}: A global correction is proposed based on literature data. For dwarf stars, a correction based on \T\ could offer more precise corrections, as the \T\ range is higher than the \g\ one. We generally advise to optimise the correction to the parameter space of the user.

\medskip
\noindent \emph{Global metallicity}: 

\smallskip
The observed \g\ bias seems to be associated with a slight \meta\ bias, presenting a similar behaviour with \T\ for \T$\la$5500~K (not related to line broadening mismatches). For hotter stars, rotational mismatches could also cause a bias.

\smallskip
{\it Proposed solution}: Global corrections are proposed based on literature data of (i) field stars and (ii) open clusters. These corrections are only significant for giant stars in the low-\g\ regime. For hot stars, the {\it vbroadM} flag (or the vbroad value) has to be checked and/or used to clean the sample.

\medskip
\noindent \emph{\alphaFe\ and individual abundances}: 

\smallskip
Atmospheric parameter biases are linked to abundance biases. In the \gspspec MatisseGauguin case, the main sources seem to be the \g\ bias and the rotational mismatch. On the contrary, the impact of the observed slight metallicity biases is probably reduced thanks to the fact that most abundances are derived with respect to iron.
In the regime \T$\la$5500~K, \alphaFe\ and individual abundance biases are of small amplitude and show a very weak trend with \T\ and/or \g. In the regime \T$\ga$5500~K, abundance biases seem dominated by rotational missmatches.

\smallskip
{\it Proposed solution}: Global corrections are proposed as a function of \g, based on a zero-point calibration to the
Local Standard of Rest at solar metallicity. Alternatively, global corrections as a function of \T\ can be implemented (as in that proposed for \alphaFe), with
very similar results in the regime of \T$\la$5500~K. For samples with a short \T\ and \g\ coverage, a constant shift to the solar
value at \meta$=$0 can be implemented if the user prefers to work with raw parameters (although the proposed calibrations are still valid). For dwarf stars, when including the hot temperature regime, a correction as a function of \T\ should be implemented. Alternatively, the sample can be cleaned using the {\it vbroadT}, {\it vbroadG,} and {\it vbroadM} (or the vbroad value). We generally advise optimisation of the correction to the parameter space of the user.



\begin{figure}[t]
    \begin{center}
        \includegraphics[width=0.35\textwidth, angle=0]{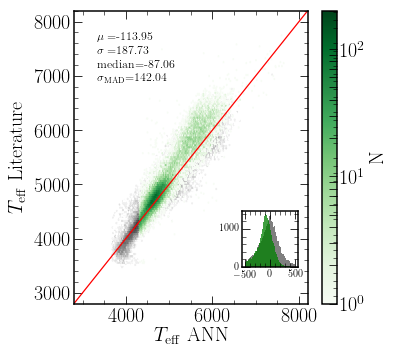}

        \includegraphics[width=0.32\textwidth, angle=0]{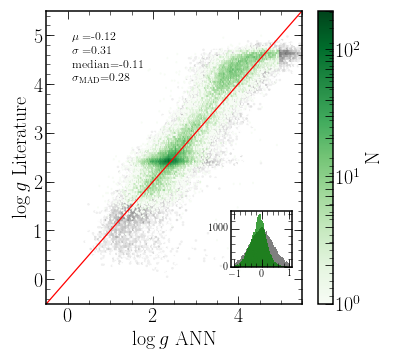}

        \includegraphics[width=0.35\textwidth, angle=0]{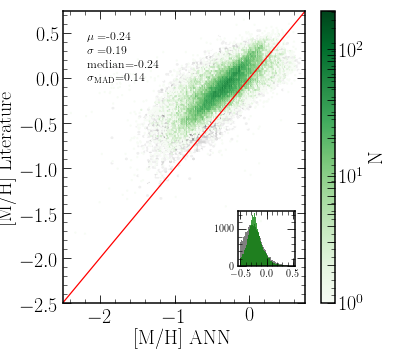}
        \caption{Same as Fig.\ref{fig:VSTs_atmospheric} but for \gspspec-ANN, published in the complementary table {\it AstrophysicalParametersSupp}. The reference high-quality subsample used for the comparison statistics is different from that shown in Fig. \ref{fig:VSTs_atmospheric}  (for \gspspec-MatisseGauguin), as imposed by the ANN quality flags.}
        \label{fig:ANN:VSTs_atmospheric}
    \end{center}
\end{figure}

\subsection{\gspspec\ ANN biases} \label{ANNBiases}
The ANN workflow produces an alternative estimation of the stellar atmospheric parameters (\T, \g, \meta,\ and \alphaFe) that can be found in the {\it AstrophysicalParametersSupp} table. In the following, the observed ANN biases with respect to the literature are presented, proceeding in a similar way to that described for MatisseGauguin results (cf. \secref{Sec:VST_parameters}). 

\begin{figure*}[t]
\includegraphics[width=0.5\textwidth,height=4.cm,angle=0]{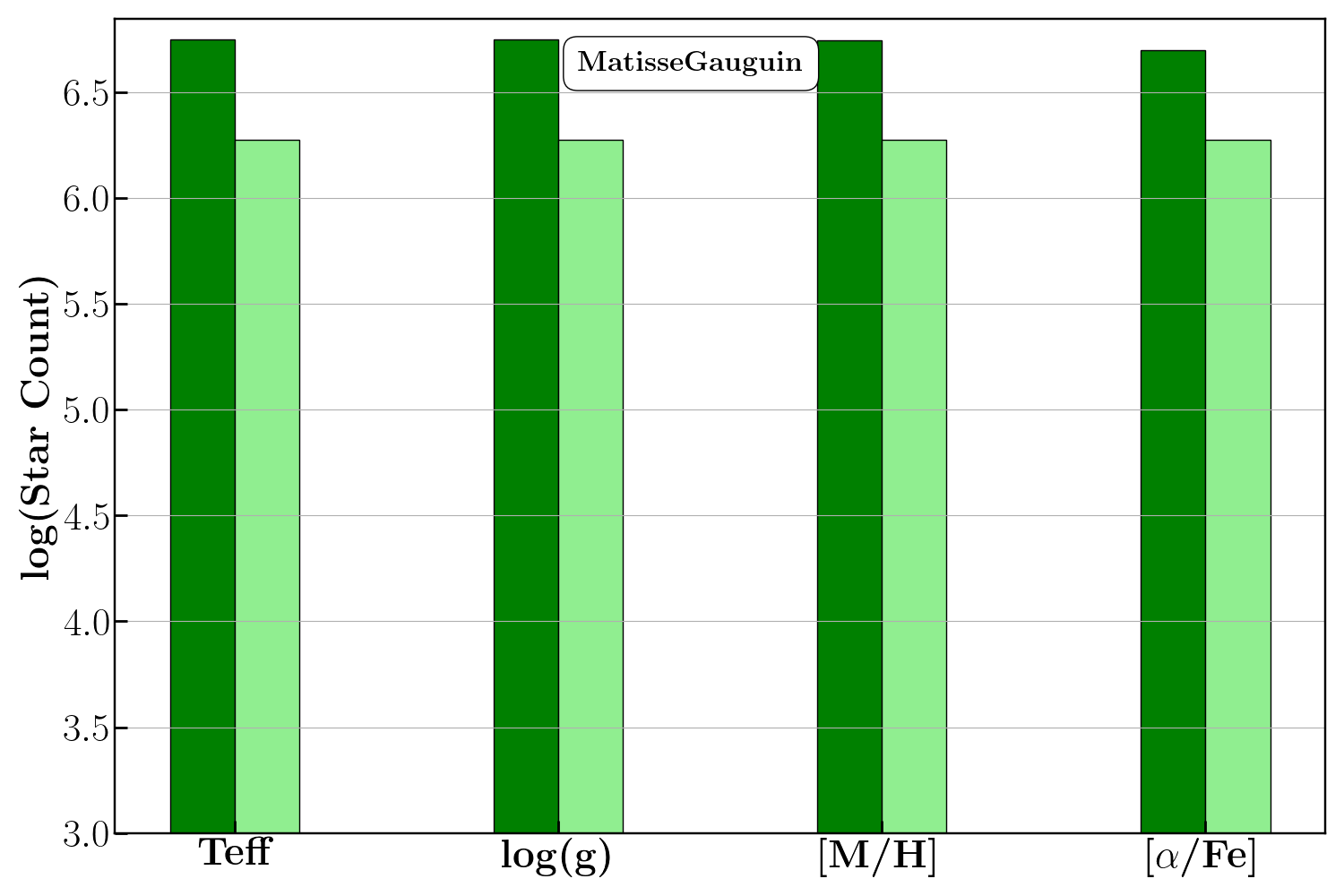}
\includegraphics[width=0.5\textwidth,height=4.cm,angle=0]{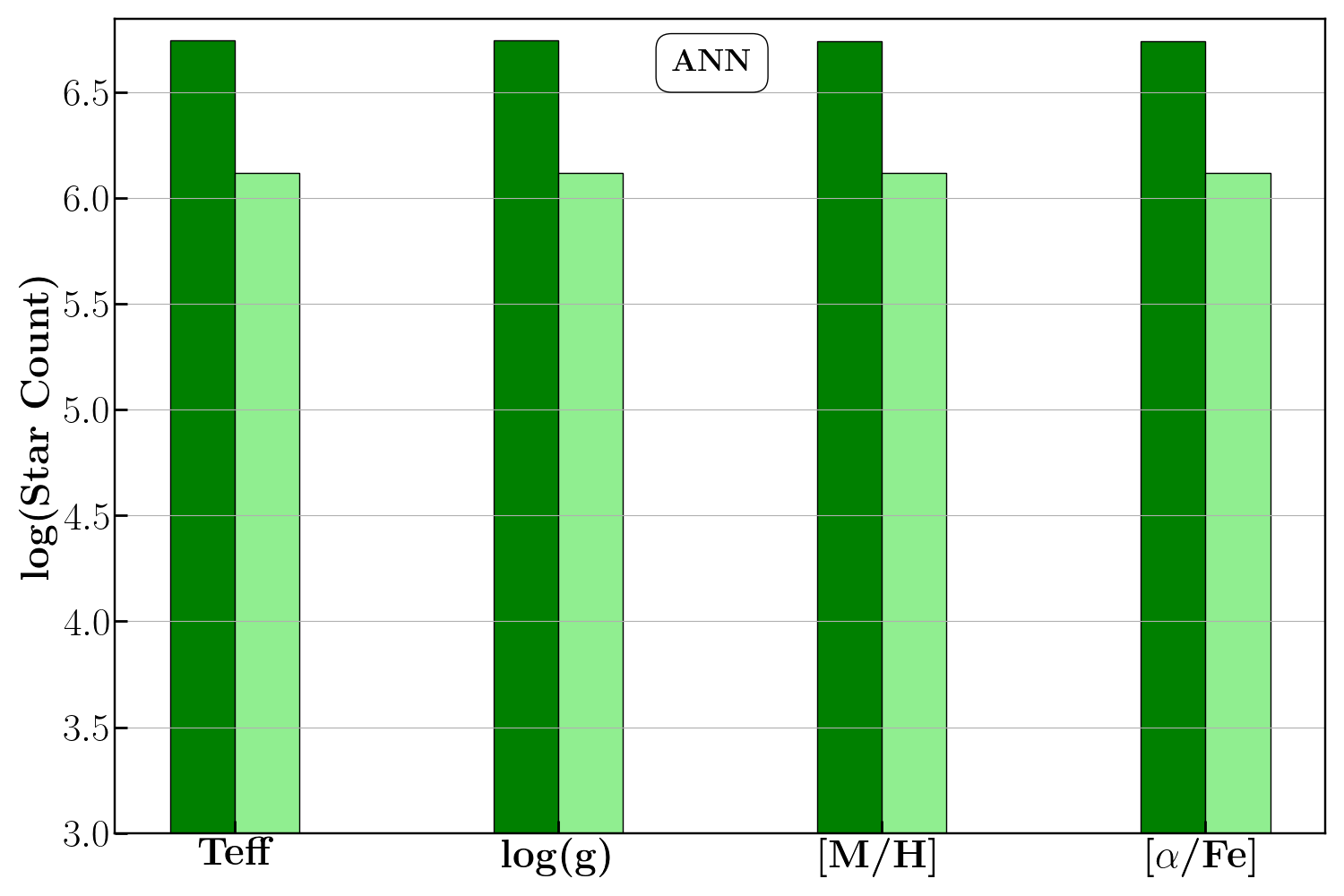}
\caption{Number of stars whose atmospheric parameters have been derived by MatisseGauguin and ANN (left and right panels, respectively). The dark green histograms refer to the whole sample whereas the light-green ones show only the very best parametrised stars with all their parameter quality flags equal to zero.}
\label{Fig.StatParam}
\end{figure*}

\begin{figure*}[t]
\includegraphics[width=1.0\textwidth,height=7.cm,angle=0]{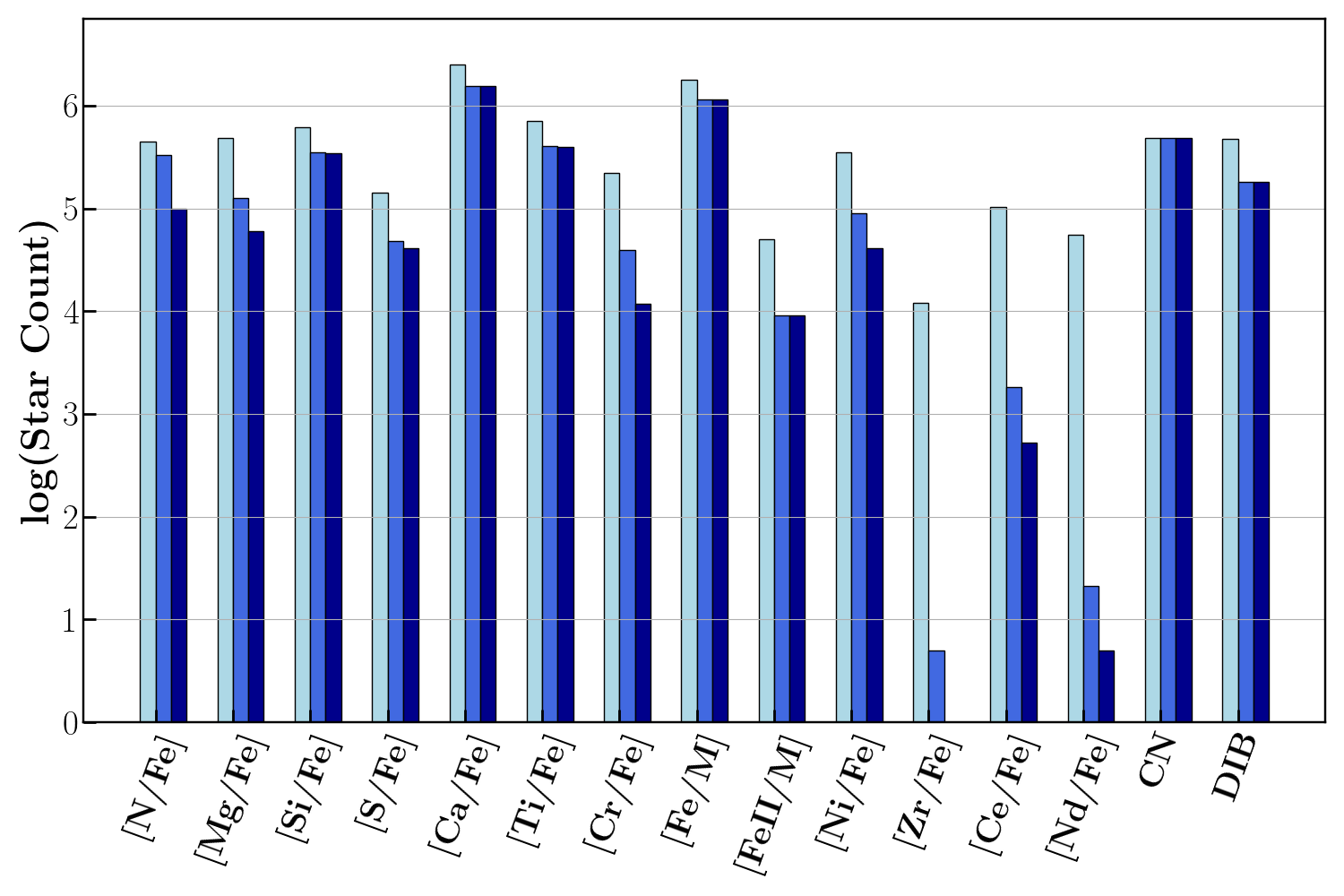}
\caption{Same as Fig.~\ref{Fig.StatParam} but for the individual abundances derived by GAUGUIN plus the CN-abundance proxy and the DIB. The light-blue histogram (left bars) refers to the whole sample. The two other sets of bars (central and right bars) show only the very best stars with all their parameter flags and their abundance uncertainty quality equal to zero. The abundance upper limit flag is lower than or equal to one and equal to zero for the medium-blue and dark-blue bars, respectively.}
\label{Fig.StatAbund}
\end{figure*}

\begin{figure*}[h]
\includegraphics[width=1.0\textwidth, height=7.cm]{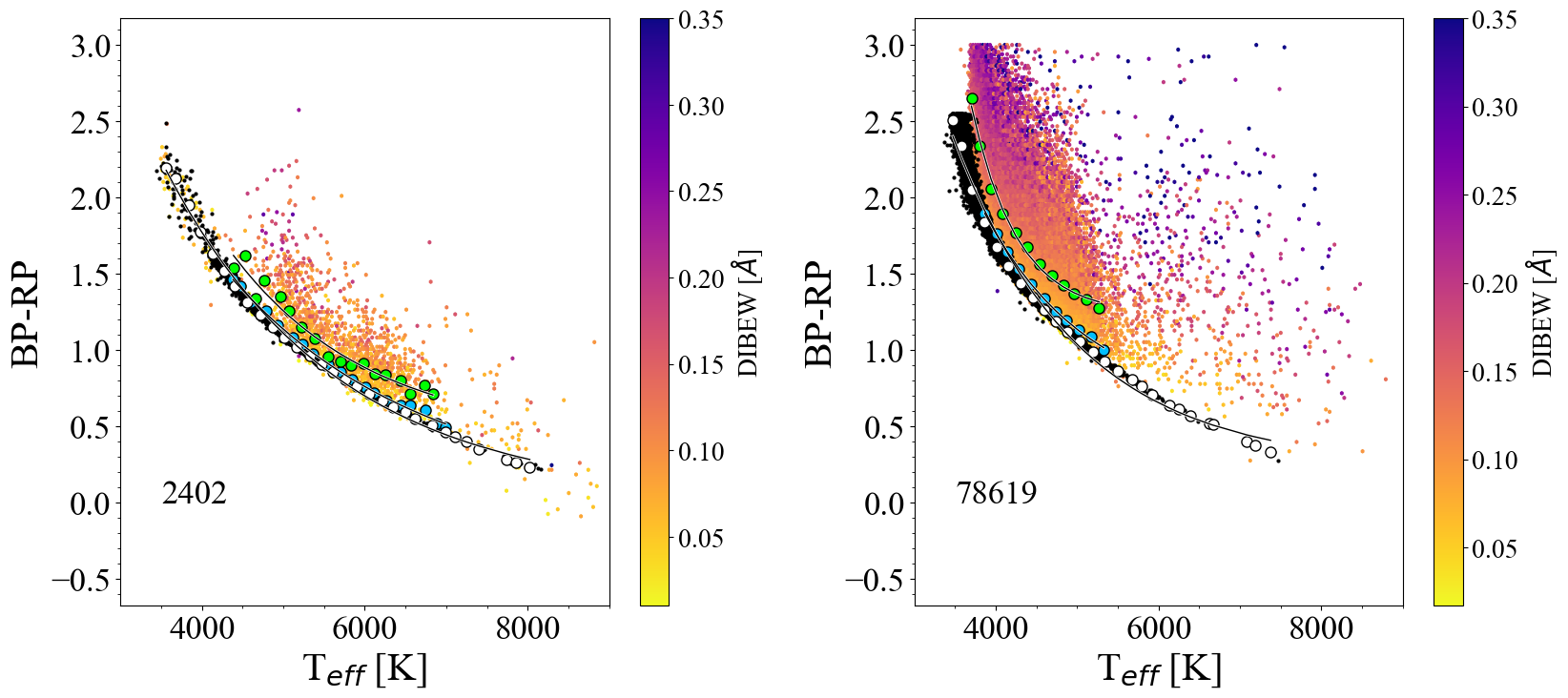}
\caption{Trend of (BP-RP) colour with \gspspec\ effective temperature produced by the MatisseGauguin workflow for dwarfs (left panel) and giants (right panel). The colour code indicates the estimated DIB EW, which increases with interstellar absorption (the DIB flag has been imposed to be equal to zero). Blue circles show the median values of the distribution for the stars with a DIB EW lower than 0.05 \AA. Green circles are the median values for stars whose DIB EW is equal to the median value of the distribution (0.07 \AA\  for dwarf stars on the left panel, and 0.12 \AA\ for giants on the right panel), plus a dispersion of $\pm$0.01~\AA. Black dots (and white circles) are the values (and their median) predicted by the \cite{Casagrande21} relation, assuming no extinction.}
\label{Fig.ColourTeffDib}
\end{figure*}

\begin{figure*}[h]
\includegraphics[width=1.0\textwidth, height=7.cm]{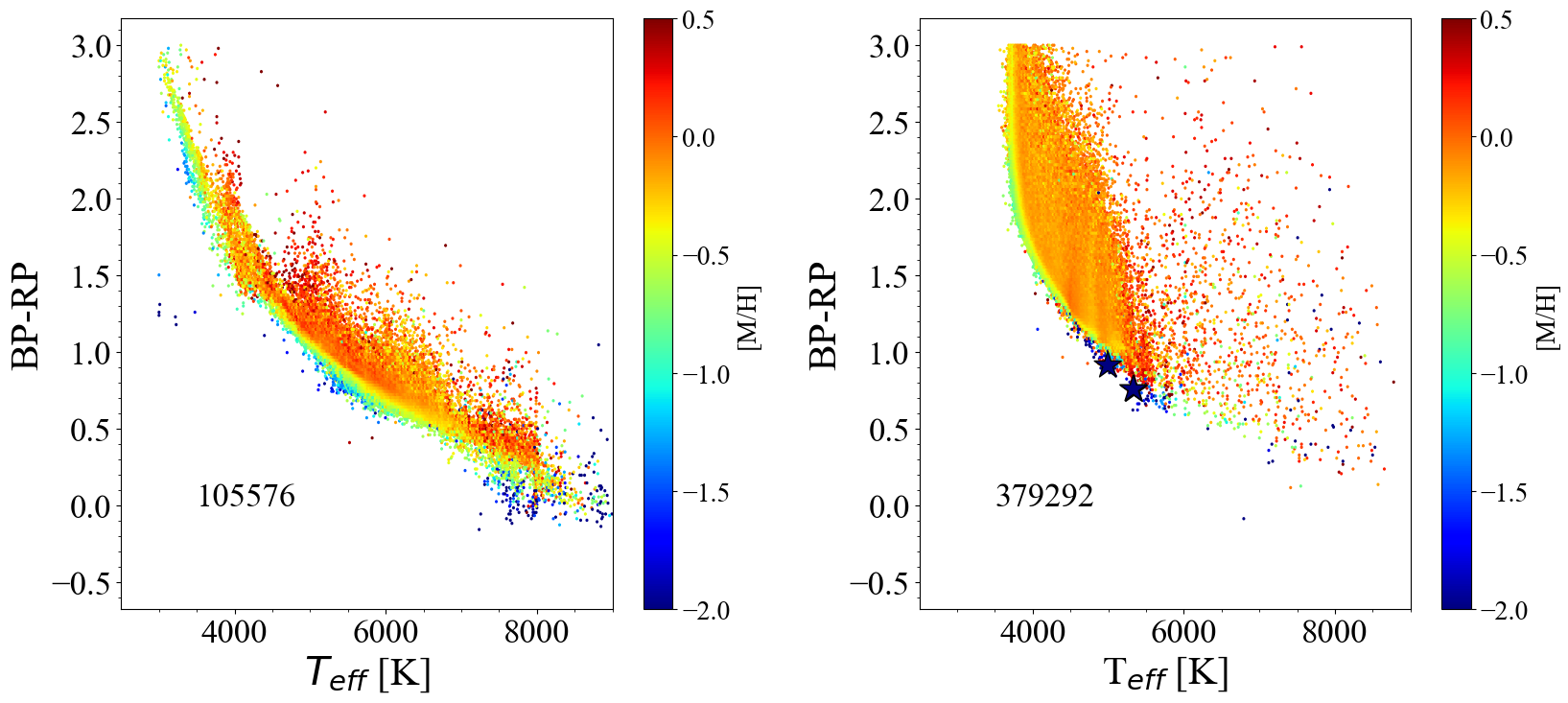}
\caption{Same as Fig.\ref{Fig.ColourTeffDib} but using the estimated stellar metallicity [M/H] as colour code. The selected stars have the first 13 quality flags in the gspspec flagging chain equal to zero. Two extremely metal-poor stars, discussed in Sect.\ref{subsec:UMPs}, are indicated by star symbols. The number of stars is indicated in each panel.}
\label{Fig.ColourTeffMeta}
\end{figure*}

\figref{fig:ANN:VSTs_atmospheric} shows the comparison with the literature for two ANN subsamples: the best quality, in green, and the medium quality, in grey.
For our best-quality sample, we selected all the sources with the first eight flags equal to zero, excluding those with broadening and radial velocity issues, higher noise uncertainties, or extrapolations. The median offsets for the 274\,592 stars of the best-quality sample are  $-114$ K, $-0.12$~dex, and $-0.24$ dex for \T, \g, and \meta, respectively, and the corresponding mean absolute deviations are $142$ K, $0.28$~dex, and $0.14$ dex. 

Compared to the literature, 
ANN results present a larger bias than MatisseGauguin in \T \ and \meta, and a slightly lower bias in \g. Nevertheless, these differences come partially from the fact that the ANN quality flags select a different  reference subsample for the comparison statistics than the one used for \gspspec-MatisseGauguin. In particular, cooler giants are outside the ANN high-quality selection. Finally, the dispersion for the ANN parameterisation is also higher than for MatisseGauguin, particularly for \T\ and \g. 

We propose simple polynomial calibrations for \T, \g,\ and \meta\ based on the above comparison with the literature using the $best$-quality sample.
It is important to note that the \T\ calibration of ANN is S/N dependent (c.f. \ref{ANN:internal}) because the ANN algorithm was trained with synthetic spectra in five S/N levels (cf. \secref{Sec:ANN}). 

We focus in the following on the high-S/N regime (S/N$_{ANN}>$50 corresponding to S/N$>$108, c.f. \tabref{tab:ann_SNReq}). For lower S/N values, we refer the reader to Sect.\ref{ANN:internal}, where the correct S/N optimisation of the algorithm is validated. We also highlight that, as the number of stars with \T\ > 6000 K in the literature is small, the proposed corrections should not be applied beyond this temperature limit. The resulting calibrations take the form of:
\begin{equation}
    X_{calibrated} = X + \sum_{i=0}^{deg} p_i \cdot X^i
,\end{equation}
where $p_{i}$ coefficients for each parameter calibration can be found in \tabref{tab:ann_main_calibrations}. Moreover, similarly to MatisseGauguin results, we also suggest a calibration for \alphaFe, independent of literature data:
\begin{equation}
    \text{\alphaFe}_{calibrated} = \text{\alphaFe} + \sum_{i=0}^3 p_i \cdot \text{\g}^i
.\end{equation}
\begin{table}[t]
\centering
    \caption{Polynomial coefficients for the calibration of ANN parameters (at S/N$_{ANN}\sim$50 for \T, see Appendix~\ref{ANN:internal} for other S/N$_{ANN}$ values).}
    \label{tab:ann_main_calibrations}
    \begin{tabular}{c|cccc}
        Parameter & $p_0$ & $p_1$ & $p_2$ & $p_3$ \\ 
        \hline
         \T & 12816 & -8.1 & 1.65E-3 & -1.07E-7
\\
         \g & -0.006 & 0.023 & & \\
         \meta & 0.092 & -0.446 & -0.07 & \\
         \alphaFe & -0.038 & 0.099 & -0.052 & 0.006 \\
        \hline
    \end{tabular}
\end{table}

In summary, although  ANN parameters 
present slightly higher biases and uncertainties than MatisseGauguin ones (and they are therefore published in the complementary table AstrophysicalParametersSupp),
their overall quality provides a methodologically different parametrisation, which could be useful, in particular, to test the MatisseGauguin classification in the low-S/N regime.

\begin{figure*}
\includegraphics[width=0.5\textwidth,angle=0]{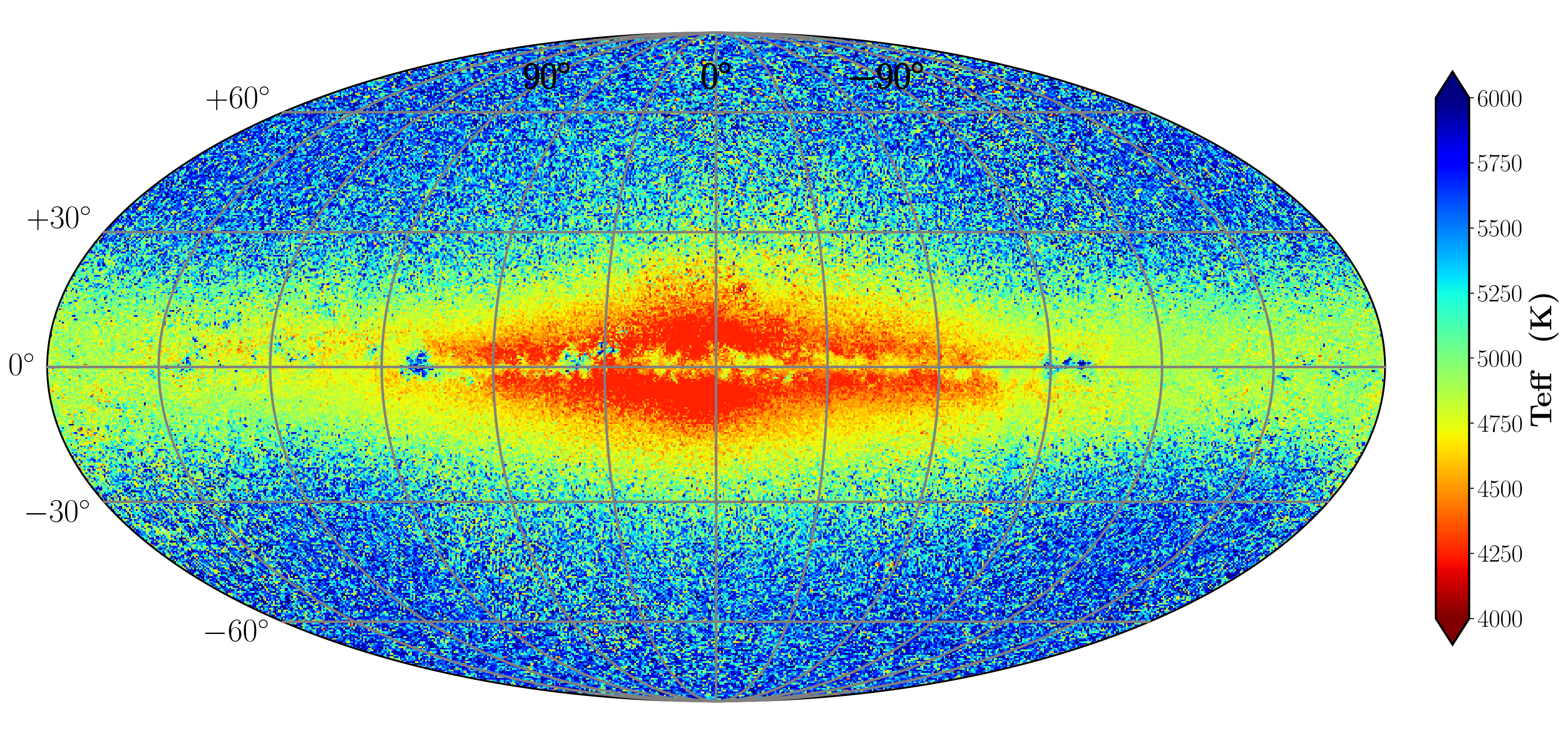}
\includegraphics[width=0.5\textwidth,angle=0]{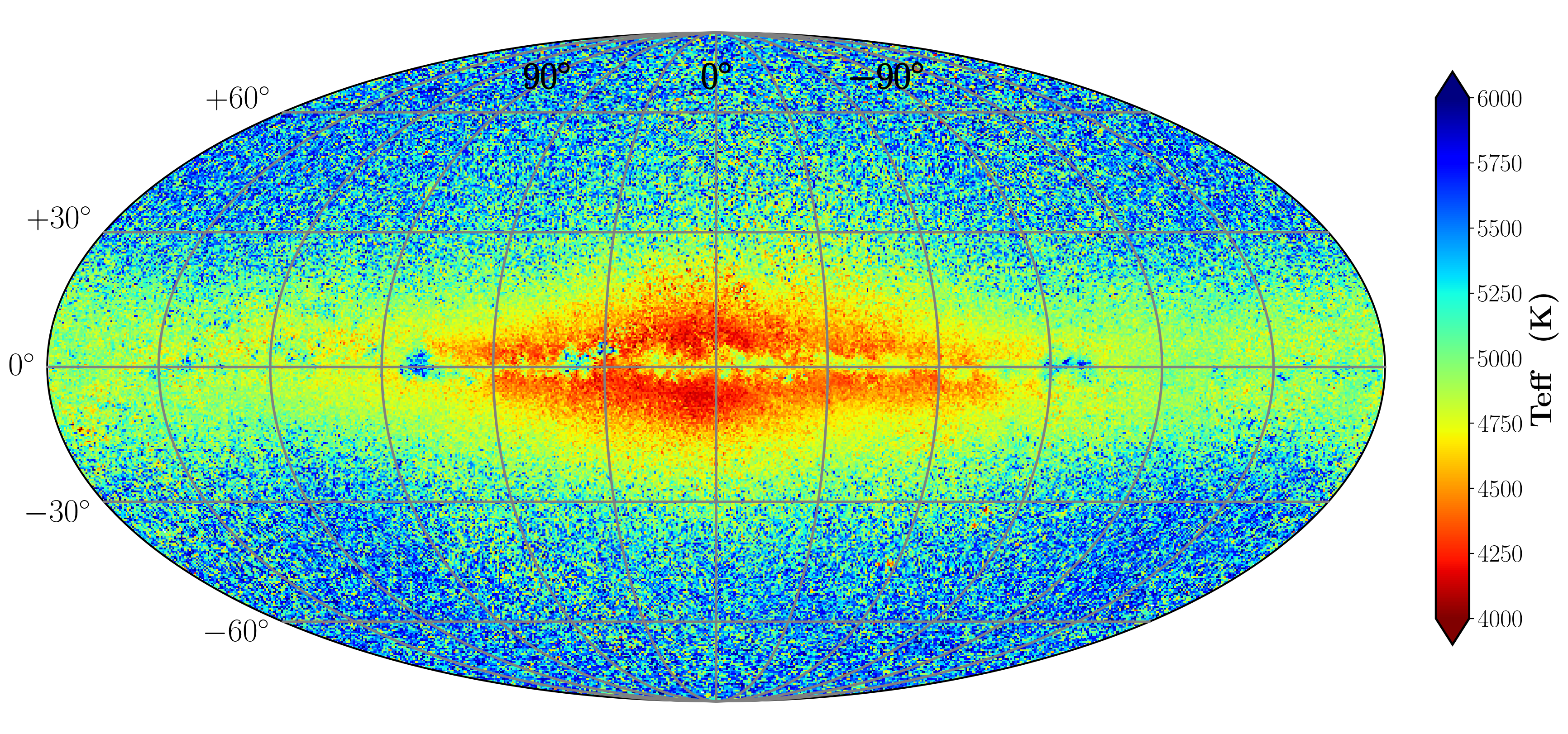}
\caption{Milky Way as revealed by the \gspspec\ effective temperature estimated by  MatisseGauguin (left) and ANN (right).
These HEALPix maps in Galactic coordinates have a spatial resolution of 0.46$^\circ$. The colour code corresponds to the median of \T\ in each pixel. }
\label{Fig.MilkyWay_Teff}
\end{figure*}

\section{Illustration of \gspspec\ results}
\label{Sec:Results}
Illustrating all DR3 \gspspec\ results is obviously out of the scope of this paper. Two performance demonstration articles exclusively based on \gspspec\ MatisseGauguin parameters show their detailed application to Galactic chemo-dynamical studies of the disc and halo populations \citep[][]{Recio22}, and interstellar medium studies through the RVS diffuse interstellar band carrier \citep[][]{Schultheis22}. The homogeneous \gspspec\ treatment of the exhaustive all-sky RVS survey enables a chemo-physical parametrisation quality  comparable to that of ground-based surveys of higher spectral resolution and wavelength coverage.
Examples of this are the precision in the estimated individual chemical abundances (including heavy elements) allowing chemo-dynamical studies of Galactic stellar populations, DIB parameter estimation from individual spectra, and the precision in the atmospheric parameters providing clear constraints on stellar evolution models (see below).

In the following, we provide a few more examples of \gspspec\ results, focusing on (i) the number of parametrised stars in different quality regimes, (ii) the colour--effective temperature relation, (iii) an illustration of the \T\ spatial distribution, (iv) the atmospheric parameters of high-S/N spectra and associated constraints on stellar evolution models, 
and (v) the parametrisation of very metal-poor stars. 
\begin{figure*}[h]
\includegraphics[width=0.33\textwidth]{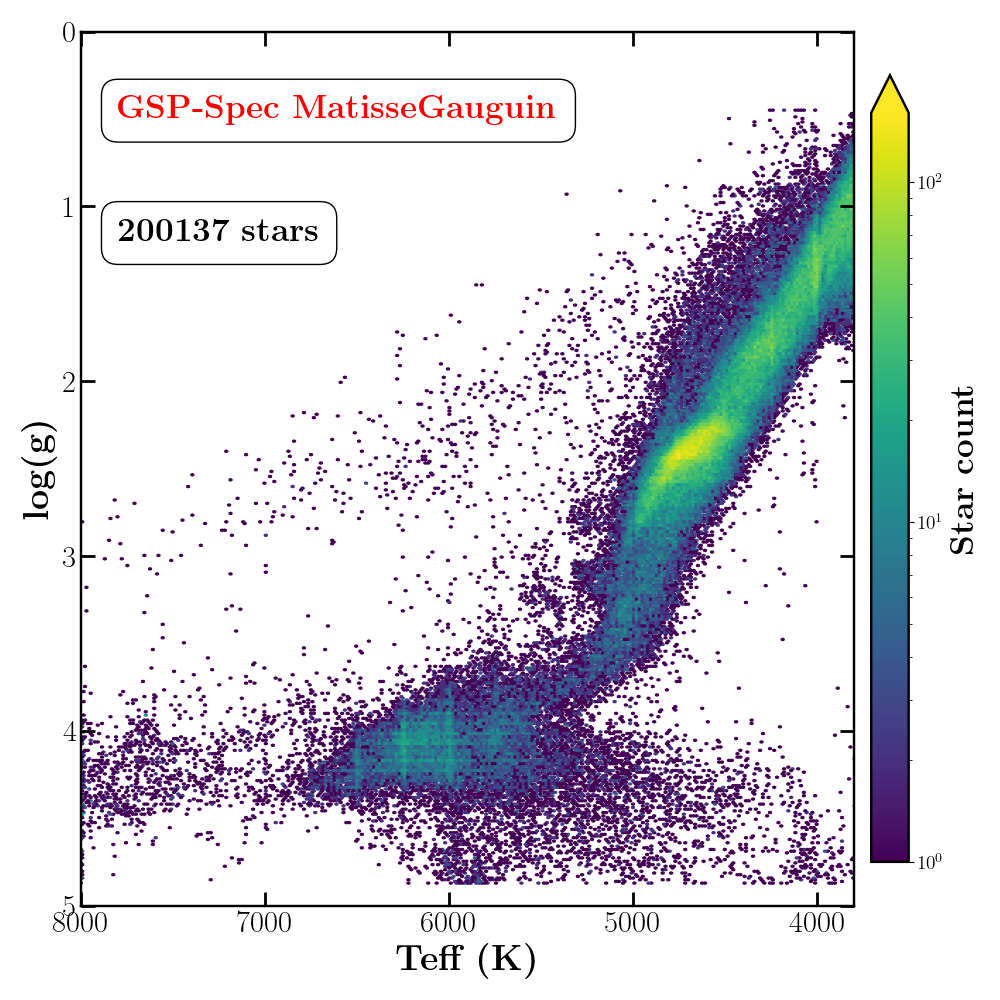}
\includegraphics[width=0.33\textwidth]{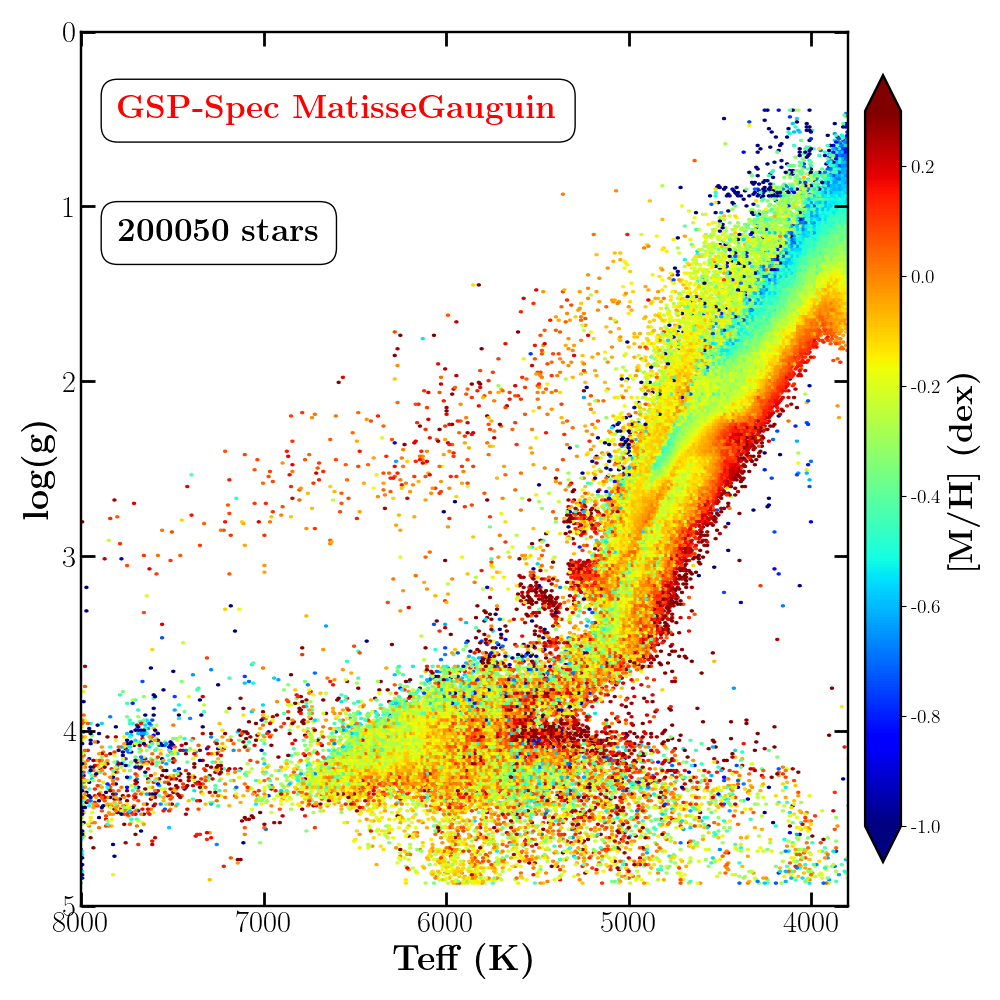}
\includegraphics[width=0.33\textwidth]{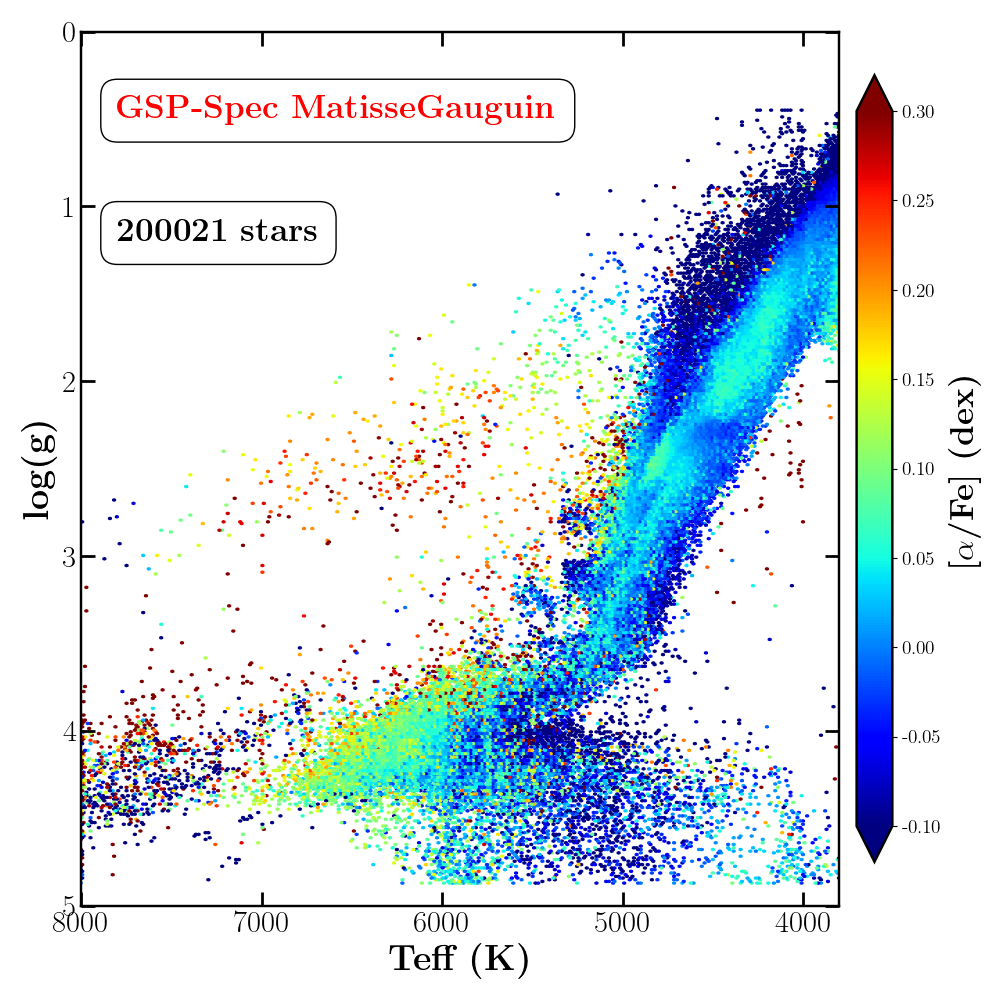}
\caption{Kiel diagrams for the MatisseGauguin output parameters (stored in the main DR3 astrophysical parameters table) for high-quality spectra (S/N>150) and excluding  high-rotating stars ($vbroadT=vbroadG=vbroadM=0$) and possibly misclassified very cool giants ($KMtypestars$=0). The colour codes of the different panels show the stellar density (left panel) and the median of \meta\ and \alphaFe\ per point (central and right panels, respectively). The proposed \g\ and \alphaFe\ calibrations are applied.}
\label{Fig.MG_Kiel}
\end{figure*}

\begin{figure}[]

\includegraphics[width=0.5\textwidth, height=8.cm]{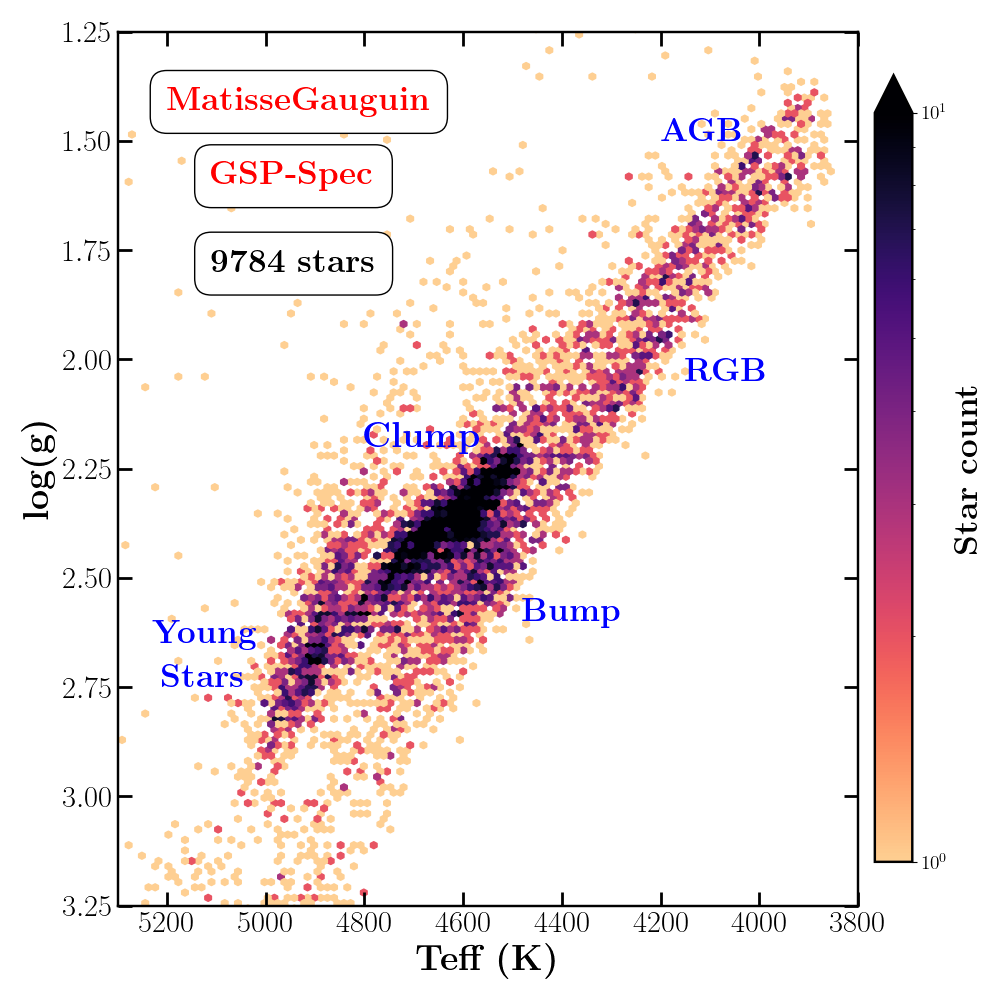}
\caption{Zoom onto the MatisseGauguin Kiel diagram for stars in a very restricted metallicity domain,  -0.05$<$\meta$<$0.00~dex, and with high-quality spectra. The RGB and AGB sequences appear as two resolved parallel tracks. The very close-by RGB bump and HB clump are also isolated. A sequence of young stars (with ages of less than $\sim$1~Gyr) can be identified in the hotter side, with an overdensity at around \T$\sim$5000~K. }
\label{Fig:Clump}
\end{figure}

\begin{figure*}[h]
    \begin{center}
\includegraphics[width=0.33\textwidth]{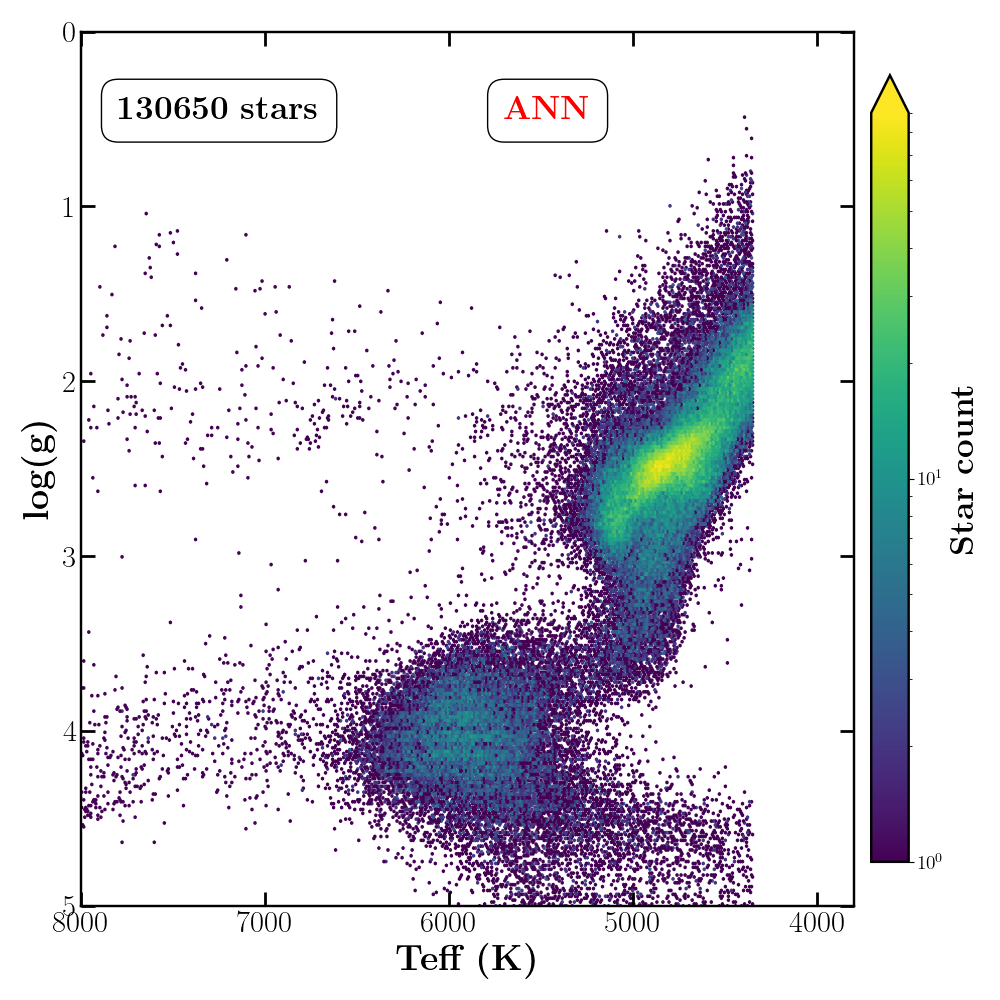}
\includegraphics[width=0.33\textwidth]{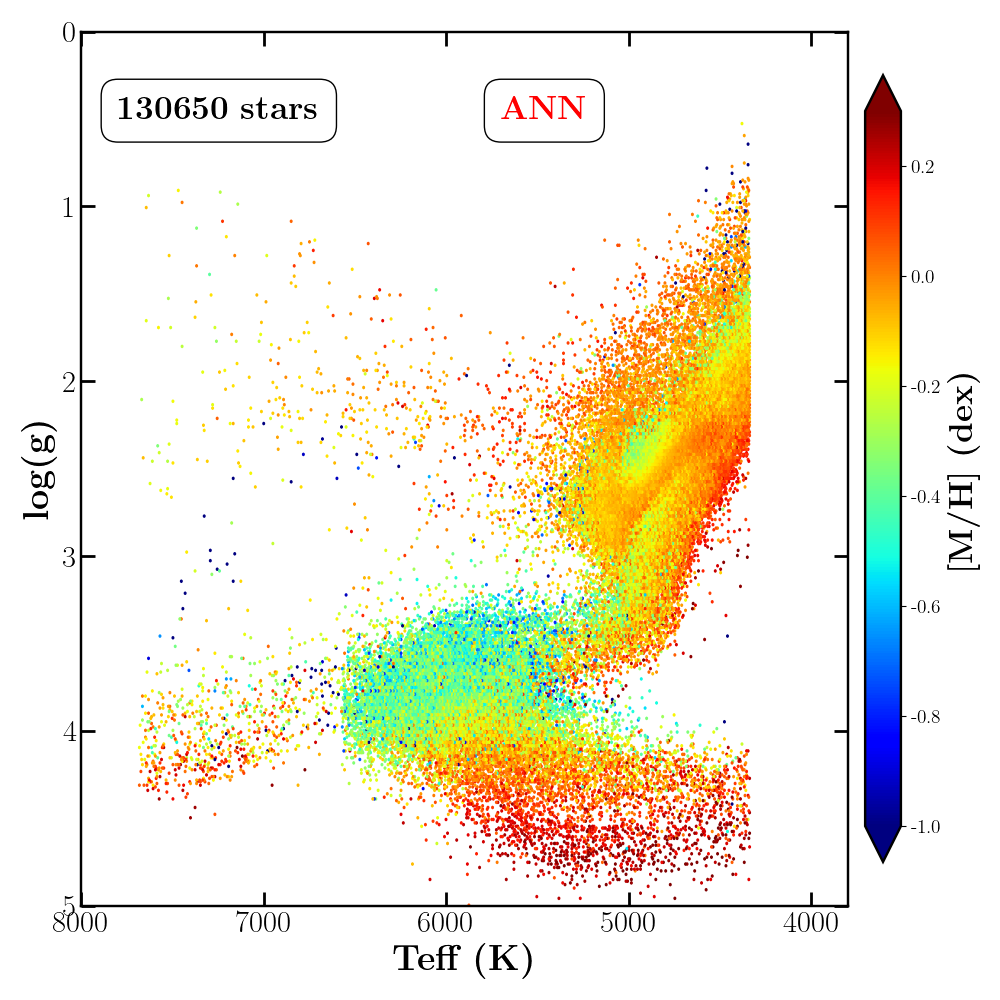}
\includegraphics[width=0.33\textwidth]{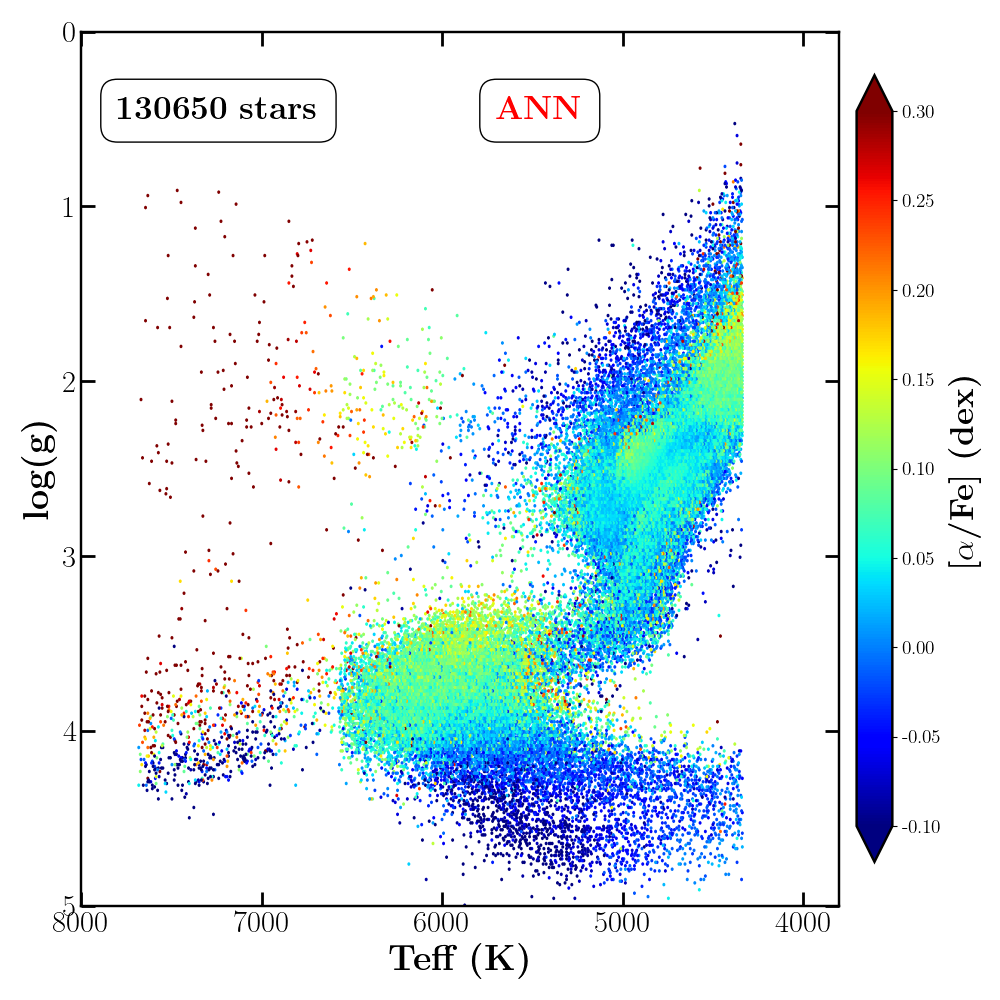}
    \caption{Same as Fig.~\ref{Fig.MG_Kiel} but for the ANN output parameters (stored in the supplementary DR3 astrophysical parameters table). In this case, we imposed that the first eight quality flags in \tabref{sec:val} be equal to zero. The calibrations proposed in Sect.~\ref{ANNBiases} were applied. Although a larger dispersion is observed with respect to MatisseGauguin, a general agreement exists, supporting the coherence of the two methodologically independent analysis.}
    \label{fig:ann_kiel}
    \end{center}
\end{figure*}

\subsection{Number of parametrised stars in different quality regimes}
As explained throughout this article, \gspspec\ has produced two 
sets of parameters (one from the MatisseGauguin workflow on the {\it AstrophysicalParameters} table, and another from the ANN workflow on the {\it AstrophysicalParametersSupp} table) for about 5.6~million stars
from their RVS spectra. The total number of derived atmospheric parameters by both workflows and the number of GAUGUIN  chemical abundances are illustrated in Fig.~\ref{Fig.StatParam} (left panel for MatisseGauguin and right panel for ANN) and Fig.~\ref{Fig.StatAbund}, respectively. In both figures, the total number of published parameters is shown together with the corresponding number for the best parametrised stars (from a high-quality selection, where all parameter flags are set to zero, including the  abundance flags from MatisseGauguin). It is important to note that imposing that the full flag chain be equal to zero corresponds to very demanding requirements, including very low associated uncertainties. This selects about two million stars for the atmospheric parameters, whereas, for the chemical abundances (cf. Fig.~\ref{Fig.StatAbund}), the number of estimates varies over several orders of magnitude from one element to another, as expected. In particular, calcium and iron (\FeI) are the most often derived species with estimates for around two millions stars, thanks to the Ca prominent lines and the numerous available iron lines. Abundances of heavy elements are derived for up to 10$^4$-10$^5$ stars, although these numbers strongly decrease when all the flags are used to filter (Ce being the heavy element with the highest number of estimates). However, we point out that this very strict quality filtering can be relaxed to increase number statistics, depending on the scientific goals of the user.

\subsection{Colour--temperature relation}
A classical way of validating effective temperature estimates is to verify their expected correlation with stellar colour. Figure~\ref{Fig.ColourTeffDib} shows the trend between the (BP-RP) colour and the \gspspec\ \T\ estimates from MatisseGauguin. To consider the effect of extinction on the (BP-RP) colour, the points are colour coded according to the EW of the DIB derived for the same stars by \gspspec\ (only stars with the DIB flag equal to zero have been selected). First, it is observed that the lower envelope of the distribution corresponds to the lower DIB EW values, as expected from the correlation between DIB absorption and extinction. To quantify this observation, the median values of the distribution for the stars with a DIB EW lower than 0.05 \AA\ (blue circles) can be compared to those whose DIB EW is equal to the median value of the distribution (0.07 \AA\  for dwarf stars in the left panel, and 0.12 \AA\ for giants in the right panel), plus a dispersion of $\pm$0.01 \AA. 
Second, the observed relation is compared to a \T\ derived from the  \cite{Casagrande21}  prescription (black dots) based on an implementation of Gaia and 2MASS photometry in the InfraRed Flux Method. No extinction has been considered in this case and the corresponding median values are shown as white circles.  The \cite{Casagrande21} predictions are in very good agreement with the low-extinction envelope of the \gspspec\ distribution (blue circles), validating the global behaviour of the estimated temperatures.

To complement this analysis, Fig.~\ref{Fig.ColourTeffMeta} presents the metallicty correlations of the colour--temperature relation for targets with all the parameter flags equal to zero. Again, the expected metallicty trend is observed in the low-extinction envelope. Interestingly, the higher extinction region above the lower envelope of the distribution is mainly occupied by metal-rich stars. This is expected from the fact that metal-rich stars are preferentially placed near the Galactic plane, where the interstellar extinction is higher.

\subsection{Sky distribution of effective temperature estimates}
Figure~\ref{Fig.MilkyWay_Teff} presents the global all-sky spatial distribution in Galactic coordinates of the stars parametrised by \gspspec, colour-coded with their MatisseGauguin effective temperature (5\,576\,282 stars, left panel) and their ANN effective temperature (5\,524\,387 stars, right panel). Both figures show the giant star population dominating the Galactic disc and bulge regions. The in-plane interstellar extinction pattern can also be noticed by its effect on the underlying parameterised populations: in higher extinction regions, cool giant stars observable at large distances become too faint in the RVS wavelength domain, and the median of the temperature distribution becomes hotter. Finally, nearby fainter dwarf stars in the foreground  dominate the regions above and below the Galactic plane, increasing the median \T\ values. It can be observed that ANN provides lower temperatures for these stars. 
For more details on the \gspspec\ selection function, we refer to \cite{Recio22} (see their Sect.~3). 

\subsection{Atmospheric parameters of high-S/N spectra}
To illustrate the \gspspec\ atmospheric parameter estimates in the high-S/N regime, we selected all the stars with 
S/N>150, excluding  high-rotating stars and potentially misclassified cool giants (imposing  {\it vbroadT=vbroadG=vbroadM}=0 and $KMtypestars$=0, respectively). This selects a sample of nearly 202\,000 stars.

Figure~\ref{Fig.MG_Kiel} presents the MatisseGauguin parametrisation of the selected objects in different Kiel diagrams colour coded according to stellar density (left panel), \meta\ (middle panel), and \alphaFe\ (right panel). We applied the \g\ calibration proposed in Sect.~\ref{sec:biases} and the \alphaFe\ calibration reported in Table~\ref{tab:calibrations} (fourth-order polynomial, without applying the suggested cuts in \g\ in order to show a complete Kiel diagram). The parameters precision can be assessed from the well-defined evolutionary sequences. For instance, the clearly distinguishable red clump presents a metallicity dependence that is independent from that of the red giant branch, as expected. Additionally, younger, more massive stars populate the hotter metal-rich sequence with logg$\la$3. These stars are located in the Milky Way spiral arms \citep[cf.][]{Recio22}. It is worth noting that, in the high-S/N regime, the algorithm shows overfitting patterns (overdensity features at the reference grid points). This can be observed in the left panel of Fig.~\ref{Fig.MG_Kiel} for the \T. The \g\ values are not affected in this figure because they have been calibrated.

The precision of the Matisse-Gauguin atmospheric parameters (without any use of astrometric inputs) can also be appreciated from Fig.~\ref{Fig:Clump}, which shows a zoom into the Kiel diagram of the stars in a very restricted metallicity domain,  -0.05$<$\meta$<$0.00~dex (defined using the upper and lower confidence values in the form \verb|mh_gspspec_upper|$<$0.00~dex and \verb|mh_gspspec_lower|$>$-0.05~dex). In addition, only stars with \T$>$3750~K, {\it KMgiantPar}=0 and \verb|logchisq_gspspec|$<$-3.75 were selected so as to avoid classification problems at the very cool end of the giant branch. It can be appreciated that the RGB bump\footnote{The RGB bump corresponds to the arrival of the narrow burning H-shell to the sharp chemical discontinuity in the H-distribution profile caused by the penetration of the convective envelope.} is resolved as an overdensity feature at \T$\sim$4600~K and \g$\sim$2.5. The very high parameter precision allows us to separate this RGB feature from the nearby horizontal branch clump visible as  a narrow elongated feature between 4500$<$\T$<$4800~K and 2.20$<$\g$<$2.50. Moreover, Fig.~\ref{Fig:Clump} shows the capability of these very high-quality \gspspec\ parameters to disentangle  the extremely close-by red giant branch and asymptotic giant branch sequences, which appear as two parallel tracks for \g$<$2.25. Finally, the overdensity located around \T$\sim$5000~K and 2.50$<$\g$<$2.80 corresponds, as mentioned above, to the evolutionary sequence of young stars of about less than 1~Gyr (cf. \cite{Recio22} for a more detailed analysis of these stars tracing the disc spiral arms). 
This will put important constraints on stellar evolution models, and specifically on the mass and metallicity dependencies of the red clump, the RGB bump, and the RGB and AGB behaviours. 

Figure~\ref{fig:ann_kiel} shows the ANN results 
for the same stars, 
after imposing that the first eight quality flags in \tabref{sec:val} be equal to zero and the four parameter calibrations proposed in Sect.~\ref{ANNBiases}. On one hand, a general agreement is observed with respect to MatisseGauguin. In particular, a well-defined Red Clump and a comparable metallicity trend for giant stars are observed, although with a higher dispersion and an underabundance of metal-poor stars  in ANN results (\meta$\le$-1.0~dex, partly explainable by the temperature cut in the cool regime due to calibration boundaries). On the other hand, 
the metallicity and \alphaFe\ distributions differ from the MatisseGauguin one for dwarf stars, presenting an unexpected trend with gravity. Despite the higher dispersion of ANN parameters, its overall agreement with MatisseGauguin brings support to the coherence of the two methodologically independent analyses. 

\begin{figure}[t]
\includegraphics[width=0.5\textwidth, height=6.cm]{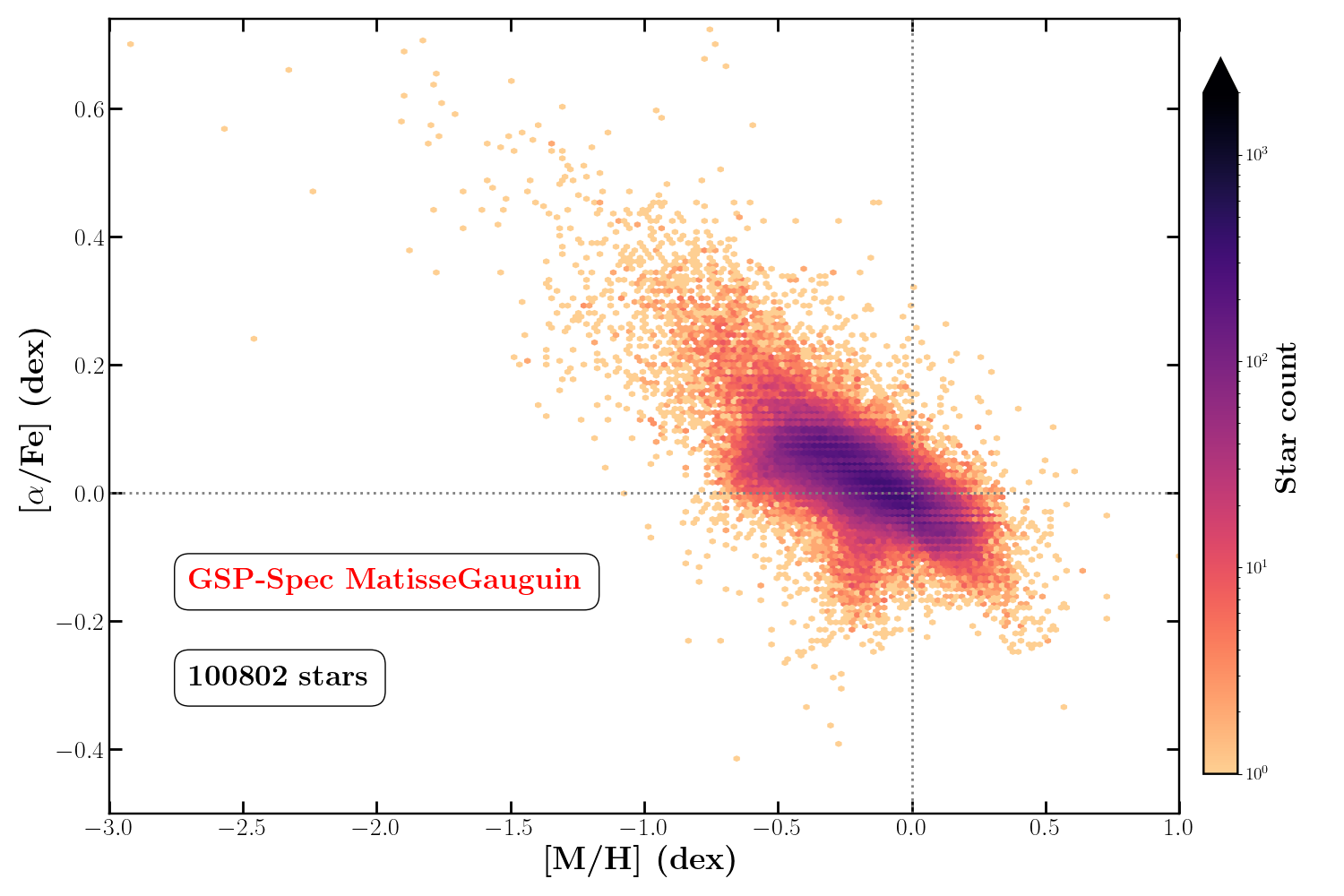}
\caption{\alphaFe\ versus \meta\ for the same MatisseGauguin stars as in Fig.\ref{Fig.MG_Kiel} but applying the recommended gravity interval for the calibration (Table~\ref{tab:calibrations}) and imposing \T$\le$6000~K and $vbroad\le$10~km.s$^{-1}$.}
\label{Fig.Alpha_Meta}
\end{figure}

Additionally, Fig.~\ref{Fig.Alpha_Meta} shows the \alphaFe\ versus \meta\ distribution from the MatisseGauguin analysis for the selected high-S/N spectra, applying the suggested cuts in \g\ for the \alphaFe\ calibration and imposing \T$\le$6000~K and $vbroad\le$10~km.s$^{-1}$. These last two filters help to control the quality of the \alphaFe\ by reducing second-order temperature trends and refining the filtering performed by the $vbroadT$, $vbroadG,$ and $vbroadM$ flags. The halo and disc sequences can be observed, with the thick disc sequence joining the thin disc one at a metallicity of around -0.4~dex. It is also worth noting that, as expected from chemical evolution models, the thin disc sequence continues to decrease at supersolar metallicities. As shown in \cite{Recio22}, the \alphaFe\ clearly correlates with the kinematical properties of stellar populations.
Moreover, the \Gaia-Enceladus sequence of accreted stars \citep{Helmi2018} is also distinguishable (lower \alphaFe\ values than those for typical thick discs and halos in the metal-poor regime). Finally, a group of low-\alphaFe\ stars at a metallicity of about \meta$\sim$-0.4~dex is also visible. This corresponds to young massive stars in the spiral arms \citep[for a discussion about the chemical properties of these stars see,][]{Recio22}. It is worth noting that, as mentioned in Sect.~\ref{sec:biases}, \gspspec\ \alphaFe\ estimates are dominated by the \CaFe\ abundance. We refer to \cite{Recio22} for a detailed illustration of individual $\alpha$-element abundances, including Mg, Ca, Si, S, and Ti, as well as other chemical species including N, iron-peak elements, and heavy elements.



\begin{figure}[t]
\includegraphics[width=0.5\textwidth,height=6.cm,angle=0]{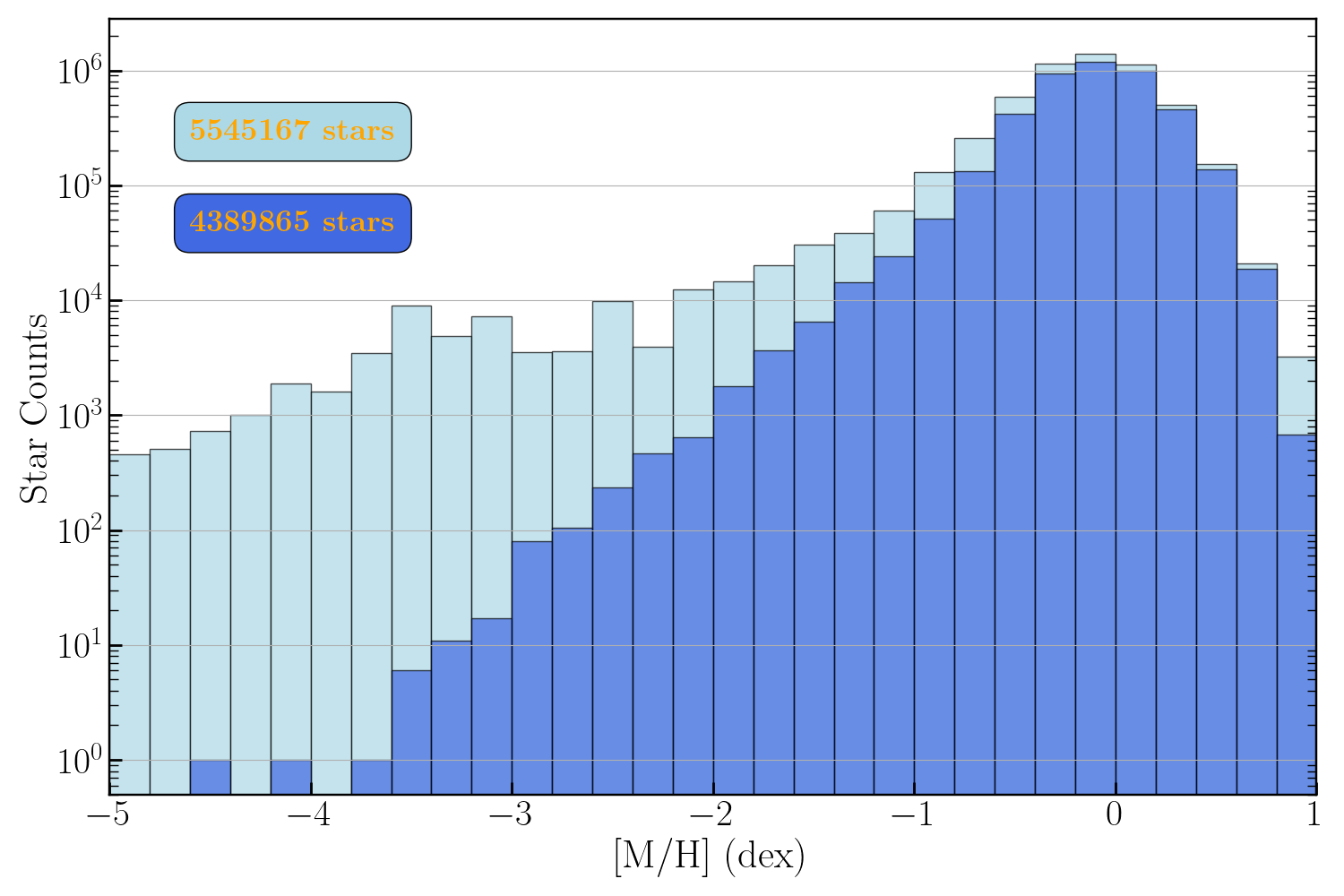}
\caption{Metallicity distributions for the MatisseGauguin parametrised stars. The light-blue histogram refers to the whole sample without any filtering. The medium-blue histogram presents a very strict filtering selecting stars with the best derived metallicities (see associated text for more details).}
\label{Fig.HistoMeta}
\end{figure}

\subsection{Parametrisation of extremely metal-poor stars}
\label{subsec:UMPs}
\begin{figure*}[h]
\includegraphics[width=0.99\textwidth, trim=4cm 0 0 0]{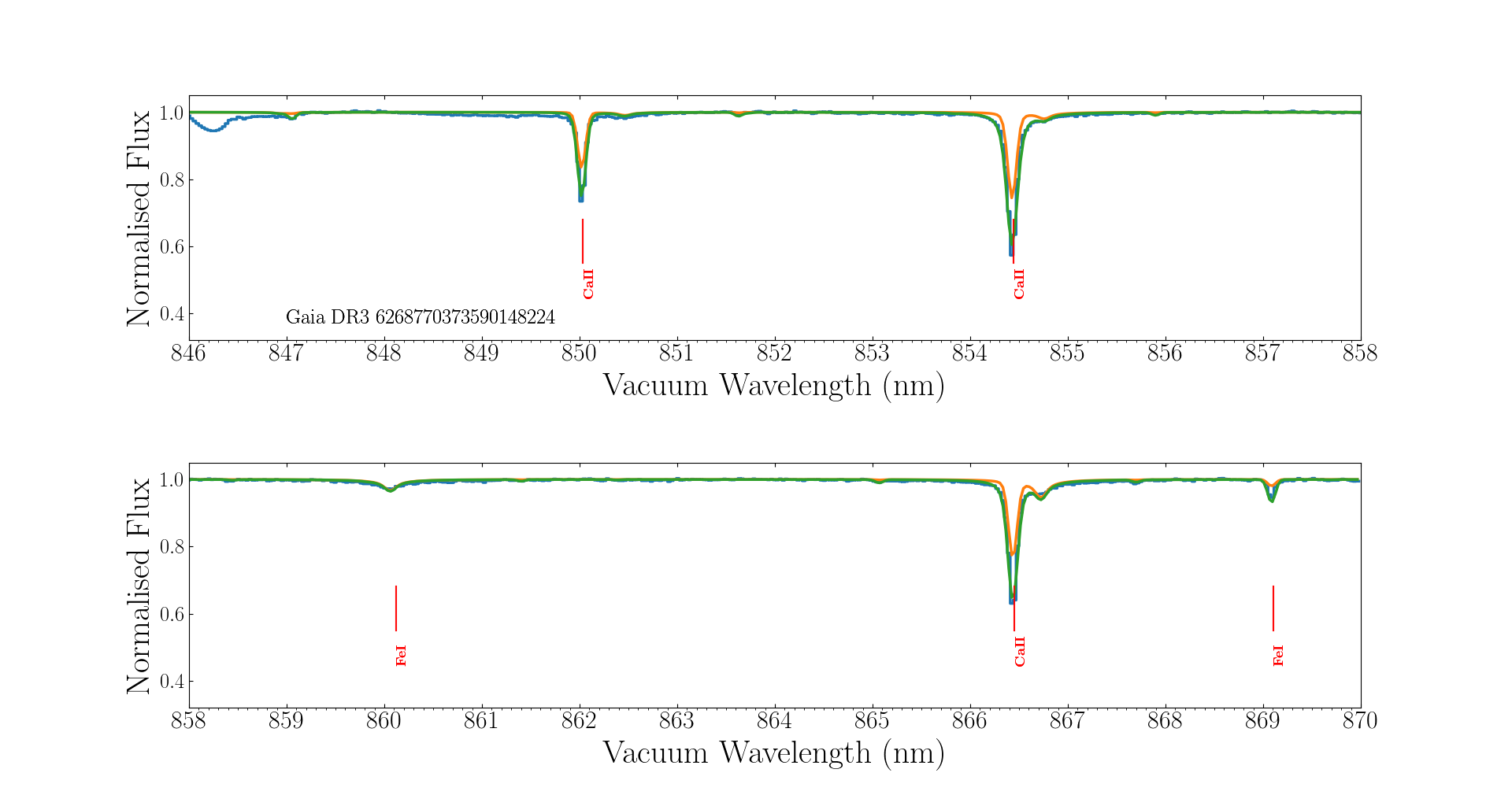}
\caption{RVS spectrum (blue histogram) of the very metal-poor star \Gaia DR3 
6268770373590148224 whose MatisseGauguin atmospheric parameters are \T=5\,331~K, \g=2.54, \meta=-3.19~dex, and \alphaFe=0.56~dex (S/N=419). The model spectra correspond to the lower and upper \meta\ values (-3.60 and -2.71~dex in orange and green, respectively). 
 No rotational profile was applied (suspected low-rotating star). See text for more details.}
\label{Fig:MetalP1}
\end{figure*}
\begin{figure*}[h]
\includegraphics[width=0.99\textwidth, trim=4cm 0 0 0]{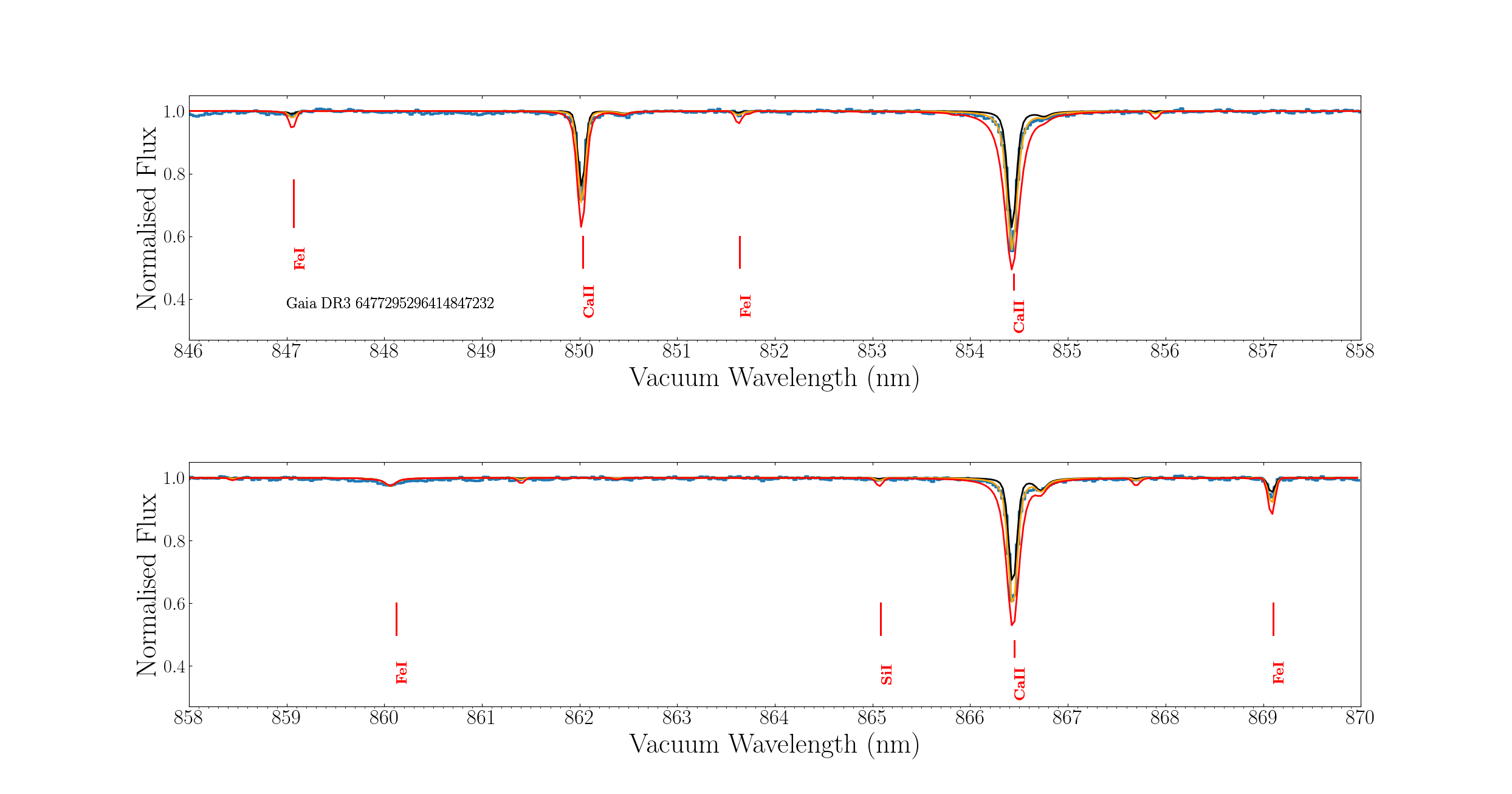}
\caption{RVS spectrum (blue histogram) of the ultra-metal-poor star \Gaia DR3 6477295296414847232 whose MatisseGauguin atmospheric parameters are \T=4\,994~K \g=2.13, \meta=-3.52~dex, and \alphaFe=0.68~dex (S/N=236). The model spectra correspond to the lower and upper \meta\ values (-3.52 and -3.07~dex in black and orange, respectively). A spectrum with \meta=-2.5~dex is also shown (red line) to definitively exclude such higher metallicities. No rotational profiles were considered.}
\label{Fig:MetalP2}
\end{figure*}

Metal-poor stars are relics of the most ancient formation epochs of the Milky Way, and in particular, they are crucial for disentangling the sequence of satellite mergers contributing to the Galaxy build-up \citep[e.g.][]{Helmi_20}. For this reason, they are the priviledged targets of several spectroscopic surveys from the ground like Pristine \citep{Pristine}. However, the lack of spectral signatures in metal-poor spectra reduces the information on the stellar parameters, increasing the uncertainties and making their parametrisation challenging. In the following, we illustrate the \gspspec\ capabilities to parametrise not only metal-poor, but also ultra-metal-poor (\meta$<$-3.0~dex) stars, providing suggestions on the necessary filters to apply.

Figure~\ref{Fig.HistoMeta} shows the \gspspec\ MatisseGauguin metallicity distribution in a logarithmic scale. The light-blue histogram refers to the complete sample without any quality filtering. In this distribution, which is useful for a rough stellar selection, there are  about 66\,000 stars with \meta$\le$-2.0~dex. It is nevertheless observed that the profile of the  histogram is  unexpectedly flat in the ultra-metal-poor regime. This is due to the \T\ limitations of the \gspspec\ reference spectra grid (\T$\le$8000~K) inducing a \T--\meta\ degeneracy. This problem can be satisfactorily resolved, as shown by the medium-blue histogram, by (i) disregarding the \meta\ values for stars with the first six characters of the \gspspec\ flagging chain $\ge$2, which limits the parameterisation biases, (ii) filtering out the metallicities of stars with {\it extrapol} flag (eighth character of the flagging chain) $\ge$3, which limits the extrapolation issues, (iii) eliminating the ultra-metal-poor stars hotter than 6000~K, which conservatively filters out metallicities with unreliable uncertainties due to border effects (in \T\ and \meta),  and (iv) filtering out possible remaining \gspspec\ misclassifications of very hot stars with stellar types O and B as estimated  by the Extended Stellar Parameteriser of Hot Stars (ESP-HS) and reported in the {\it AstrophysicalParameters} table as {\it spectraltype\_esphs}. To complete the previous filters, which are optimised for the very metal-poor regime, the medium-blue histogram of Fig.\ref{Fig.HistoMeta} filters out the metallicities of stars with \T$<$3500 or \g$>4.9$ or {\it KMgiantPar}$>$0. The filtering implemeted by the {\it KMgiantPar} flag, which controls the quality of the parameterisation of very cool giants, can be slightly extended, as reported in \citet{Recio22}, disregarding the metallicity of stars with \T$<$4150~K and 2.4$<$\g$<$3.8.  Thanks to these different quality filters, the medium-blue histogram presented in Fig.~\ref{Fig.HistoMeta} recovers the expected decrease in the number of stars in the very metal-poor regime, reporting only very reliable results within the corresponding uncertainties. Among these, there are about 
300 with \meta$<$-2.5~dex and about 40 stars with \meta$<$-3.0~dex.


To confirm that \gspspec\ is indeed able to correctly estimate the parameters of very metal-poor and ultra-metal-poor stars, we show the RVS spectrum of two of them, which were randomly chosen among the highest S/N spectra (Figs.~\ref{Fig:MetalP1} and~\ref{Fig:MetalP2}). The very few lines present in these spectra are extremely weak (except those of \CaII) and, as a consequence, no individual abundances were derived for both stars. \meta\ is therefore estimated only from very few available weak calcium or iron lines. The careful visual inspection of the synthetic spectra fit of the corresponding RVS spectra corroborates the very metal-poor nature of both stars.


We first validated the atmospheric parameters of \Gaia DR3 
6268770373590148224 (Fig.~\ref{Fig:MetalP1}, \meta$=$-3.19~dex) by computing several synthetic spectra with parameters found within the upper and lower MatisseGauguin uncertainties, i.e. between 5\,097 and 5\,456~dex for \T\ and [2.38, 2.76] for \g. 
The parameter values are confirmed within the uncertainties. The global fit is excellent and \meta\ is indeed found between the published range [-3.6, -2.7]. We note again that this rather large range of \meta\ is caused by the quasi absence of lines in its RVS spectrum. We also checked the literature for this star and found it to be known as the peculiar star HD~140283, already studied in several articles, the first one going back about 70 years \citep{HD140283}. Its most recent published parameters seem to converge towards a slightly hotter star with the majority of published \meta\ being found around -2.5~dex, in agreement with our estimates taking into account the literature error bars. Its \alphaFe\ value of 0.56~dex and its kinematical parameters \citep[a Galactic azimuthal velocity of $\sim$29~km.s$^{-1}$, as taken from][]{Recio22} make this star a typical halo representative.

The second example, \Gaia DR3 6477295296414847232, is shown in Fig.~\ref{Fig:MetalP2}. This star has \meta$=$-3.52~dex (and confidence values between -3.94~dex and -3.17~dex), a \T\ of 4994~K (with confidence values between 4781~K and 5071~K) and a \g\ of 2.13 (with confidence values between 1.5 and 2.26). Again, this parameterisation is confirmed by visual inspection of the spectrum fit in Fig.~\ref{Fig:MetalP2}. Additionally, to exclude \meta\ values higher than the reported upper confidence level (\meta$=$-3.07~dex),  Fig.~\ref{Fig:MetalP2} includes a synthetic spectrum (in orange) corresponding to \meta$=$-2.5~dex, which clearly overestimates the CaII and Fe lines depth and width and confirms that a lower metallicity  is required, as in the one estimated by \gspspec. Finally, the literature inspection of this object, identified as the peculiar star HD~2000654, leads to 23 different metallicity estimates with a median value of -2.9~dex$\pm$0.25~dex, in agreement with our results. It is interesting to note that \cite{Roederer14} and \cite{Hansen18} classify this star has an r-process enhanced object of r-I type, with a metallicity of -3.13~dex and -2.91~dex, respectively. In addition, based on high-resolution spectra,  \cite{Roederer14} reports a \CaFe$=$0.47~dex$\pm$0.17 in agreement with our calibrated \alphaFe=$0.62$~dex, a carbon abundance of [C/Fe]$=$0.31~dex, and a nitrogen abundance of [N/Fe]$=$-0.4~dex. More recently, \cite{Placco18}, using medium-resolution spectra, classify this star as a CEMP-II object, with a carbon enhancement of [C/Fe]$=$0.71~dex. It is important to remark that no sign of carbon enhancement, perturbing the metallicity estimate, is present in the RVS spectra due to the absence of CH molecular lines. In summary, the literature results again validate the \gspspec\ parameterisation of this ultra-metal-poor star, including its \alphaFe\  estimates.

\section{Summary and conclusions}\label{sec:Conclusions}
Here, we summarise the stellar parametrisation of \Gaia\ RVS spectra performed by the \gspspec\ module and published as part of \Gaia\ DR3. The goals, the input data, the used methodologies, and the validation are presented in detail. The resulting catalogues are published in the {\it AstrophysicalParameters} table (for the \gspspec\ MatisseGauguin workflow, including stellar atmospheric parameters, individual chemical abundances, a cyanogen differential EW, and DIB feature parameters), and in the {\it AstrophysicalParametersSupp} (for the ANN workflow providing atmospheric parameters). The \gspspec\ catalogue flags are also carefully defined and guidance for their use is illustrated with examples.
We highly recommend future users of the \gspspec\ parameters to adopt these flags for their specific science cases.

With about 5.6 million stars, the \Gaia DR3 \gspspec\ all-sky catalogue is the largest compilation of stellar chemo-physical parameters ever published and the first of its kind based on data acquired in space. The extreme homogeneity of the analysis combined with continuous data collection for almost three years enable a careful spectroscopic data reduction, a detailed modelling of systematic errors, and consequently, higher number statistics and a parametrisation quality that is comparable to that of ground-based surveys of higher spectral resolution and wavelength coverage.

\gspspec\ parameters open new horizons in stellar, Galactic, and insterstellar medium studies. In addition to the scientific performance analysis of \gspspec\ data published in \cite{Recio22} and \cite{Schultheis22}, we illustrate the  precision of the parameters here with (i) the colour--temperature relation, (ii) the Kiel diagrams and the \alphaFe\  vs. \meta\ distribution in the high-S/N regime (S/N$>$150, more than 2 million stars), (iii) our ability to disentangle different evolutionary stages of giant stars that are extremely close-by in the parameter space (RGB/AGB, bump/clump), and finally, (iv) a demonstration of the capability of \gspspec\  in the challenging parametrisation of metal-poor and extremely metal-poor stars.

Finally, it is worth noting that, as \gspspec\ is one of the parametrisation modules activated at the end of the DPAC analysis chain, this \Gaia third data release is actually the first \gspspec\ data release. The  acquired experience will benefit future releases, for which the number of parametrised stars will be a factor of ten larger ($\sim$50 million stars) as a result of the spectra S/N increase with observing time. It is important to note that the present data set is already at least a factor 8 larger than previous individual ground-based catalogues and a factor 3 larger than their very heterogeneous joint compilation. \gspspec\ is therefore exploring Galactic regions that we had previously only hypothesised from models (based on low number statistics). Thanks to the \Gaia RVS \gspspec\ chemo-physical parametrisation, we now have a privileged view of the sky from beyond Earth. 

\begin{acknowledgements}
This work presents results from the European Space Agency (ESA) space mission \Gaia  (https://www.cosmos.esa.int/gaia). \Gaia\ data are processed by the \Gaia\ Data Processing and Analysis Consortium (DPAC). Funding for the DPAC is provided by national institutions, in particular the institutions participating in the \Gaia\ MultiLateral Agreement (MLA). The \Gaia\ archive website is \url{https://archives.esac.esa.int/gaia}. Acknowledgments from the financial institutions are given in Appendix~\ref{Appendix:MERCI}. 
\\
We sincerely thank the stellar atmosphere group in Uppsala for providing the MARCS model atmospheres, B. Plez for having developped and maintaining the TURBOSPECTRUM package and, M. Bergemann for providing before publication the adopted relation between microturbulent velocity and atmospheric parameters. We also thank A. Bragaglia 
for her comments on the manuscript and the anonymous referee for very useful comments and suggestions.
\\
Finally, part of the calculations have been performed with the high-performance computing facility SIGAMM, hosted by the Observatoire de la Côte d'Azur. 
We acknowledge financial supports from the french space agency (CNES), Agence National de la Recherche (ANR 14-CE33-014-01) and  Programmes Nationaux de Physique Stellaire \& Cosmologie et Galaxies (PNPS \& PNCG) of CNRS/INSU.
ES, ARB, PdL, GK and MS acknowledge funding from the European Union’s Horizon 2020 research and innovation program under SPACE-H2020 grant agreement number 101004214 (EXPLORE project).

\end{acknowledgements}

\bibliographystyle{aa}  
\bibliography{biblio} 

\begin{appendix}

\section{\gspspec\ and radial velocities}
\label{Sect:rv}

The number of stars missing radial velocities (\Vrad) for different \gspspec\ parameters are provided in Table \ref{table:Vrad}.  Parameters not listed are not missing any \Vrad.

\begin{table}[h]
\caption{\gspspec\ and radial velocity statistics.}          
\label{table:Vrad}      
\centering                       
\begin{tabular}{l r}          
\hline\hline                     
 \gspspec\ parameter  & \# stars missing \Vrad \\     
\hline  
\verb|teff_gspspec| & 95 \\
\verb|logg_gspspec| & 95 \\
\verb|mh_gspspec| & 84 \\
\verb|alphafe_gspspec| & 83 \\
\verb|fem_gspspec| & 9 \\
\verb|sife_gspspec| & 6 \\
\verb|cafe_gspspec| & 6 \\
\verb|mgfe_gspspec| & 2 \\
\verb|feiim_gspspec| & 2 \\
\verb|sfe_gspspec| & 6 \\
\hline
\end{tabular}
\end{table}

\section{Atomic lines selected for the chemical analysis}
\label{Appendix:lines}

\begin{table}
\caption{ List of the atomic lines adopted for the determination of individual chemical abundances by \gspspec . Col.~2 refers to the reference wavelength of the analysed lines (see text for details). The abundance determination window corresponds to the interval $[\lambda_{ab}^{-}, \lambda_{ab}^{+}]$ (third and fourth column, respectively) while the refined normalisation window includes the wavelength range $[\lambda_{norm}^{-}, \lambda_{norm}^{+}]$ (fifth and sixth column, respectively). All the wavelengths are in nanometres and in the vacuum.}
\begin{tabular}{ |c|c|c|c|c|c| }
\hline
Elt & $\lambda$ & $\lambda_{ab}^{-}$ & $\lambda_{ab}^{+}$ & $\lambda_{norm}^{-}$ &  $\lambda_{norm}^{+}$\\
\hline
N~{\sc i} & 863.161 & 863.071 & 863.281 & 862.891 & 863.371\\
N~{\sc i} & 868.579 & 868.489 & 868.699 & 868.309 & 868.939\\
\hline
Mg~{\sc i} & 847.602 & 847.512 & 847.692 & 847.212 & 847.812\\
\hline
Si~{\sc i} & 853.851 & 853.731 & 853.941 & 853.371 & 854.961\\
$^{*}$Si~{\sc i} & 855.916 & 855.856 & 856.036 & 855.376 & 856.156\\
Si~{\sc i} & 868.872 & 868.782 & 868.992 & 868.602 & 869.232\\
\hline
$^{*}$S~{\sc i} & 867.258 & 866.988 & 867.378 & 866.898 & 867.998\\
$^{*}$S~{\sc i} & 869.701 & 869.551 & 869.821 & 869.281 & 869.971\\
\hline
Ca~{\sc i} & 863.631 & 863.511 & 863.691 & 863.361 & 863.931\\
\hline
Ca~{\sc ii} & 849.856 & 849.706 & 849.976 & 849.586 & 850.276\\
Ca~{\sc ii} & 850.216 & 850.156 & 850.276 & 849.886 & 850.306\\
Ca~{\sc ii} & 854.264 & 854.114 & 854.384 & 853.544 & 854.864\\
Ca~{\sc ii} & 854.624 & 854.564 & 854.744 & 854.294 & 854.804\\
Ca~{\sc ii} & 866.272 & 866.152 & 866.332 & 866.002 & 866.572\\
Ca~{\sc ii} & 866.632 & 866.512 & 866.692 & 866.302 & 866.782\\
\hline
$^{*}$Ti~{\sc i} & 852.069 & 851.979 & 852.129 & 851.799 & 852.249\\
Ti~{\sc i} & 857.209 & 857.119 & 857.269 & 856.999 & 857.359\\
Ti~{\sc i} & 869.472 & 869.382 & 869.562 & 869.292 & 869.832\\
\hline
Cr~{\sc i} & 855.118 & 855.058 & 855.208 & 854.878 & 855.478\\
Cr~{\sc i} & 864.567 & 864.447 & 864.627 & 864.207 & 864.867\\
\hline
$^{*}$Fe~{\sc i} & 848.296 & 848.206 & 848.446 & 847.666 & 848.896\\
$^{*}$Fe~{\sc i} & 851.641 & 851.551 & 851.851 & 851.281 & 852.001\\
$^{*}$Fe~{\sc i} & 852.901 & 852.691 & 853.081 & 852.481 & 853.321\\
Fe~{\sc i} & 857.416 & 857.296 & 857.506 & 856.876 & 858.166\\
Fe~{\sc i} & 858.462 & 858.312 & 858.612 & 858.132 & 858.762\\
Fe~{\sc i} & 862.397 & 862.277 & 862.517 & 862.127 & 862.697\\
Fe~{\sc i} & 867.713 & 867.593 & 867.863 & 867.443 & 868.013\\
$^{*}$Fe~{\sc i} & 869.101 & 868.891 & 869.191 & 868.441 & 869.821\\
\hline
Fe~{\sc ii} & 858.794 & 858.764 & 858.824 & 858.254 & 859.274\\
\hline
Ni~{\sc i} & 863.937 & 863.847 & 864.027 & 863.697 & 864.147\\
\hline
Zr~{\sc ii} & 852.748 & 852.658 & 852.838 & 852.388 & 853.018\\
\hline
$^{*}$Ce~{\sc ii} & 851.375 & 851.285 & 851.465 & 851.015 & 851.555\\
\hline
Nd~{\sc ii} & 859.389 & 859.299 & 859.479 & 859.209 & 859.689\\
\hline
\end{tabular}
\label{Table_linelist}
\end{table}

 This Appendix introduces the list of selected lines used in the determination of the chemical abundances. In Table \ref{Table_linelist}, we summarise the reference wavelength value of each atomic line, as well as the wavelength ranges considered for its abundance determination and second normalisation windows.  We note that the reference wavelength (col.~2) can differ from the vacuum wavelength of the analysed atomic line in case of multiplets or broad lines. For instance, for the \CaII\ IR triplet transitions at 850.036, 854.444, and 866.452~nm, two Ca abundances have been derived from the wings of each line to avoid the line core that could  not be well modelled. In those cases, col.~2 refers to one of the \CaII\ wings.
 
 The following lines (denoted by an asterisk in Table \ref{Table_linelist}) have multiple lines within the same abundance determination window: (855.913, 855.916) 
 for Si~{\sc i}; (867.082, 867.258, 867.297, 867.366) and (869.632, 869.701) for S~{\sc i}; (852.037, 852.069) for Ti~{\sc i}; (848.283, 848.296, 848.431), (851.641, 851.745, 851.751), (852.738, 852.901, 853.020) and (868.916, 869.101) for Fe~{\sc i}\footnote{Only the strongest transitions are provided for iron.}; (851.368, 851.381, 851.375) for Ce~{\sc ii}. For the \FeII~line measured in hot star spectra  
 (see Sect.~\ref{SecFe2}), some blends of weak \FeI\ transitions may be present in cooler star spectra. 
 
 \clearpage
 

\section{Definition of the \gspspec\ flags}
\label{Appendix:flags}
\begin{figure*}[h]
\centering
\includegraphics[width=17cm,height=10cm]{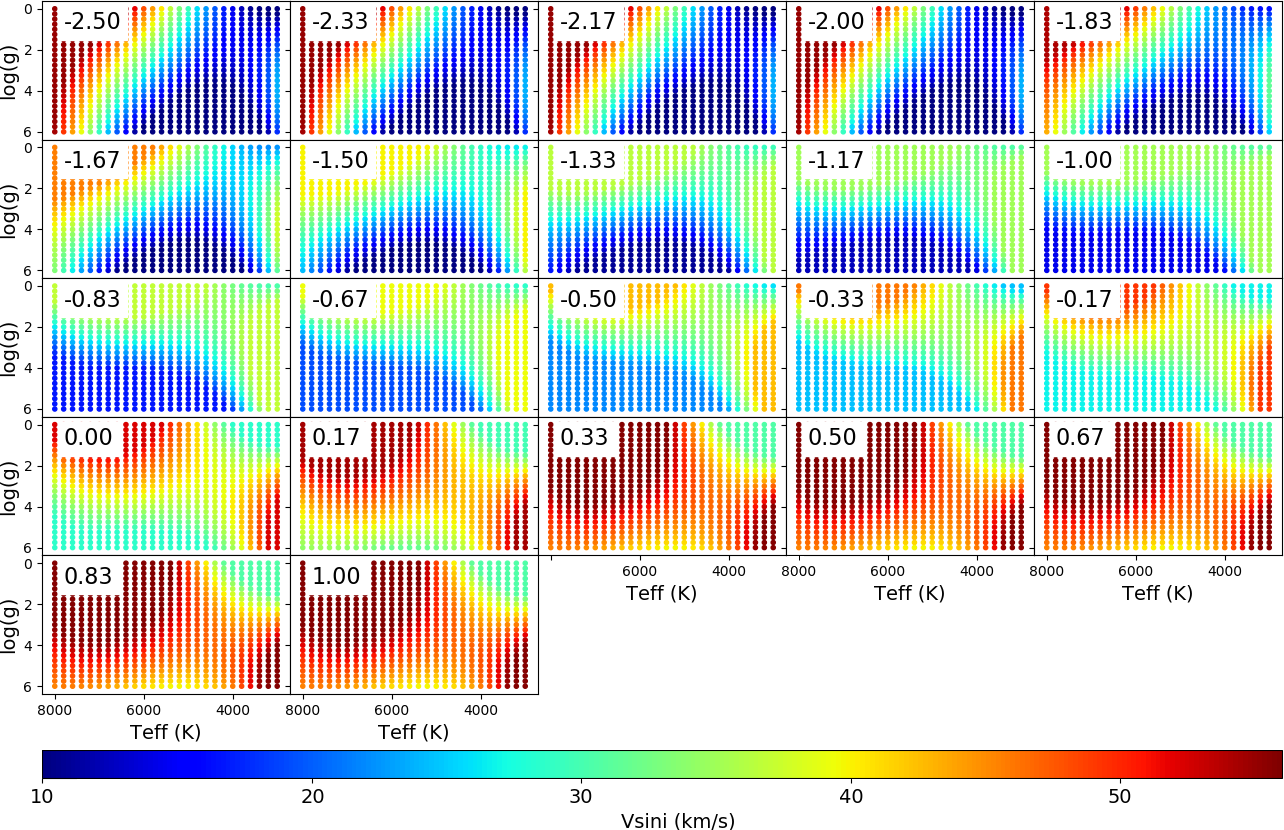}
\caption{Limiting \Vsini\ values (colour code) leading to a bias of 250<$\Delta$\T$\leq$500~K in the \gspspec\ parametrisation. This has been used to estimate the third-order polynomial with \T, \g,\ and \meta\ as variables used to define the $vbroad$T flag (equal to 1 in this example). The \meta\ values for each panel are indicated in their upper right corner.}
\label{Fig:FlagVsini}
\end{figure*}

\begin{figure}[h]
    \centering
    \includegraphics[width=0.5\textwidth]{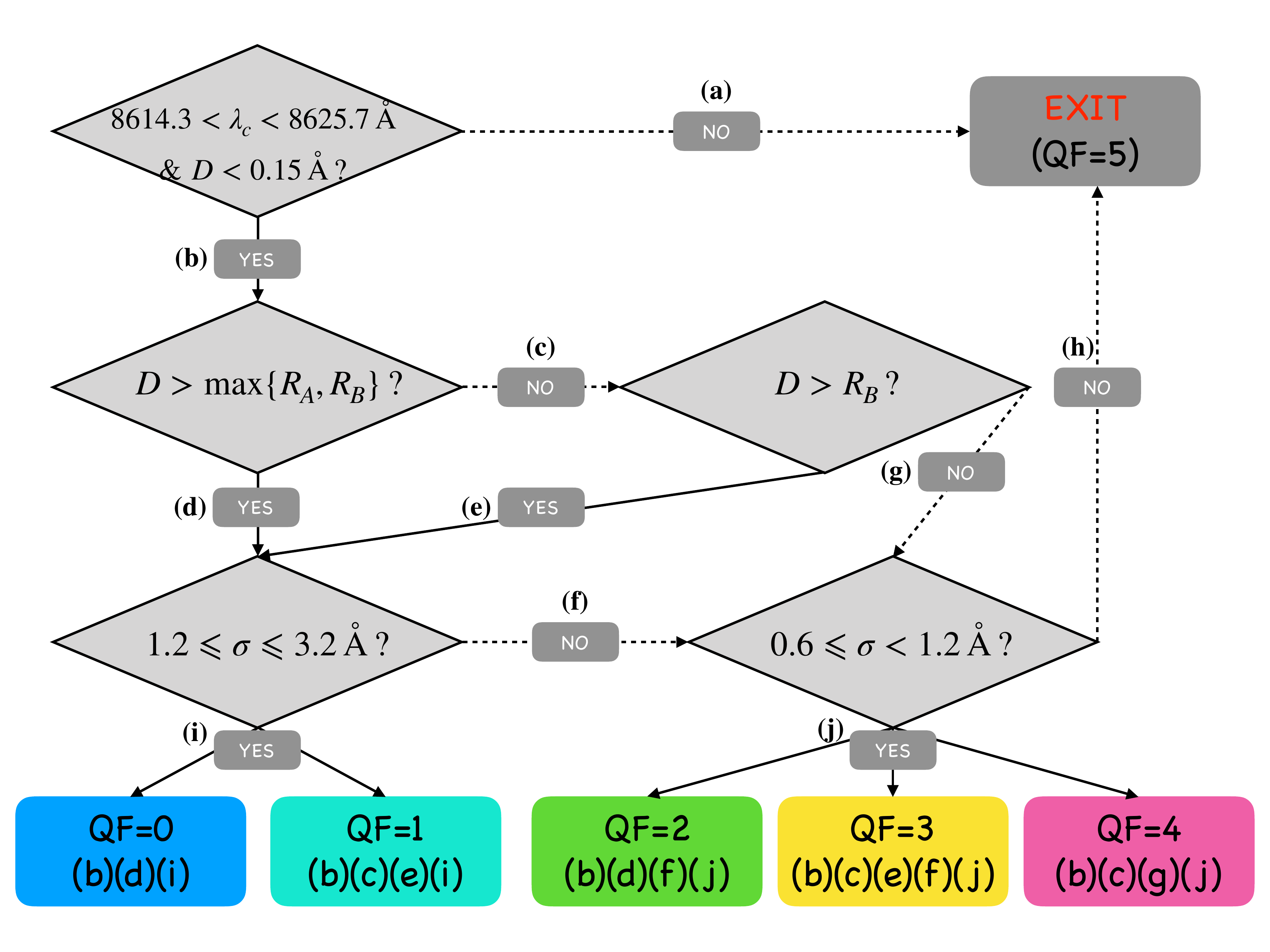}
    \caption{Flow chart of the different values for the DIB quality flag. See associated text in Sect.~\ref{subsec:DIBqf}.} 
    \label{QFdib}
\end{figure}

The following tables include the detailed definition of the individual characters in the \gspspec\ quality flag chain presented in Table~\ref{tab:MG_QFchain}. 
In addition, Fig.~\ref{Fig:FlagVsini} illustrates the implemented modelling of parameter biases induced by rotational broadening, leading to the definition of $vbroad$T, $vbroad$G and $vbroad$M quality flags (cf. Sec.~\ref{subsec:rotation} and Table~\ref{tab:vsiniQF}). The particular case of effective temperature biases is illustrated.
Finally, Fig.~\ref{QFdib} presents the validation flow chart associated with the definition of quality flags for the DIB parametrisation (cf. Sect.~\ref{subsec:DIBqf} and Table~\ref{tab:DIBflag}).

\begin{table}[h]
    \caption{Definition of the parameter flags considering potential biases due to rotational velocity and/or macroturbulence. These flags are part of the  {\it flags$\_$gspspec} string chain defined in Table~\ref{tab:MG_QFchain}.}
\centering
\begin{tabular}{| c | c |  c |}
\hline
\hline
\multirow{2}{*}{ \bf Flag }  & \multirow{3}{*}{\bf Condition} & \multirow{3}{*}{\bf Flag value}  \\
\multirow{2}{*}{ \bf name} & &   \\
& & \\
\hline
\hline 
\multirow{5}{*}{\color{teal} \bf vbroadT} &  \multirow{2}{*}{$\Delta$Teff$>$2000 K} &  \multirow{2}{*}{{\color{red} Filtered all}} \\
 & & \\
\cline{2-3}
              & 500<$\Delta$Teff$\leq$2000 K & Flag 2  \\
\cline{2-3}
              & 250<$\Delta$Teff$\leq$500 K & Flag 1\\
\cline{2-3}
             & $\Delta$Teff$\leq$250 K & Flag 0 \\

\hline
\multirow{6}{*}{\color{teal} \bf vbroadG} &  \multirow{3}{*}{$\Delta$logg$>$2 dex} &  {\color{red}Filter all \footnotesize \it except Teff } \\
& & {\color{red} \footnotesize \it  \& DIB if Teff>7000 K}    \\
     &         & Flag 9  \\
\cline{2-3}
              & 1<$\Delta$logg$\leq$2 dex & Flag 2  \\
\cline{2-3}
              & 0.5<$\Delta$logg$\leq$1 dex & Flag 1 \\
\cline{2-3}
             & $\Delta$logg$\leq$0.5 dex & Flag 0  \\
\hline
\multirow{5}{*}{\color{teal} \bf vbroadM} &  \multirow{2}{*}{$\Delta$[M/H]$>$2 dex} &  {\color{red} Filter [M/H] \& [X/Fe]} \\
     & & Flag 9  \\
\cline{2-3}
              & 0.5<$\Delta$[M/H]$\leq$2 dex & Flag 2 \\
\cline{2-3}
              & 0.25<$\Delta$[M/H]$\leq$0.5 dex & Flag 1 \\
\cline{2-3}
             & $\Delta$[M/H]$\leq$0.25 dex & Flag 0  \\
\hline
\hline
   \end{tabular}
    \label{tab:vsiniQF}
\end{table}

\begin{table}[h]
\centering
    \caption{Same as Table~\ref{tab:vsiniQF} but for the potential biases due to uncertainties in the radial velocity shift correction (see also Table~\ref{tab:MG_QFchain}).}
\begin{tabular}{| c | c |  c |}
\hline
\hline
\multirow{2}{*}{ \bf Flag }  & \multirow{3}{*}{\bf Condition} & \multirow{3}{*}{\bf Flag value}  \\
\multirow{2}{*}{ \bf name} & &   \\
& & \\
\hline
\hline
\multirow{4}{*}{\color{teal} \bf vradT} &  \multirow{2}{*}{$\Delta$Teff$>$2000 K} &  \multirow{2}{*}{{\color{red} Filter all}} \\
 & & \\
\cline{2-3}
              & 500<$\Delta$Teff$\leq$2000 K & Flag 2 \\
\cline{2-3}
              & 250<$\Delta$Teff$\leq$500 K & Flag 1 \\
\cline{2-3}
             & $\Delta$Teff$\leq$250 K & Flag 0  \\

\hline
\multirow{6}{*}{\color{teal} \bf vradG} &  \multirow{3}{*}{$\Delta$logg$>$2 dex} & {\color{red}Filter all \footnotesize \it except Teff }  \\
& &  {\color{red} \footnotesize \it \& DIB if Teff>7000 K}  \\
     & & Flag 9  \\
\cline{2-3}
              & 1<$\Delta$logg$\leq$2 dex & Flag 2  \\
\cline{2-3}
              & 0.5<$\Delta$logg$\leq$1 dex & Flag 1 \\
\cline{2-3}
             & $\Delta$logg$\leq$0.5 dex & Flag 0 \\
\hline
\multirow{5}{*}{\color{teal} \bf vradM} &  \multirow{2}{*}{$\Delta$[M/H]$>$2 dex} &  {\color{red} Filter [M/H] \& [X/Fe]} \\
    & & Flag 9 \\
\cline{2-3}
              & 0.5<$\Delta$[M/H]$\leq$2 dex & Flag 2 \\
\cline{2-3}
              & 0.25<$\Delta$[M/H]$\leq$0.5 dex & Flag 1 \\
\cline{2-3}
             & $\Delta$[M/H]$\leq$0.25 dex & Flag 0 \\
\hline
\hline
   \end{tabular}
    \label{tab:vradQF}
\end{table}

\begin{table}[h]
\centering
\caption{Definition of the parameter flags considering potential biases due to uncertainties in the RVS flux (MatisseGauguin parametrisation; see also Table~\ref{tab:MG_QFchain}).}
\begin{tabular}{| c | c |  c |}
\hline
\hline
\multirow{2}{*}{ \bf Flag }  & \multirow{3}{*}{\bf Condition} & \multirow{3}{*}{\bf Flag value}  \\
\multirow{2}{*}{ \bf name} & &   \\
& & \\
\hline
\hline
\multirow{25}{*}{\color{teal} \bf fluxNoise} & $\sigma$Teff$>$2000 K or & {\color{red} Filter all} \\
                         &  $\sigma$logg$>$2 dex &  Flag 9\\
\cline{2-3}                
                         & $\sigma$Teff$\leq$2000 K and  & {\color{red} Filter [M/H]} \\
                         & $\sigma$logg$\leq$2 dex  and  & {\color{red}\& [X/Fe]} \\
                         & $\sigma$[M/H]$>$2 dex  & Flag 5 \\
\cline{2-3}                
                         & $\sigma$Teff$\leq$2000 K and  &  {\color{red} Filter [$\alpha$/Fe]}\\
                         & $\sigma$logg$\leq$2 dex  and  & {\color{red}\& [X/Fe]} \\
                         & $\sigma$[M/H]$\leq$2 dex and  & Flag 4 \\  
                         & $\sigma$[$\alpha$/Fe]$>$0.8 dex  &   \\ 
\cline{2-3}                           
                         & 500<$\sigma$Teff$\leq$2000 K and   &\multirow{4}{*}{Flag 3} \\
                         & 1<$\sigma$logg$\leq$2 dex  and  & \\
                         & 0.5<$\sigma$[M/H]$\leq$2 dex and & \\  
                         & 0.2<$\sigma$[$\alpha$/Fe]$\leq$0.8 dex  &   \\  
\cline{2-3}                           
                         & 250<$\sigma$Teff$\leq$500 K and   &\multirow{4}{*}{Flag 2} \\
                         & 0.5<$\sigma$logg$\leq$1 dex  and  &  \\
                         & 0.25<$\sigma$[M/H]$\leq$0.5 dex and & \\  
                         & 0.1<$\sigma$[$\alpha$/Fe]$\leq$0.2 dex  &   \\ 
\cline{2-3}         
                        & 100<$\sigma$Teff$\leq$250 K and   &\multirow{4}{*}{Flag 1} \\
                         & 0.2<$\sigma$logg$\leq$0.5 dex  and  &  \\
                         & 0.1<$\sigma$[M/H]$\leq$0.25 dex and & \\  
                         & 0.05<$\sigma$[$\alpha$/Fe]$\leq$0.1 dex  &   \\    
\cline{2-3}         
                        & $\sigma$Teff$\leq$100 K and   &\multirow{4}{*}{Flag 0}\\
                         & $\sigma$logg$\leq$0.2 dex  and  &  \\
                         & $\sigma$[M/H]$\leq$0.1 dex and & \\  
                         & $\sigma$[$\alpha$/Fe]$\leq$0.05 dex  &   \\                            
\hline
\hline 
 \end{tabular}
    \label{tab:noiseQF}
\end{table}

\begin{table}[h]
\centering
\caption{Same as Table~\ref{tab:noiseQF} but for the ANN parametrisation.}
\begin{tabular}{|l|c|}
    \hline
    \hline 
    \multicolumn{2}{|c|}{\color{teal} \bf ANN fluxNoise} \\
    \hline
    \multicolumn{1}{|c|}{\bf Condition} & {\bf Value} \\
    \hline
    \hline 
    \begin{tabular}{l}
        $\sigma$\T> 525~K or $\sigma$\g>0.9 or\\
        $\sigma$\meta>0.5~or $\sigma$\alphaFe< 0.16~dex
    \end{tabular} & Flag 9 \\
    \hline
    \begin{tabular}{l}
        225<$\sigma$\T$\leq$525~K and\\
        0.4<$\sigma$\g$\leq$0.9 or\\
        0.19<$\sigma$\meta$\leq$0.5~dex or\\
        0.09<$\sigma$\alphaFe$\leq$0.16~dex
    \end{tabular} & Flag 3 \\
    \hline
    \begin{tabular}{l}
        140<$\sigma$\T$\leq$225~K and\\
        0.3<$\sigma$\g$\leq$0.4 or\\
        0.12<$\sigma$\meta$\leq$0.19~dex or\\
        0.06<$\sigma$\alphaFe$\leq$0.09~dex
    \end{tabular} & Flag 2 \\
    \hline
    \begin{tabular}{l}
        100<$\sigma$\T$\leq$140~K and\\
        0.24<$\sigma$\g$\leq$0.3 or\\
        0.09<$\sigma$\meta$\leq$0.12~dex or\\
        0.05<$\sigma$\alphaFe$\leq$0.06~dex
    \end{tabular} & Flag 1 \\
    \hline
    \begin{tabular}{l}
        $\sigma$\T$\leq$100~K and $\sigma$\g$\leq$0.24 and\\
        $\sigma$\meta$\leq$0.09~dex and $\sigma$\alphaFe$\leq$0.05 dex
    \end{tabular} & Flag 0 \\
    \hline 
    \hline 
\end{tabular}
\label{tab:noiseQFANN}
\end{table}

\begin{table}[h]
\centering
 \caption{Definition of parameter flags considering potential biases due to extrapolated parameters (MatisseGauguin parametrisation, see also Table~\ref{tab:MG_QFchain}).}
\begin{tabular}{| c | c |  c |}
\hline
\hline
\multirow{2}{*}{ \bf Flag }  & \multirow{3}{*}{\bf Condition} & \multirow{3}{*}{\bf Flag value}  \\
\multirow{2}{*}{ \bf name} & &   \\
& & \\
\hline
\hline
\multirow{31}{*}{\color{teal} \bf extrapol} & $gof$=NaN and &{\color{red} Filter all } \\
                         &  (Teff$>$9000 K or   & {  \footnotesize \it \color{red} except DIB }\\
                         &  Teff$<$2500 K  or  & {  \footnotesize \it \color{red} if Teff>7000 K} \\
                         &  logg$>$6 or& Flag 9 \\
                         & logg$<$-1 )  & \\
\cline{2-3}                
                         & $gof$=NaN and & \\
                         & 2500$\leq$Teff$\leq$9000 K and  & {\color{red} Filter}\\
                         & -1$\leq$logg$\leq$6 and  &{\color{red} [M/H],[X/Fe]} \\
                         & ([M/H]$<$-6 dex or & Flag 4 \\
                         & [M/H]$>$1.5 dex ) & \\
\cline{2-3}    
                         & $gof$=NaN and & \\
                         & 2500$\leq$Teff$<$9000 K and  & \\
                         & -1$\leq$logg$\leq$6 and  &{\color{red} Filter [X/Fe]} \\
                         & -6$\leq$[M/H]$\leq$1.5 dex and & Flag 3 \\
                         & [$\alpha$/Fe] {\footnotesize out from standard by} $\pm$ 0.8 & \\
\cline{2-3}    
                         & $gof$=NaN and & \multirow{5}{*}{Flag 2} \\
                         & 2500$\leq$Teff$\leq$9000 K and  & \\
                         & -1$\leq$logg$\leq$6 dex and  &  \\
                         & -6$\leq$[M/H]$\leq$1.5 and &  \\
                         & [$\alpha$/Fe] {\footnotesize within} $\pm$ 0.8 {  \footnotesize from standard } & \\     
\cline{2-3}    
                         & $gof\neq$NaN and &\multirow{5}{*}{Flag 1} \\
                         & (Teff$\geq$7625 or Teff$\leq$3500 K or & \\
                         & logg$\geq$4.75 or logg$\leq$0.25 or &  \\
                         & [M/H]$\leq$-3  or [M/H]$\geq$0.75 dex or &  \\
                         & [$\alpha$/Fe] {\footnotesize out from standard by} $\pm$ 0.35)  & \\
\cline{2-3}  
                         & $gof\neq$NaN and & \multirow{5}{*}{Flag 0} \\
                         & 3500$<$Teff$<$7625 K and  & \\
                         & 0.25$<$logg$<$4.75 and  &  \\
                         & -3$<$[M/H]$<$0.75 dex and &  \\
                         & [$\alpha$/Fe] {\footnotesize within} $\pm$ 0.35 {  \footnotesize from standard } &\\   
\hline
\hline 
 \end{tabular}
    \label{tab:extrapol}
\end{table} 

\begin{table}[h]
\centering
\caption{Same as Table~\ref{tab:extrapol} but for the ANN parametrisation.}
\begin{tabular}{|l|c|}
    \hline
    \hline
    \multicolumn{2}{|c|}{\color{teal} \bf ANN extrapol} \\
    \hline
    \multicolumn{1}{|c|}{\bf Condition} & {\bf Value} \\
    \hline
    \hline
    \begin{tabular}{l}
        \T>8320~K or \T<3680~K or \\
        \g>6.0 or \g<-1.0 or \\
        \meta>1.3~dex or \meta<-5.3~dex or \\
        \alphaFe>1.0~dex or \alphaFe<-0.6~dex
    \end{tabular} & Flag 9 \\
    \hline
    \begin{tabular}{l}
        7680<\T$\leq$8320~K or\\
        3680$\leq$\T<4320~K or\\
        4.9<\g$\leq$6.0 or\\
        -1.0$\leq$\g<0.06 or\\
        0.6<\meta$\leq$1.4~dex or\\
        -5.3$\leq$\meta<-4.6~dex or\\
        0.6<\alphaFe$\leq$1.0~dex or\\
        -0.5$\leq$\alphaFe<0.2~dex
    \end{tabular} & Flag 1 \\
    \hline
    \begin{tabular}{l}
        4320$\leq$\T$\leq$7680~K and\\
        0.06$\leq$\g$\leq$4.9 and\\
        -4.6$\leq$\meta$\leq$0.6~dex and\\
        -0.2$\leq$\alphaFe$\leq$0.6~dex
    \end{tabular} & Flag 0 \\ 
    \hline
    \hline 
\end{tabular}
\label{tab:extrapolANN}
\end{table}

\begin{table}[h]
\centering
    \caption{Definition of parameter flags considering RVS flux issues or emission line probability (see also Table~\ref{tab:MG_QFchain}).}
\begin{tabular}{| c | c |  c |}
\hline
\hline
\multirow{2}{*}{ \bf Flag }  & \multirow{3}{*}{\bf Condition} & \multirow{3}{*}{\bf Flag value}  \\
\multirow{2}{*}{ \bf name} & &   \\
& & \\
\hline
\hline
\multirow{3}{*}{\color{teal} \bf nanFlux}  & \multirow{3}{*}{Flux$=$NaN} & {\color{red}Filter all}\\
                &  &{  \footnotesize \it \color{red} except DIB if Teff>7000 K}  \\
               & & Flag 9  \\                
\hline
\hline
\multirow{3}{*} {\color{teal} \bf emission}  & \multirow{3}{*}{{\it CU6$\_$is$\_$emission}} & {\color{red} Filter all} \\
               &  &{  \footnotesize \it \color{red} except DIB if Teff>7000 K}  \\
               & & Flag 9  \\
\hline
\hline
\multirow{6}{*}{\color{teal} \bf negFlux}  & \multirow{2}{*}{$>$ 2 $wlp$} & {\color{red} Filter all} \\
            & \multirow{2}{*}{with flux$<$0}  & {  \footnotesize \it \color{red} except DIB if Teff>7000 K}  \\
            & & Flag 9 \\
\cline{2-3}
            & 1 or 2 $wlp$ & \multirow{2}{*}{Flag 1}  \\
            & with flux$<$0 &   \\
\cline{2-3}
            & flux>0 & Flag 0  \\
            
\hline
\hline
\multirow{4}{*}{\color{teal} \bf nullFluxErr} &  $\sigma$Teff$=$0 K or   &\\
                         & $\sigma$logg$=$0 dex  or  & {\color{red} Filter all  } \\
                         & $\sigma$[M/H]$=$0 dex or & Flag 9\\  
                         & $\sigma$[$\alpha$/Fe]$=$0 dex  &   \\ 
\hline
\hline    
   \end{tabular}
    \label{tab:FluxProb}
\end{table}

\begin{table}[h]
\centering
    \caption{Definition of the parameter flag considering problems in the paramerisation of KM-type giants. $F_{\rm min}$ is the minimum flux value in the corresponding RVS spectrum (see also Table~\ref{tab:MG_QFchain}).}
\begin{tabular}{| c | c |  c |}
\hline
\hline
\multirow{2}{*}{ \bf Flag }  & \multirow{3}{*}{\bf Condition} & \multirow{3}{*}{\bf Flag value}  \\
\multirow{2}{*}{ \bf name} & &   \\
& & \\
\hline
\hline
\multirow{9}{*}{\color{teal} \bf  KM-type} & \T<4000~K and \g<3.5 & Flag 2 \\
\multirow{9}{*}{\color{teal} \bf stars}     & and ($gof$>-3.0 and $F_{\rm min}$>0.22) & {\color{red} Filter \alphaFe}  \\
     \cline{2-3}
     & \T<4000 K and \g<3.5 & \\
     & and [ (-3.4<$gof$<-3.0)   & Flag 1  \\
     & or ($gof$>-3.0 and $F_{\rm min}$<0.22)  & {\color{red} Filter \alphaFe} \\
     & or ($gof$<-3.4 and $F_{\rm min}$>0.22) ]& \\
     \cline{2-3}
     & [ \T<4000 K and \g<3.5 & \multirow{3}{*}{Flag 0} \\
     & and ($gof$<-3.4 or $F_{\rm min}$<0.22) ] &  \\
     & or \T>4000 or \g>3.5 &  \\
\hline
\hline    
   \end{tabular}
    \label{tab:KMflag}
\end{table}

\begin{table*}[h]
\centering
    \caption{Definition of individual abundance upper limit flags (see Table~\ref{tab:MG_QFchain}).  Xfe$\_$gspspec$\_$upper is the upper confidence value of the abundance (corresponding to the 84th quantile of the Monte-Carlo distribution). $\sigma$[X/Fe] is the 84th-16th interquantile abundance uncertainty. XfeUpperLimit is the mean value of the abundance upper limit for the considered lines of the X-element in the spectrum (depending on the mean S/N in the line $wlp$ and the stellar parameters). X\_MAD\_UpperLimit is the median absolute deviation of upper limit in the line $wlp$. Finally, the c-coefficients are reported in \tabref{tab:XFeCoefs}.}
\begin{tabular}{| c | c |  c |}
\hline
\hline
\multirow{2}{*}{ \bf Flag }  & \multirow{3}{*}{\bf Condition} & \multirow{3}{*}{\bf Flag value}  \\
\multirow{2}{*}{ \bf name} & &   \\
& & \\
\hline
\hline
 & vbroadT$\ge$2 or vbroadG$\ge$2 & \\
& or  vbroadM$\ge$2 or $\sigma$[X/Fe] = 0 &  \\ 
& or (Xfe$\_$gspspec$\_$upper - Xfe$\_$gspspec) = 0 &  \\
 & or Teff$\le$c1 or Teff$\ge$c2  & Flag 9   \\
 & or logg$\le$c3 or logg$\ge$c4 & {\color{red} Filter [X/Fe]}  \\
 & or ( (2 - XfeUpperLimit) / $\sigma$[X/Fe] )$\le$c5 & \\
 & or (S/N$\le$c6 and gof$\ge$c7) &  \\
\multirow{20}{*}{\color{blue} \bf  XUpLim} & or (Xfe$\_$gspspec + [M/H])$\le$c8   \\
      \cline{2-3}
 & vbroadT<2 and  vbroadG<2  \\
& and  vbroadM<2 and $\sigma$[X/Fe] != 0 &  \\
& and (Xfe$\_$gspspec$\_$upper - Xfe$\_$gspspec) != 0 &  \\
 & and c1<Teff<c2  &    \\
 & and c3<logg<c4   & Flag 2  \\
 & and ( (2 - XfeUpperLimit) / $\sigma$[X/Fe] )>c5  & \\
 & and (S/N>c6 or gof<c7) &  \\
 & and (Xfe$\_$gspspec + [M/H])<c8 &  \\
 & and ( (Xfe$\_$gspspec -  XfeUpperLimit) / (1.48$\cdot$X\_MAD\_UpperLimit) )<1.5 & \\
      \cline{2-3}     
        & vbroadT<2 and  vbroadG<2 & \\
& and vbroadM<2 and $\sigma$[X/Fe] != 0 &  \\
& and ( Xfe$\_$gspspec$\_$upper- Xfe$\_$gspspec) != 0 &  \\
 & and c1<Teff<c2  &    \\
 & and c3<logg<c4   & Flag 1  \\
 & and ( (2 - XfeUpperLimit) / $\sigma$[X/Fe] )>c5  & \\
 & and (S/N>c6 or gof<c7) &  \\
 & and (Xfe$\_$gspspec+[M/H])<c8 &  \\
 & and 1.5$\le$( (Xfe$\_$gspspec - XfeUpperLimit) / (1.48$\cdot$X\_MAD\_UpperLimit) )<2.5 & \\
      \cline{2-3}     
   & vbroadT<2 and  vbroadG<2 & \\
& and vbroadM<2 and $\sigma$[X/Fe] != 0 &  \\
& and (Xfe$\_$gspspec$\_$upper - Xfe$\_$gspspec) != 0 &  \\
 & and c1<Teff<c2  &    \\
 & and c3<logg<c4   & Flag 0  \\
 & and ( (2 - XfeUpperLimit) / $\sigma$[X/Fe] )>c5  & \\
 & and (S/N>c6 or gof<c7) &  \\
 & and ( Xfe$\_$gspspec + [M/H])<c8 &  \\
 & and ( (Xfe$\_$gspspec - XfeUpperLimit) / (1.48$\cdot$X\_MAD\_UpperLimit) )$\ge$2.5 & \\   
\hline
\hline    
   \end{tabular}
    \label{tab:XUpLim}
\end{table*}

\begin{table*}[h]
\centering
    \caption{Definition of individual abundance uncertainty flags (see Table~\ref{tab:MG_QFchain}).  Xfe$\_$gspspec$\_$upper is the upper confidence value of the abundance (corresponding to the 84th quantile of the Monte-Carlo distribution). $\sigma$[X/Fe] is the 84th-16th interquantile abundance uncertainty. XfeUpperLimit is the mean value of the abundance upper limit for the considered lines of the X-element in the spectrum (depending on the mean S/N in the line $wlp$ and the stellar parameters). 
    Finally, the c-coefficients are reported in \tabref{tab:XFeCoefs}.}
\begin{tabular}{| c | c |  c |}
\hline
\hline
\multirow{2}{*}{ \bf Flag }  & \multirow{3}{*}{\bf Condition} & \multirow{3}{*}{\bf Flag value}  \\
\multirow{2}{*}{ \bf name} & &   \\
& & \\
\hline
\hline
 &vbroadT$\ge$2 or vbroadG$\ge$2 & \\
& or vbroadM$\ge$2 or $\sigma$[X/Fe]$=$0 &  \\
& or (Xfe$\_$gspspec$\_$upper - Xfe$\_$gspspec) = 0  &  \\
 & or Teff$\le$c1 or Teff$\ge$c2   & Flag 9   \\
 & or logg$\le$c3 or logg$\ge$c4   & {\color{red} Filter [X/Fe]}  \\
 & or ( (2-XfeUpperLimit) / $\sigma$[X/Fe] )$\le$c5  & \\
 & or (S/N$\le$c6 and gof$\ge$c7) &  \\
 & or (Xfe$\_$gspspec + [M/H])$\le$c8 &  \\
      \cline{2-3}
 &vbroadT<2 and vbroadG<2 & \\
& and vbroadM<2 and $\sigma$[X/Fe] != 0 &  \\
& and (Xfe$\_$gspspec$\_$upper - Xfe$\_$gspspec) != 0  & Flag 2  \\
 & and c1<Teff<c2 and c3<logg<c4  &   \\
 & and c5<( (2 - XfeUpperLimit) / $\sigma$[X/Fe] )<7  & \\
  & and (S/N>c6 or gof<c7) &  \\
\color{blue} \bf  XUncer  & and (Xfe$\_$gspspec + [M/H])<c8 &  \\
      \cline{2-3}     
 &vbroadT<2 and vbroadG<2 & \\
& and vbroadM<2 and $\sigma$[X/Fe] != 0 &  \\
& and (Xfe$\_$gspspec$\_$upper - Xfe$\_$gspspec) != 0 &  \\
 & and c1<Teff<c2 and c3<logg<c4  & Flag 1  \\
 & and 7 $\le$ ( (2 - XfeUpperLimit) / $\sigma$[X/Fe] ) < 10  & \\
 & and (S/N>c6 or gof<c7) &  \\
 & and (Xfe$\_$gspspec + [M/H])<c8 &  \\
      \cline{2-3}     
  &vbroadT<2 and vbroadG<2 & \\
& and vbroadM<2 and $\sigma$[X/Fe] != 0 &  \\
& and (Xfe$\_$gspspec$\_$upper - Xfe$\_$gspspec) != 0 &  \\
 & and c1<Teff<c2 and c3<logg<c4 & Flag 0  \\
 & and ( (2 - XfeUpperLimit) / $\sigma$[X/Fe] ) $\ge$ 10  & \\
 & and (S/N>c6 or gof<c7) &  \\
 & and (Xfe$\_$gspspec + [M/H])<c8 &  \\     
 \hline
\hline    
   \end{tabular}
    \label{tab:XUncer}
\end{table*}

\begin{table*}[h]
\centering
\caption{Coefficients for individual chemical abundance filtering (see [X/Fe] upperLimit flag and $\sigma$[X/Fe] quality flag in \tabref{tab:XUpLim} and \ref{tab:XUncer}, respectively).}

\begin{tabular}{| l | c |  c | c | c| c | c | c| c|}
\hline
\hline
\multirow{2}{*}{\bf Chemical abundance}  & \multirow{2}{*}{\bf c1} & \multirow{2}{*}{\bf c2} & \multirow{2}{*}{\bf c3} & \multirow{2}{*}{\bf c4} & \multirow{2}{*}{\bf c5} & \multirow{2}{*}{\bf c6} & \multirow{2}{*}{\bf c7} & \multirow{2}{*}{\bf c8}  \\ 
 & & & & & & & & \\
\hline
\hline
\NFe & 4200 & 8000  & ~0.0 & 5.5 & 4.5 & 100 & -3.6  &99 \\
\hline
\MgFe & 3500 & 8000 & -1.0 & 5.5 & 5.5 & 80 & -3.5 &99\\
\hline 
\SiFe & 4000 & 8000 & -1.0 & 5.5 & 6.0 & 110 & -3.8 &99\\
\hline
\SFe & 5500 & 8000 & ~3.0 & 5.5 & 5.0 & 120 & -3.7 &99\\
\hline
\CaFe & 3500 & 8000  & -1.0 & 5.5 & 10.0 & 60 & -3.2 &99\\
\hline
\TiFe & 4000 & 6500 & -1.0 & 5.5 & 6.0 & 110 & -3.65 &99\\
\hline
\CrFe & 3500 & 6000 & -1.0 & 5.5 & 6.0 & 1000 & -3.65 & 1.5\\
\hline
\FeIM & 3500 & 8000 & -1.0 & 5.5 & 5.0 & 1000 & -3.4 & 1.5\\
\hline
\FeIIM & 5700 & 8000 & ~3.5 & 5.5 & 5.0 & 70 & -3.5 & 1.5\\
\hline
\NiFe & 4000 & 6500 & -1.0 & 5.5 & 6.0 & 100 & -3.6 & 1.5\\
\hline
\ZrFe & 3500 & 8000 & -1.0 & 5.5 & 1.0 & 100 & -3.4 &99\\
\hline
\CeFe & 3500 & 8000 & -1.0 & 5.5 & 5.0 & 100 & -3.5 &99\\
\hline
\NdFe & 3500 & 5500 & -1.0 & 5.5 & 2.0 & 100 & -3.5 &99\\
\hline
\hline
\end{tabular}
\label{tab:XFeCoefs}
\end{table*}

\begin{table}[h]
\centering
    \caption{Definition of the quality flag of the CN differential EW with respect to the solar C and N abundances.
    }
\begin{tabular}{| c | c |  c |}
\hline
\hline
\multirow{2}{*}{ \bf Flag }  & \multirow{3}{*}{\bf Condition} & \multirow{3}{*}{\bf Flag value}  \\
\multirow{2}{*}{ \bf name} & & \\
& & \\
\hline
\hline
\multirow{12}{*}{
\bf DeltaCNq } & vbroadT$\ge$1 or vbroadG$\ge$1 & \\
 & or vbroadM$\ge$1 or CN$\_$EW$\_$err = 0 &  \\
 & or S/N $\le$ 80 or $gof$$\ge$-3.5 & Flag 9  \\
 & or \T$\ge$4800 K or \g$\ge$3.8  &{\color{red} Filter}  \\
 & or |CN$\_$p1 - 849.037|$\ge$0.05  & {\color{red} CN} \\
 & or CN$\_$p2$\ge$0.25 &  \\
 \cline{2-3} 
 & vbroadT<1 and vbroadG<1 & \multirow{6}{*}{Flag 0}\\
 & and vbroadM<1 and CN$\_$EW$\_$err != 0 &  \\
 & and S/N > 80 and $gof$<-3.5 &   \\
 & and \T<4800 K and \g<3.8  &  \\
 & and |CN$\_$p1 - 849.037|<0.05 &  \\
 & and CN$\_$p2<0.25 &  \\
\hline
\hline    
   \end{tabular}
    \label{tab:CNqflag}
\end{table}

\begin{table}[h]
\centering
  \caption{Definition of the quality flag for the DIB parameterisation. See Sect.~\ref{subsec:DIBqf} for the definition of $Ra$ and $Rb$.}
\begin{tabular}{| c | c |  c |}
\hline
\hline
\multirow{2}{*}{ \bf Flag }  & \multirow{3}{*}{\bf Condition} & \multirow{3}{*}{\bf Flag value}  \\
\multirow{2}{*}{ \bf name} & &   \\
& & \\
\hline
\hline
\multirow{20}{*}{
\bf DIBq}  & S/N$\le$70 or Vrad$\_$err > 5 km.s$^{-1}$ & \\
 & or \T<3\,500 K or \T>10$^{5}$ K & Flag 9  \\
 & or $\Sigma_{i}$ RVS$\_$Flux$_{i}$ < 0 &{\color{red} Filter DIB}  \\
 & or p$_{0}$ < 3/(S/N) &  \\
\cline{2-3} 
    & $p_1$ < 861.66~nm   &\multirow{3}{*}{Flag 5} \\
                         & or $p_1$ > 862.81~nm   &  \\
                         & or $p_0$ > 0.015~nm  & \\  
                         
\cline{2-3}         
                        & 861.66~nm < $p_1$ < 862.81~nm    &\multirow{3}{*}{Flag 4} \\
                         & and $p_0$ < $R_{b}$   &  \\
                         & and 0.6 < $p_2$ < 1.2 & \\  
\cline{2-3}         
                        & 861.66~nm < $p_1$ < 862.81~nm   & \multirow{3}{*}{Flag 3}\\
                         &  and $p_0 > R_{b}$   &  \\
                         & and $0.6 < p_2 < 1.2$ & \\  

\cline{2-3}         
                        & 861.66~nm < $p_1$ < 862.81~nm   &\multirow{3}{*}{Flag 2} \\
                         & and $p_0 > max(R_{a},R_{b})$   &  \\
                         & and  $0.6 < p_2 < 1.2$ & \\  

\cline{2-3}         
                        & 861.66~nm < $p_1$ < 862.81~nm   &\multirow{3}{*}{Flag 1} \\
                         & and $p_0 > R_{b}$   &  \\
                         & and $1.2 < p_2 < 3.2$ & \\  

\cline{2-3}         
                        & 861.66~nm < $p_1$ < 862.81~nm   &\multirow{3}{*}{Flag 0} \\
                         & and $p_0 > max(R_{a},R_{b})$   &  \\
                         & and $1.2 < p_2 < 3.2$ & \\  

\hline
\hline
 \end{tabular}
  \label{tab:DIBflag}
\end{table}

\clearpage

\section{Bias comparisons per survey for MatisseGauguin parameters}
\label{appendix:param_biases_surveys}

 \begin{figure}[h]
    \centering
    \includegraphics[width=0.4\textwidth]{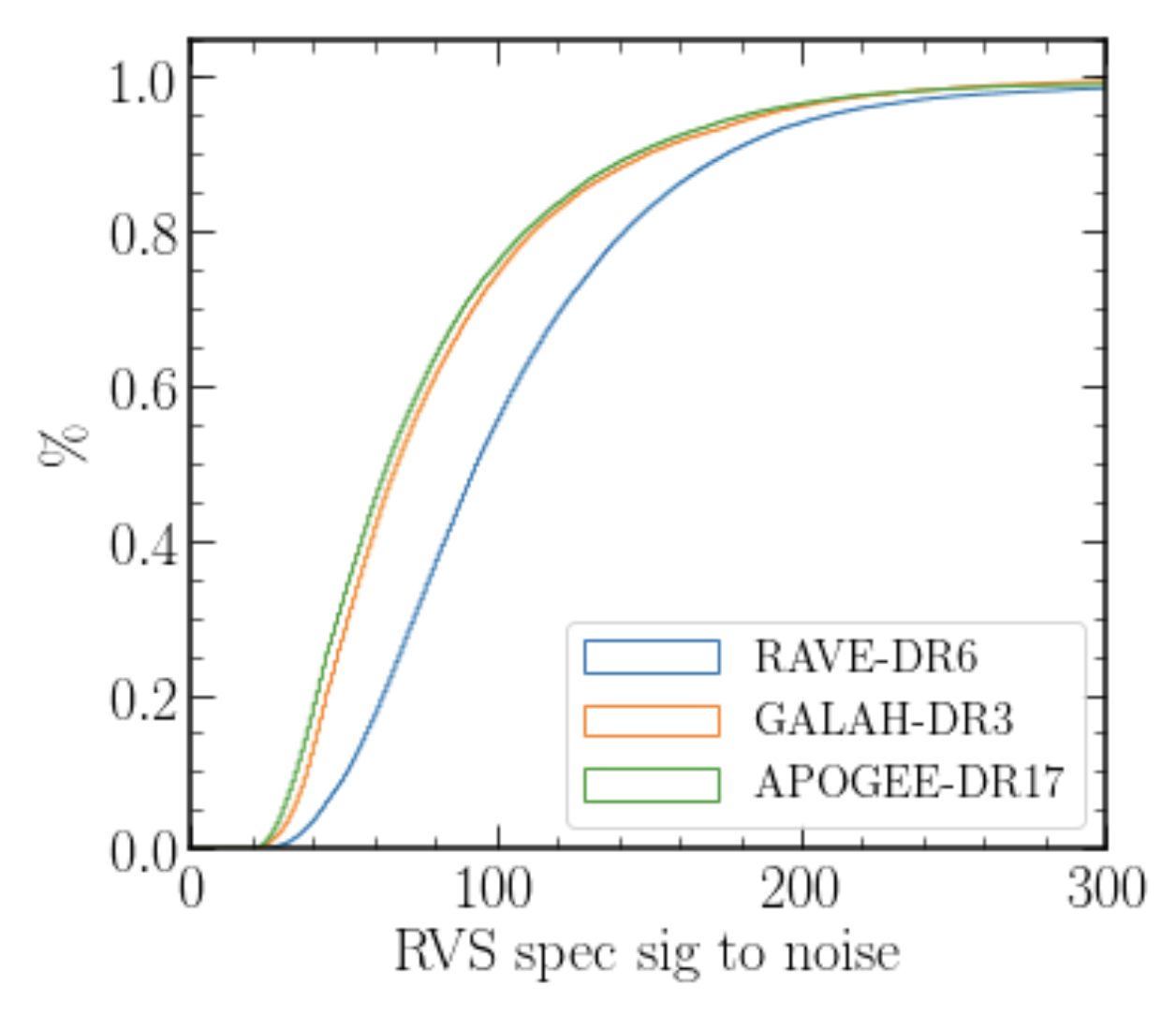}
    \caption{Cumulative histogram of RVS S/N for the selected comparison sample between \gspspec\ MatisseGauguin and three ground-based surveys: RAVE-DR6, GALAH-DR3, APOGEE-DR17. Table~\ref{Tab:survey_offsets} provides the median values and standard deviation per survey. }
    \label{fig:SNRsurveys}
\end{figure}

\begin{figure}[h]
\begin{center}
\includegraphics[width=0.45\textwidth]{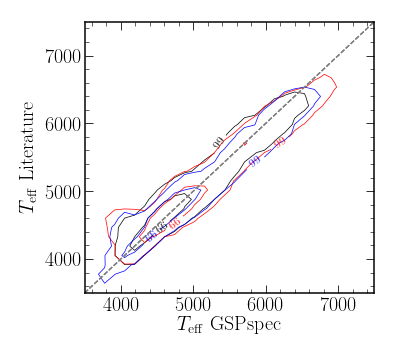}
\includegraphics[width=0.45\textwidth]{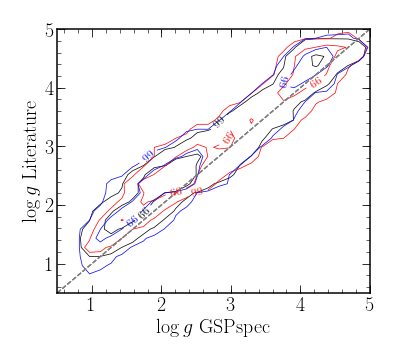}
\includegraphics[width=0.45\textwidth]{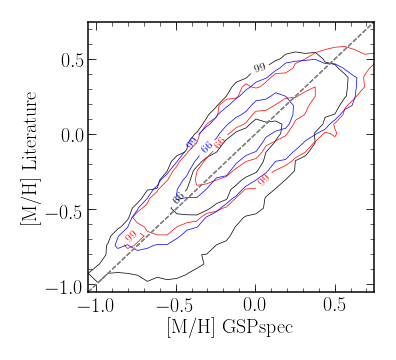}
\caption{Similar to Fig.~\ref{fig:VSTs_atmospheric} but showing only the contour lines of the 99th and 66th percentiles for RAVE-DR6 (black), GALAH-DR3 (red), and APOGEE-DR17 (blue).  }
\label{fig:VSTs_atmospheric_Per_Survey}
\end{center}
\end{figure}

Here, we perform a similar analysis to that shown in Sect.~\ref{Sec:VST_parameters}, but we investigate how the individual surveys compare with   \gspspec. 
First of all, Figure~\ref{fig:SNRsurveys} presents a cumulative histogram of the RVS spectra S/N for the selected comparison samples between \gspspec\ MatisseGauguin and RAVE-DR6, GALAH-DR3, and APOGEE-DR17. As expected from the selection functions of the different ground-based surveys, RAVE-DR6 targets have RVS spectra with higher S/N  values than GALAH-DR3 or APOGEE-DR17.

Figure~\ref{fig:VSTs_atmospheric_Per_Survey} is the equivalent to Fig.~\ref{fig:VSTs_atmospheric}, showing only the 99 and 66 percent contour lines for RAVE-DR6 (in black), GALAH-DR3 (in red), and APOGEE-DR17 in blue. Table~\ref{Tab:survey_offsets} quantifies the comparisons, by showing the median offset (\gspspec -- reference) as well as the robust sigma before and after the calibration for each survey. One can see that trends are similar no matter the reference catalogue, and that the biases are significantly decreased when using the calibrated values. It is important to note here that RVS-RAVE targets benefit from a higher S/N with respect to those of RVS-GALAH and RVS-APOGEE. 

\begin{figure*}[h]
    \centering
    \includegraphics[width=0.9\textwidth]{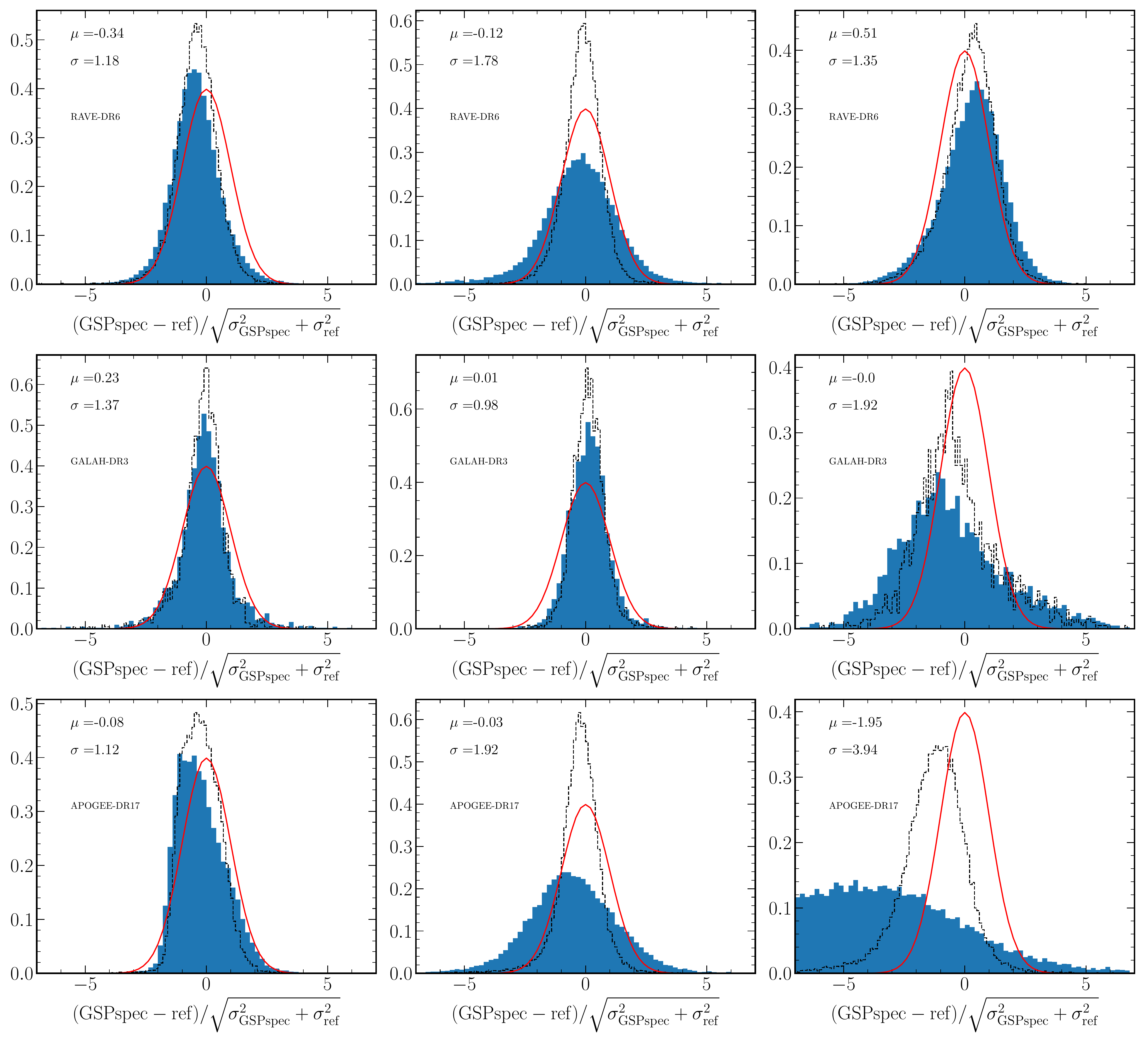}
    \caption{Distributions of parameter differences, normalised with respect to the reported \gspspec\ and literature uncertainties. From left to right: \T, \g,\ and \meta\ differences. From up to bottom: Differences with respect to RAVE-DR6, GALAH-DR3, and APOGEE-DR17. Dotted histograms correspond to the same distributions inflating the uncertainties (\gspspec\ and literature) by a factor of 4. The red curve shows a normal distribution of unit dispersion and zero mean. An unbiased parameter estimation with correct uncertainties should follow this distribution. } 
    \label{fig:SurveysUncertainties}
\end{figure*}

Finally, we investigated how \gspspec\ MatisseGauguin and the literature uncertainties compare to the observed parameter differences. To this purpose, Figure~\ref{fig:SurveysUncertainties} shows, in blue, the histograms of the \T\ (left column), \g\ (middle column), and \meta\ (right column) differences with respect to RAVE-DR6 (upper row), GALAH-DR3 (middle row), and APOGEE-DR17 (lower row). \gspspec\ MatisseGauguin \g\ and \meta\ values are calibrated. These parameter differences are normalised by the total uncertainty (defined as the quadratic sum of the \gspspec\ and the survey's uncertainties). The dotted histograms show the same distributions inflating the reported uncertainties by a factor of 4. Additionally, the red curves show a normal distribution of unit dispersion and zero mean. An unbiased parameter estimation with correct uncertainties should follow this distribution.  Regarding the effective temperature, the reported uncertainties from both \gspspec\ and the literature seem to correspond to the observed differences (with some uncertainty overestimation for the RVS-GALAH sample). Regarding \g , the situation differs from one survey to another. While the agreement between \gspspec\ MatisseGauguin and GALAH is good and the reported uncertainties appear overestimated  again, the comparison with RAVE and APOGEE suggests that the reported uncertainties are underestimated by a factor of 2 or 3 (the factor 4 is excluded by the normal distribution). 
Finally, the right column histograms show that \meta\ uncertainties are coherent with the observed differences between \gspspec\ and RAVE. However, \meta\ uncertainties from \gspspec\ or the GALAH/APOGEE reference or both seem underestimated by about a factor of 4. While in these examples, we only illustrate the impact of artificially inflating \gspspec\ uncertainties (through the dotted histograms), it cannot be excluded that the disagreement with respect to the normal distribution is caused by an underestimation of the uncertainties reported by the literature, as possibly suggested by the variety of situations that exist, for the same atmospheric parameter, when comparing to different surveys.

This analysis illustrates the complexity of comparing parametrisation results from different sources, each of them with its own uncertainty definitions, methodological and theoretical trends, and underlying selection functions. Once again, the importance of using a homogeneous catalogue for scientific purposes rather than a  compilation of different sources (even after re-calibrations) is highlighted.

\begin{table*}[h]
\centering
  \caption{Median offsets and robust sigma between \gspspec\ and individual surveys}
  \label{Tab:survey_offsets}
\begin{tabular}{c|ccc|cc|c}
 & \T & \g & \meta & \g$_{\rm calibrated}$ & \meta$_{\rm calibrated}$ & RVS S/N\\  \hline
   RAVE-DR6 & (-12; 93)  & (-0.28; 0.19)& (-0.05; 0.11)& (-0.003; 0.18) & (-0.05; 0.09) & (94; 64)\\ 
 GALAH-DR3 & (20;87)  & (-0.26; 0.21)& (0.01; 0.10)& (0.003; 0.18) &  (-0.001; 0.10) & (68; 53) \\ 
   APOGEE-DR17 & (-32; 86) & (-0.32; 0.17)& (0.04; 0.12)& (-0.005; 0.15) & (0.06; 0.12) & (65; 80)\\ \hline
\end{tabular}
\end{table*} 

\clearpage

\section{Illustration of polynomial corrections for MatisseGauguin  chemical abundances and quantification of uncertainties}
\label{AppendixBiases}
\begin{table*}[h]
    \centering
 \caption{Uncertainties on the polynomial coefficients of Table~\ref{tab:calibrations}. }
    \begin{tabular}{l|cccccc}
   \hline
    Element  &  $\delta p_0$  &  $\delta p_1$  &  $\delta p_2$  &  $\delta p_3$  &  $\delta p_4$ & {\it extrapol} flag\\ \hline 
    & \multicolumn{5}{c}{\cellcolor{gray!20} As a function of \g} & \cellcolor{gray!20}\\
    \hline
 \alphaFe  &    0.0965  &    0.0960  &   0.0302  &   0.0031  &  0.0000  & 0\\
    \CaFe  &    0.0750  &    0.0747  &   0.0236  &   0.0024  &  0.0000 & 0\\
    \MgFe  &    0.2002  &    0.2252  &   0.0819  &   0.0095  &  0.0000 & 0\\
     \SFe  &    6.5357  &    4.7804  &   1.1639  &   0.0943  &  0.0000 & 0\\
    \SiFe  &    0.1491  &    0.1562  &   0.0522  &   0.0056  &  0.0000 & 0\\
    \TiFe  &    0.0729  &    0.0807  &   0.0286  &   0.0032  &  0.0000 & 0\\
    \CrFe  &    0.0951  &    0.1310  &   0.0571  &   0.0077  &  0.0000 & 0\\
    \FeIH  &    0.1254  &    0.1260  &   0.0401  &   0.0041  &  0.0000 & 0\\
  \FeIIH  &   13.9985  &   10.1731  &   2.4608  &   0.1981  &  0.0000 & 0\\
    \NiFe  &    0.1829  &    0.1979  &   0.0692  &   0.0078  &  0.0000 & 0\\
     \NFe  &    0.0580  &    0.0716  &   0.0283  &   0.0035  &  0.0000 & 0\\
     \hline
 \alphaFe   &   0.0646   &   0.0971  &   0.0524  &   0.0119  &   0.0010 & $\le$1\\
    \CaFe   &   0.1000  &   0.1419  &   0.0726  &   0.0158  &   0.0012 & $\le$1\\
    \hline
    & \multicolumn{5}{c}{\cellcolor{gray!20} As a function of $t$=\T/5750}& \cellcolor{gray!20} \\ 
    \hline
 \alphaFe  &   0.5852   &   1.8825  &  2.0028   &  0.7049   &   0.0000 & $\le$1\\
 \CaFe  &   0.7270   &   2.3284  &  2.4674  &  0.8655   &   0.0000 & $\le$1\\ 
 \SFe  &   0.0323   & 0.0300 &  0.0000  &  0.0000 &   0.0000 & $\le$1\\  
\hline
\end{tabular}
    \label{tab:calibrations_errors}
\end{table*}

\begin{figure*}
\begin{center}
\includegraphics[width=\linewidth, angle=0]{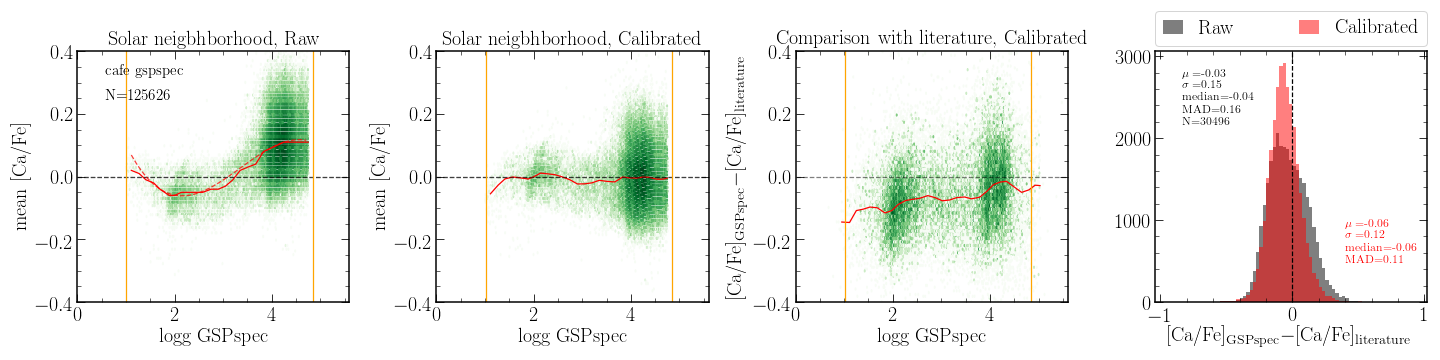}

\includegraphics[width=\linewidth, angle=0]{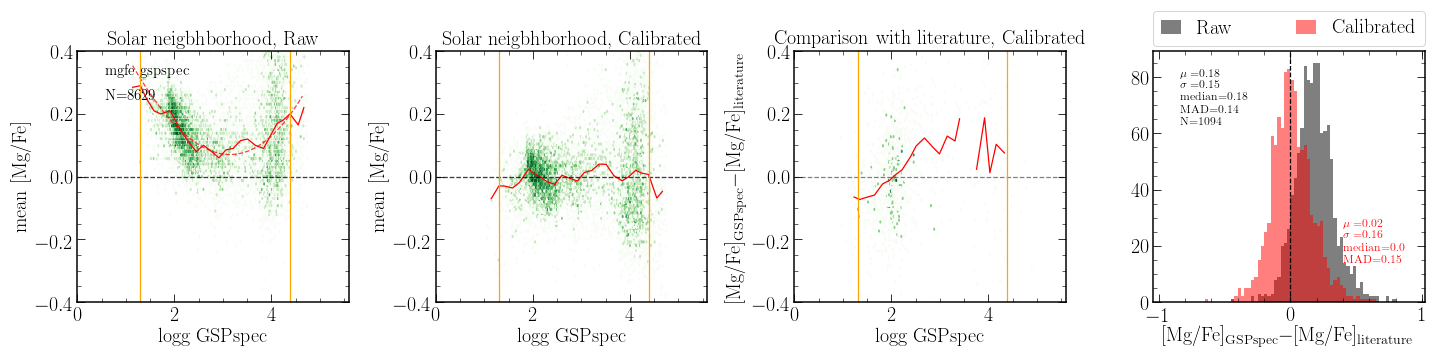}

\includegraphics[width=\linewidth, angle=0]{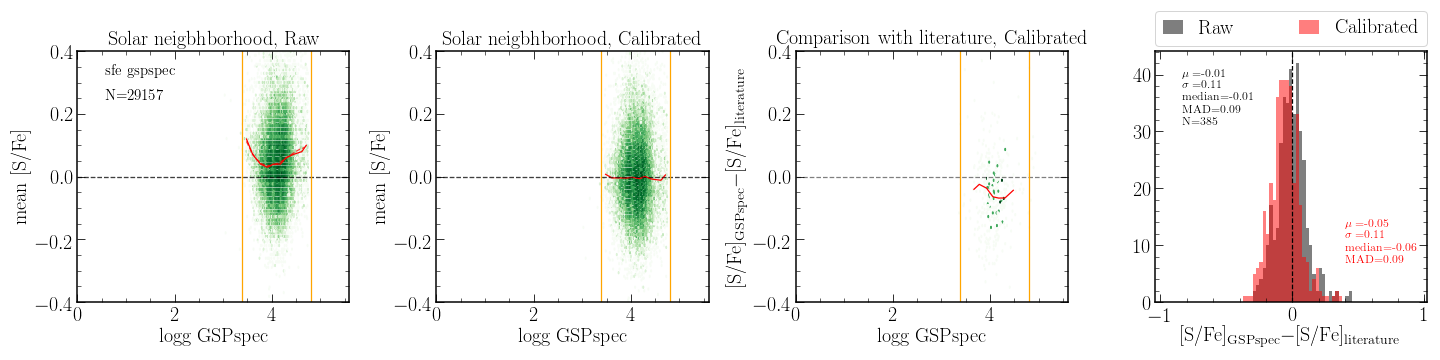}

\includegraphics[width=\linewidth, angle=0]{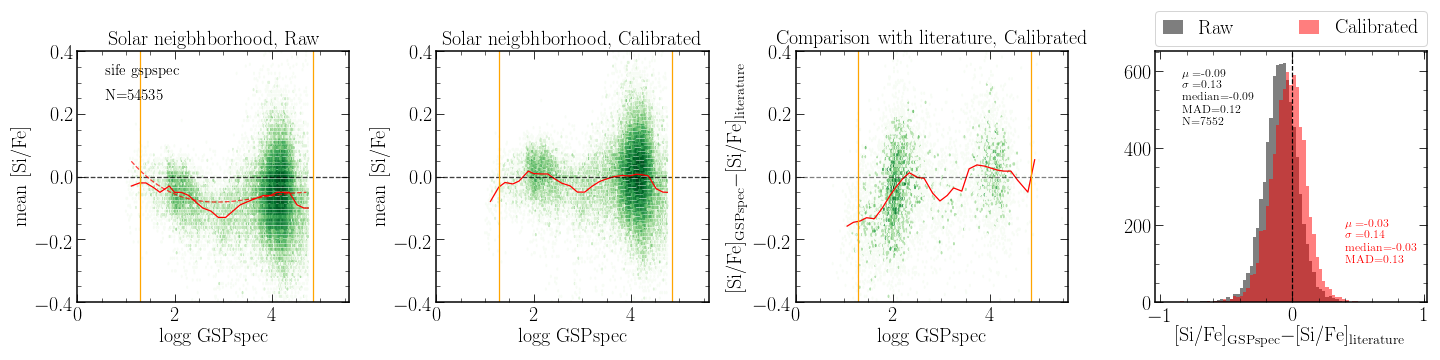}

\includegraphics[width=\linewidth, angle=0]{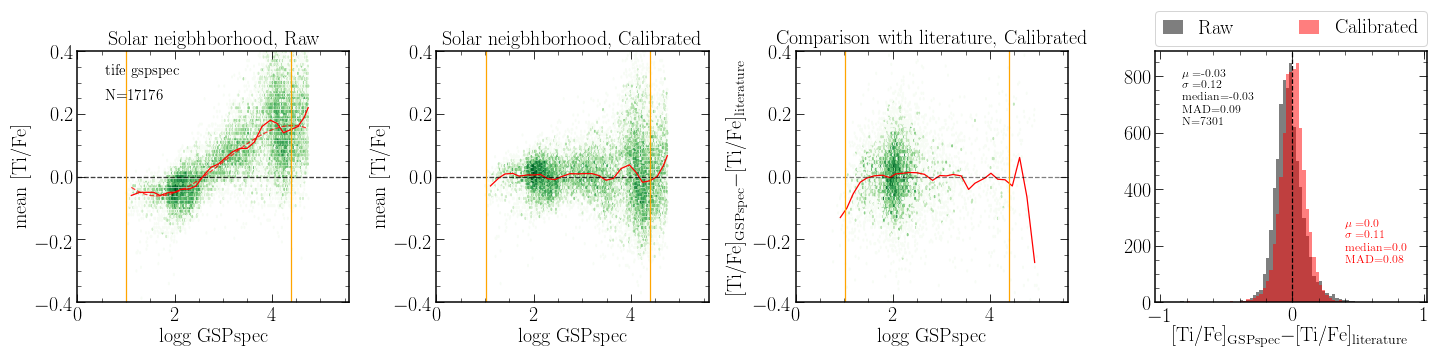}
\caption{Same as Fig.~\ref{fig:calib_alphaFe_only}, but for individual $\alpha$-elements. }cs.Metalm
\label{fig:calibs_alphas}
\end{center}
\end{figure*}

\begin{figure*}
\begin{center}
\includegraphics[width=\linewidth, angle=0]{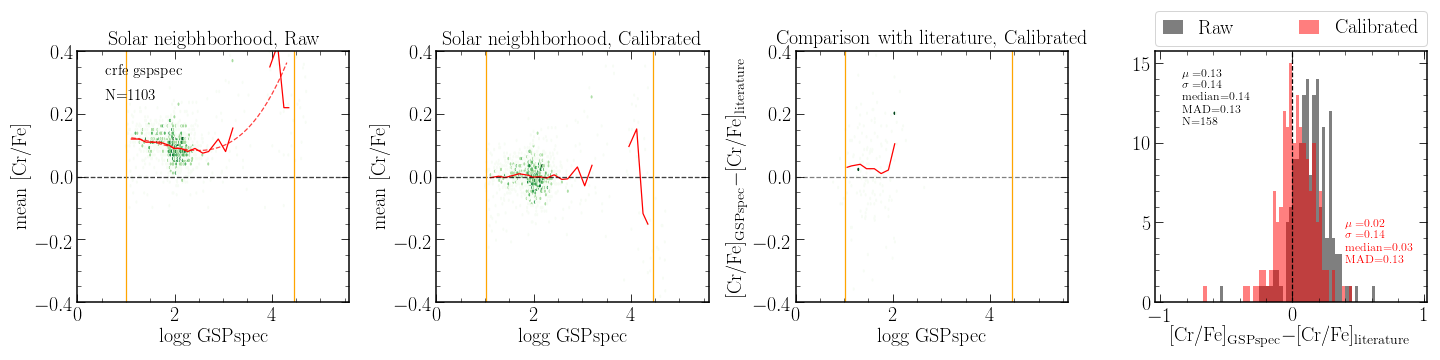}

\includegraphics[width=\linewidth, angle=0]{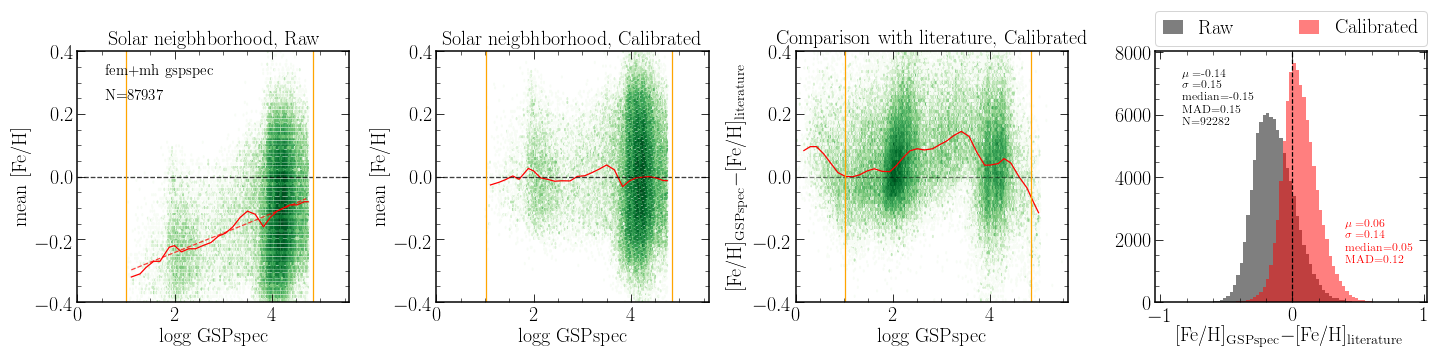}

\includegraphics[width=\linewidth, angle=0]{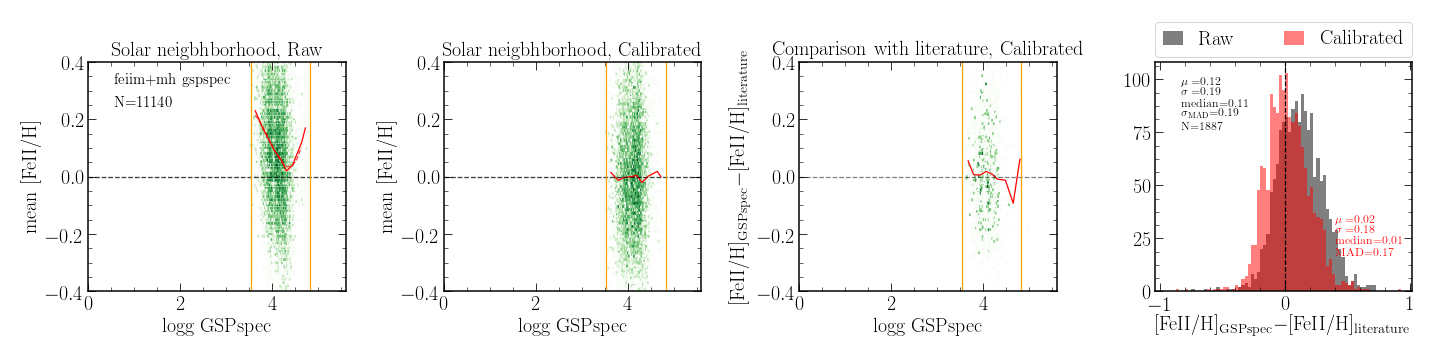}

\includegraphics[width=\linewidth, angle=0]{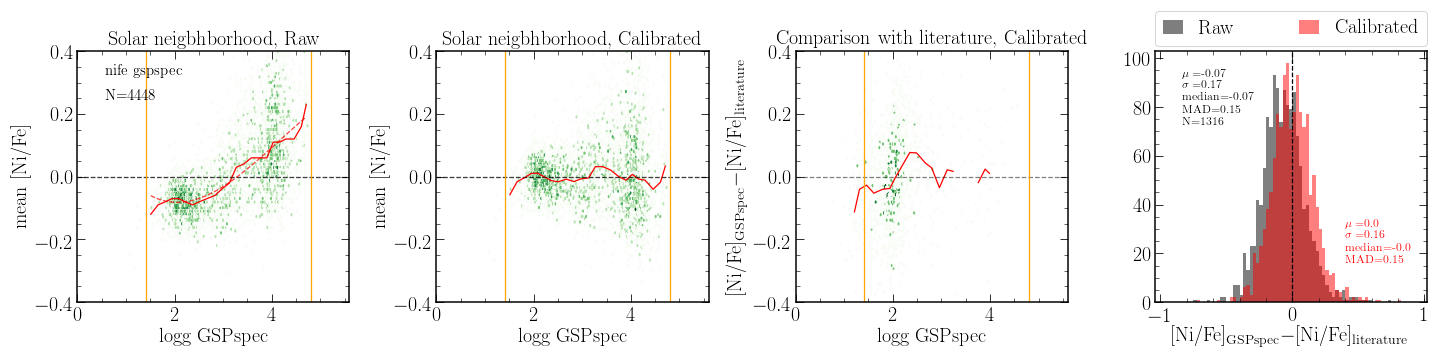}
\caption{Same as Fig.~\ref{fig:calib_alphaFe_only}, but for individual iron-peak elements. }
\label{fig:calibs_irons}
\end{center}
\end{figure*}

Figures~\ref{fig:calibs_alphas}~and~\ref{fig:calibs_irons}  illustrate the calibrations for individual chemical abundances and the comparison with literature data presented in Sect.~\ref{sec:Elemental_calibrations}. In addition, the uncertainties in the polynomial coefficients $p_1$, $p_2$, $p_3$, and $p_4$ provided in Table~\ref{tab:calibrations} are presented in  Table~\ref{tab:calibrations_errors}.

%

\clearpage 
\section{Validation of ANN biases and uncertainties as a function of S/N.} \label{ANN:internal}
As explained in Sect~\ref{Sec:ANN}, the ANN algorithm is trained with noisy spectra to optimise the parametrisation in different S/N regimes. For this reason, it is important to validate the correct behaviour of internal and external errors as a function of S/N.

\begin{figure}[h]

    \begin{center}
        \includegraphics[width=0.45\textwidth, angle=0]{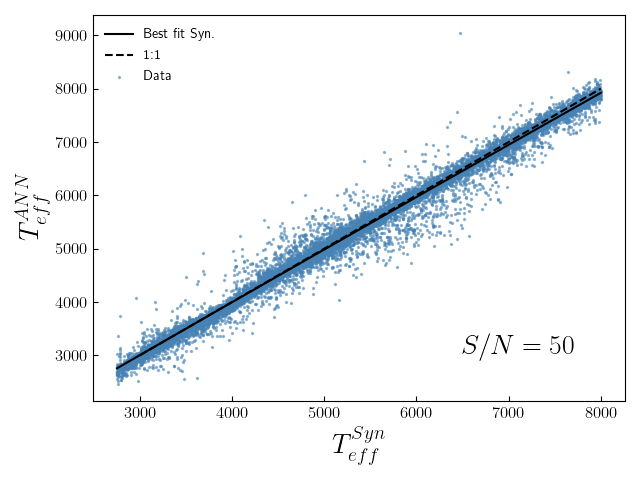}
            \caption{Illustration of the ANN tests with synthetic spectra to evaluate internal biases and uncertainties, for S/N$_{ANN}=$50. The estimated \T$^{ANN}$ as a function of the true parameter \T$^{Syn}$ is shown, including the polynomial fit modelling the observed behaviour. Similar analyses were performed for \g, \meta,\ and \alphaFe. }
        \label{fig:ann_bias_synth}
    \end{center}
\end{figure}

To study the internal biases and uncertainties, a parametrisation test with a random sample of 10\,000 synthetic spectra in the three S/N regimes listed in \tabref{tab:ann_SNReq} was performed.
 First of all, we studied the global behaviour of the bias as a function of S/N and a possible dependency of the bias on the parameters themselves by fitting the obtained parameter $X^{ANN}$ as a function of the true parameter $X^{Syn}$ (Fig.~\ref{fig:ann_bias_synth}). To model this behaviour, we use three different functions: a simple straight line, a parabola, and a piecewise first-order polynomial function (two, three, and five degrees of freedom, respectively) selecting as the best function the one with the smallest Bayesian information criterion (BIC). This process is repeated for each S/N.

\begin{figure}[h]
    \begin{center}
        \includegraphics[width=0.45\textwidth, angle=0]{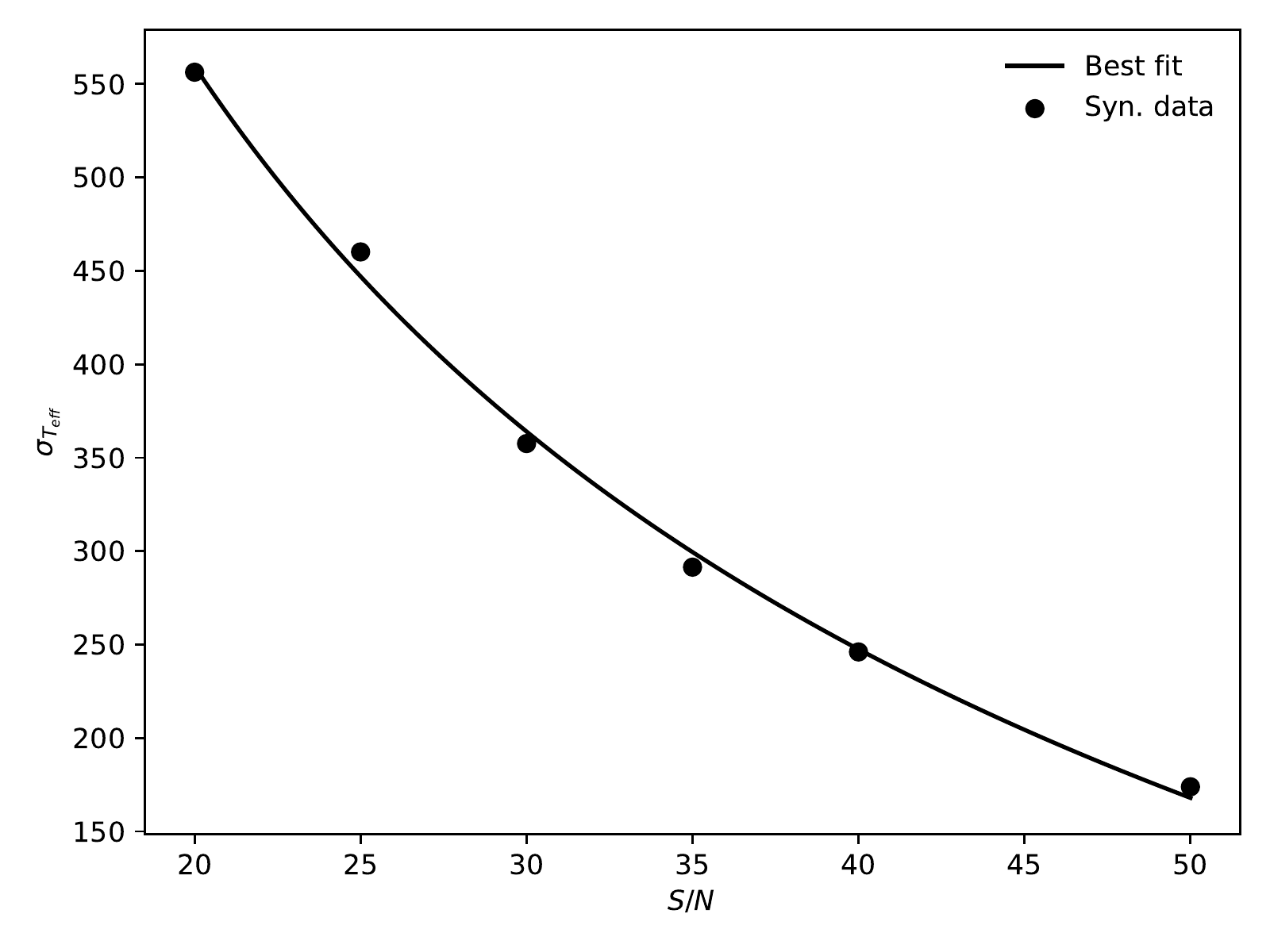}
            \caption{Illustration of the estimated trends on the internal ANN $T_{eff}$ uncertainty with S/N. The best fit to this trend is also shown. A similar analysis was performed for log(g), [M/H], and \alphaFe.}
        \label{fig:ann_width_synth}
    \end{center}
\end{figure}

In addition, for each S/N, the internal uncertainty ($\sigma_{inter}$) on each parameter was estimated from the standard deviation of the distribution $X^{ANN} - X^{Syn}$. 
 The internal uncertainty trends with S/ N\ are shown in \figref{fig:ann_width_synth} together with the function that best fits these points. To find this best-fit function, two possible functional relationships were considered: simple parabolic and an inverse square root of the S/N, selecting once again the function with the minimum BIC. It is worth noting that the preferred function is the inverse square root function in all S/N bins, confirming the consistency of the estimations and leading to the following equations:
\begin{equation}
    \sigma_{inter\_Teff}=-505 + 4763/\sqrt{S/N} 
,\end{equation}    
\begin{equation}
    \sigma_{inter\_logg}=-0.9 + 8/\sqrt{S/N} 
,\end{equation}
\begin{equation}
    \sigma_{inter\_[M/H]}=-0.6 + 5/\sqrt{S/N} 
,\end{equation}
\begin{equation}
    \sigma_{inter\_[\alpha/Fe]}=-0.6 + 5/\sqrt{S/N}
.\end{equation}


\tabref{tab:ann_internal} summarises 
the estimated internal biases and uncertainties as a function of S/N. As expected, internal biases are negligible. 


\begin{table}[t]
\centering
\caption{ANN internal biases and uncertainties (from the mean absolute deviation) in the different S/N regimes  considered for the ANN training.}
\begin{tabular}{cc|c|c|c|c|}
    \cline{3-6}
    \multicolumn{1}{l}{\scriptsize S/N} & & \T & \g & \meta & \alphaFe \\ 
    \hline
    \hline
    \multicolumn{1}{|c|}{50} &
        \makecell[c]{\scriptsize Bias\\ \scriptsize MAD} &
        \makecell[r]{-19 K\\ 102 K} &
        \makecell[r]{-0.02 dex\\ 0.16 dex} &
        \makecell[r]{0.03 dex\\ 0.13 dex} &
        \makecell[r]{-0.01 dex\\ 0.06 dex} \\ 
    \hline
    \multicolumn{1}{|c|}{40} &
        \makecell[c]{\scriptsize Bias\\ \scriptsize MAD} &
        \makecell[r]{30 K\\ 149 K} & 
        \makecell[r]{-0.07 dex\\ 0.25 dex} & 
        \makecell[r]{-0.01 dex\\ 0.18 dex} & 
        \makecell[r]{0.02 dex\\ 0.09 dex} \\
    \hline
    \multicolumn{1}{|c|}{35} &
        \makecell[c]{\scriptsize Bias\\ \scriptsize MAD} &
        \makecell[r]{39 K\\ 179 K} & 
        \makecell[r]{-0.01 dex\\ 0.31 dex} & 
        \makecell[r]{0.07 dex\\ 0.21 dex} & 
        \makecell[r]{0.02 dex\\ 0.11 dex} \\
    \hline
    \multicolumn{1}{|c|}{30} &
        \makecell[c]{\scriptsize Bias\\ \scriptsize MAD} &
        \makecell[r]{-22 K\\ 233 K} & 
        \makecell[r]{-0.01 dex\\ 0.41 dex} & 
        \makecell[r]{-0.05 dex\\ 0.28 dex} & 
        \makecell[r]{0.01 dex\\ 0.13 dex} \\
    \hline
    \multicolumn{1}{|c|}{25} &
        \makecell[c]{\scriptsize Bias\\ \scriptsize MAD} &
        \makecell[r]{-2 K\\ 318 K} & 
        \makecell[r]{-0.01 dex\\ 0.56 dex} & 
        \makecell[r]{0.03 dex\\ 0.38 dex} & 
        \makecell[r]{0.01 dex\\ 0.16 dex} \\
        \hline
\end{tabular}
\label{tab:ann_internal}
\end{table}

\begin{table}[t]
\centering
\caption{ANN biases and mean absolute deviations with respect to the literature, in the different S/N regimes considered for the ANN training.}
\begin{tabular}{cc|c|c|c|c|}
    \cline{3-6}
    \multicolumn{1}{l}{\scriptsize S/N} & & \T & \g & \meta &  \alphaFe \\ 
    \hline 
    \hline
    \multicolumn{1}{|c|}{50} &
        \makecell[c]{\scriptsize Bias\\ \scriptsize MAD} &
        \makecell[r]{-162 K\\ 155 K} & 
        \makecell[r]{-0.13 dex\\ 0.21 dex} & 
        \makecell[r]{-0.29 dex\\ 0.15 dex} & 
        \makecell[r]{0.07 dex\\ 0.05 dex} \\ 
    \hline
    \multicolumn{1}{|c|}{40} &
        \makecell[c]{\scriptsize Bias\\ \scriptsize MAD} &
        \makecell[r]{-89 K\\ 139 K} & 
        \makecell[r]{-0.10 dex\\ 0.29 dex} & 
        \makecell[r]{-0.26 dex\\ 0.13 dex} & 
        \makecell[r]{0.10 dex\\ 0.07 dex} \\
    \hline
    \multicolumn{1}{|c|}{35} &
        \makecell[c]{\scriptsize Bias\\ \scriptsize MAD} &
        \makecell[r]{-38 K\\ 148 K} & 
        \makecell[r]{0.005 dex\\ 0.32 dex} & 
        \makecell[r]{-0.25 dex\\ 0.15 dex} & 
        \makecell[r]{0.12 dex\\ 0.08 dex} \\
    \hline
    \multicolumn{1}{|c|}{30} &
        \makecell[c]{\scriptsize Bias\\ \scriptsize MAD} &
        \makecell[r]{-6 K\\ 192 K} & 
        \makecell[r]{-0.04 dex\\ 0.42 dex} & 
        \makecell[r]{-0.32 dex\\ 0.19 dex} & 
        \makecell[r]{0.11 dex\\ 0.11 dex} \\
    \hline
    \multicolumn{1}{|c|}{25} &
        \makecell[c]{\scriptsize Bias\\ \scriptsize MAD} &
        \makecell[r]{-98 K\\ 253 K} & 
        \makecell[r]{-0.02 dex\\ 0.47 dex} & 
        \makecell[r]{-0.43 dex\\ 0.25 dex} & 
        \makecell[r]{0.09 dex\\ 0.12 dex} \\
        \hline
\end{tabular}
\label{tab:ann_external}
\end{table}

\begin{figure}
    \begin{center}
        \includegraphics[width=0.45\textwidth, angle=0]{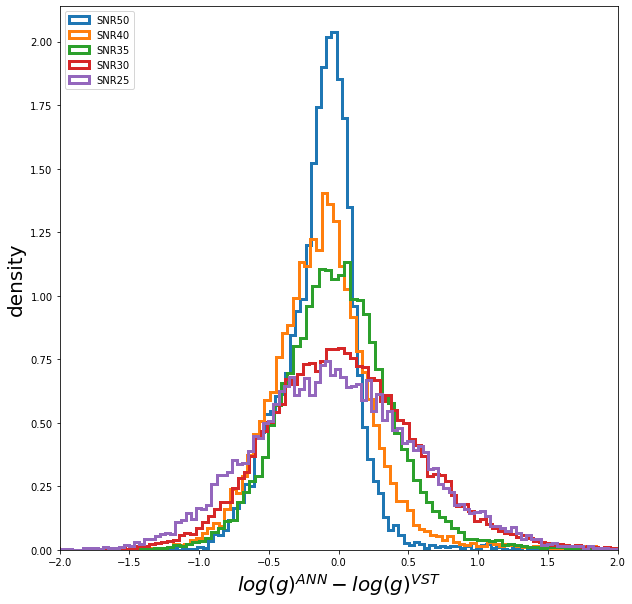}
       \caption{Error distributions for ANN estimations with respect to the literature (Validation Source Table - VST).}
        \label{fig:ann_distr_vst}
    \end{center}
\end{figure}

Finally, to complete the previous validation of the trend of the ANN estimates with S/N, the differences with respect to the literature (see Sect.~\ref{ANNBiases}) were examined. As a significant proportion of results from the three reference surveys have an important S/N dependence, with the lower resolution RAVE survey dominating for brighter sources in the high-S/N regime, we decided to validate the uncertainty behaviour with S/N with APOGEE DR16 and GALAH DR3 exclusively\footnote{These two surveys, as expected from their higher wavelength coverage and resolution, also show a better agreement with ANN parameters for sources with S/N$>$50.}. Figure \ref{fig:ann_distr_vst} illustrates the distribution of \T\ differences with respect to the literature for the five S/N regimes of the ANN training. Similar analyses were performed for the other three atmospheric parameters and the estimated biases and mean absolute deviations are reported in \tabref{tab:ann_external}. The expected increase in the spread for lower S/N regimes can be seen, validating the S/N optimisation of the ANN algorithm.

After the analysis of the uncertainties, we realised that there is a direct relation between \T\ and S/N and so we decided to propose different calibrations depending on the S/N ranges defined in \tabref{tab:ann_SNReq}. Furthermore, we observed that the number of stars with \T\ > 6000 K in the literature is statistically insignificant, and so the calibration beyond this limit should not be applied. For \g, \meta,\ and \alphaFe, although there is an intrinsic relation with S/N, the global calibration proved to be the best solution.
We provide the calibration of \T\ for S/N$_{ANN}\sim$50 in \secref{ANNBiases} and we give the polynomial coefficients for lower S/N in Table~\ref{tab:ann_calibrations}.

%



\begin{table}[t]
    \centering
    \caption{Polynomial coefficients for the \T\ calibration at different S/N$_{ANN}$ values.}
    \begin{tabular}{cc|cccc}
        & $S/N_{ANN}$ & $p_0$ & $p_1$ & $p_2$ & $p_3$ \\ 
        \hline
        \hline
        \multirow{4}{*}{ \bf \T } & 30 & -834 & 3.13E-1 & -2.53E-5 & \\
                                  & 35 & -1344 & 4.95E-1 & -4.26E-5 & \\
                                  & 40 & 3182 & -2.34 & 5.38E-4 & -3.84E-8 \\
                                  & 50 & 12816 & -8.1 & 1.65E-3 & -1.07E-7 \\
        \hline
        \hline
    \end{tabular}
    \label{tab:ann_calibrations}
\end{table}


\section{Query examples from the Gaia Archive}

\subsection{MatisseGauguin  parameters from the AstrophysicalParameters table}

\lstset{language=SQL}

\begin{lstlisting}[caption={\texttt{ADQL} query example with simple cuts in the limiting parameters.},captionpos=b]
SELECT source_id
FROM user_dr3int6.astrophysical_parameters
WHERE ((teff_gspspec>=3800) OR (logg_gspspec>=3.5)) AND ((teff_gspspec>=4150) OR (logg_gspspec>=3.6) OR (logg_gspspec<=2.4))
\end{lstlisting}

\begin{lstlisting}[caption={\texttt{ADQL} query example including conditions on the parameter flags (c.f. Table~\ref{tab:MG_QFchain})},captionpos=b]
SELECT source_id 
FROM user_dr3int6.astrophysical_parameters 
WHERE  (teff_gspspec>3500) AND (logg_gspspec>0) AND (logg_gspspec<5) AND ((teff_gspspec_upper-teff_gspspec_lower)<750) AND ((logg_gspspec_upper-logg_gspspec_lower)<1.) AND ((mh_gspspec_upper-mh_gspspec_lower)<.5) AND (teff_gspspec>=3800 OR logg_gspspec<=3.5) AND (teff_gspspec>=4150 OR logg_gspspec<=2.4 OR logg_gspspec>=3.6 ) AND  ((flags_gspspec LIKE "____________0%") OR (flags_gspspec LIKE "____________1%")) AND ((flags_gspspec LIKE "0%") OR (flags_gspspec LIKE "1%")) AND  ((flags_gspspec LIKE "_0%") OR (flags_gspspec LIKE "_1%")) AND  ((flags_gspspec LIKE "__0%") OR (flags_gspspec LIKE "__1%")) AND  ((flags_gspspec LIKE "___0%") OR (flags_gspspec LIKE "___1%")) AND  ((flags_gspspec LIKE "____0%") OR (flags_gspspec LIKE "____1%")) AND  ((flags_gspspec LIKE "_____0%") OR (flags_gspspec LIKE "_____1%")) AND  ((flags_gspspec LIKE "______0%") OR (flags_gspspec LIKE "______1%") OR (flags_gspspec LIKE "______2%") OR (flags_gspspec LIKE "______3%")) AND  ((flags_gspspec LIKE "_______0%") OR (flags_gspspec LIKE "_______1%") OR (flags_gspspec LIKE "_______2%"))
\end{lstlisting}

\subsection{ANN parameters from the  AstrophysicalParametersSupp table}

\lstset{language=SQL}

\begin{lstlisting}[caption={Best quality sources, no S/N dependency ($\sim$1.3 M sources)},captionpos=b]
SELECT source_id, teff_gspspec_ann, logg_gspspec_ann, mh_gspspec_ann, alphafe_gspspec_ann, flags_gspspec_ann
FROM user_dr3int6.astrophysical_parameters_supp
WHERE TO_BIGINT(flags_gspspec_ann) < 10000
\end{lstlisting}

\begin{lstlisting}[caption={Best quality sources with S/N > 108 (S/N$_{ANN}$ 50) ($\sim$275 k sources)},captionpos=b]
SELECT ann.source_id, teff_gspspec_ann, logg_gspspec_ann, mh_gspspec_ann, alphafe_gspspec_ann, flags_gspspec_ann, rv_expected_sig_to_noise
FROM user_dr3int6.gaia_source as gaia RIGHT JOIN 
    (
    SELECT source_id, teff_gspspec_ann, logg_gspspec_ann, mh_gspspec_ann, alphafe_gspspec_ann, flags_gspspec_ann
    FROM user_dr3int6.astrophysical_parameters_supp
    WHERE TO_BIGINT(flags_gspspec_ann) < 10000
    ) as ann USING(source_id)
WHERE rv_expected_sig_to_noise > 108
\end{lstlisting}

\section{Acknowledgements}
\label{Appendix:MERCI}
({\it Funding}) The \Gaia\ mission and data processing have financially been supported by, in alphabetical order by country:
\\ -- the Algerian Centre de Recherche en Astronomie, Astrophysique et G\'{e}ophysique of Bouzareah Observatory;
\\ -- the Austrian Fonds zur F\"{o}rderung der wissenschaftlichen Forschung (FWF) Hertha Firnberg Programme through grants T359, P20046, and P23737;
\\ -- the BELgian federal Science Policy Office (BELSPO) through various PROgramme de D\'{e}veloppement d'Exp\'{e}riences scientifiques (PRODEX)
      grants, the Research Foundation Flanders (Fonds Wetenschappelijk Onderzoek) through grant VS.091.16N,
      the Fonds de la Recherche Scientifique (FNRS), and the Research Council of Katholieke Universiteit (KU) Leuven through
      grant C16/18/005 (Pushing AsteRoseismology to the next level with TESS, GaiA, and the Sloan DIgital Sky SurvEy -- PARADISE);  
\\ -- the Brazil-France exchange programmes Funda\c{c}\~{a}o de Amparo \`{a} Pesquisa do Estado de S\~{a}o Paulo (FAPESP) and Coordena\c{c}\~{a}o de Aperfeicoamento de Pessoal de N\'{\i}vel Superior (CAPES) - Comit\'{e} Fran\c{c}ais d'Evaluation de la Coop\'{e}ration Universitaire et Scientifique avec le Br\'{e}sil (COFECUB);
\\ -- the Chilean Agencia Nacional de Investigaci\'{o}n y Desarrollo (ANID) through Fondo Nacional de Desarrollo Cient\'{\i}fico y Tecnol\'{o}gico (FONDECYT) Regular Project 1210992 (L.~Chemin);
\\ -- the National Natural Science Foundation of China (NSFC) through grants 11573054, 11703065, and 12173069, the China Scholarship Council through grant 201806040200, and the Natural Science Foundation of Shanghai through grant 21ZR1474100;  
\\ -- the Tenure Track Pilot Programme of the Croatian Science Foundation and the \'{E}cole Polytechnique F\'{e}d\'{e}rale de Lausanne and the project TTP-2018-07-1171 `Mining the Variable Sky', with the funds of the Croatian-Swiss Research Programme;
\\ -- the Czech-Republic Ministry of Education, Youth, and Sports through grant LG 15010 and INTER-EXCELLENCE grant LTAUSA18093, and the Czech Space Office through ESA PECS contract 98058;
\\ -- the Danish Ministry of Science;
\\ -- the Estonian Ministry of Education and Research through grant IUT40-1;
\\ -- the European Commission’s Sixth Framework Programme through the European Leadership in Space Astrometry (\href{https://www.cosmos.esa.int/web/gaia/elsa-rtn-programme}{ELSA}) Marie Curie Research Training Network (MRTN-CT-2006-033481), through Marie Curie project PIOF-GA-2009-255267 (Space AsteroSeismology \& RR Lyrae stars, SAS-RRL), and through a Marie Curie Transfer-of-Knowledge (ToK) fellowship (MTKD-CT-2004-014188); the European Commission's Seventh Framework Programme through grant FP7-606740 (FP7-SPACE-2013-1) for the \Gaia\ European Network for Improved data User Services (\href{https://gaia.ub.edu/twiki/do/view/GENIUS/}{GENIUS}) and through grant 264895 for the \Gaia\ Research for European Astronomy Training (\href{https://www.cosmos.esa.int/web/gaia/great-programme}{GREAT-ITN}) network;
\\ -- the European Cooperation in Science and Technology (COST) through COST Action CA18104 `Revealing the Milky Way with \Gaia\ (MW-Gaia)';
\\ -- the European Research Council (ERC) through grants 320360, 647208, and 834148 and through the European Union’s Horizon 2020 research and innovation and excellent science programmes through Marie Sk{\l}odowska-Curie grant 745617 (Our Galaxy at full HD -- Gal-HD) and 895174 (The build-up and fate of self-gravitating systems in the Universe) as well as grants 687378 (Small Bodies: Near and Far), 682115 (Using the Magellanic Clouds to Understand the Interaction of Galaxies), 695099 (A sub-percent distance scale from binaries and Cepheids -- CepBin), 716155 (Structured ACCREtion Disks -- SACCRED), 951549 (Sub-percent calibration of the extragalactic distance scale in the era of big surveys -- UniverScale), and 101004214 (Innovative Scientific Data Exploration and Exploitation Applications for Space Sciences -- EXPLORE);
\\ -- the European Science Foundation (ESF), in the framework of the \Gaia\ Research for European Astronomy Training Research Network Programme (\href{https://www.cosmos.esa.int/web/gaia/great-programme}{GREAT-ESF});
\\ -- the European Space Agency (ESA) in the framework of the \Gaia\ project, through the Plan for European Cooperating States (PECS) programme through contracts C98090 and 4000106398/12/NL/KML for Hungary, through contract 4000115263/15/NL/IB for Germany, and through PROgramme de D\'{e}veloppement d'Exp\'{e}riences scientifiques (PRODEX) grant 4000127986 for Slovenia;  
\\ -- the Academy of Finland through grants 299543, 307157, 325805, 328654, 336546, and 345115 and the Magnus Ehrnrooth Foundation;
\\ -- the French Centre National d’\'{E}tudes Spatiales (CNES), the Agence Nationale de la Recherche (ANR) through grant ANR-10-IDEX-0001-02 for the `Investissements d'avenir' programme, through grant ANR-15-CE31-0007 for project `Modelling the Milky Way in the \Gaia\ era’ (MOD4Gaia), through grant ANR-14-CE33-0014-01 for project `The Milky Way disc formation in the \Gaia\ era’ (ARCHEOGAL), through grant ANR-15-CE31-0012-01 for project `Unlocking the potential of Cepheids as primary distance calibrators’ (UnlockCepheids), through grant ANR-19-CE31-0017 for project `Secular evolution of galaxies' (SEGAL), and through grant ANR-18-CE31-0006 for project `Galactic Dark Matter' (GaDaMa), the Centre National de la Recherche Scientifique (CNRS) and its SNO \Gaia\ of the Institut des Sciences de l’Univers (INSU), its Programmes Nationaux: Cosmologie et Galaxies (PNCG), Gravitation R\'{e}f\'{e}rences Astronomie M\'{e}trologie (PNGRAM), Plan\'{e}tologie (PNP), Physique et Chimie du Milieu Interstellaire (PCMI), and Physique Stellaire (PNPS), the `Action F\'{e}d\'{e}ratrice \Gaia ' of the Observatoire de Paris, the R\'{e}gion de Franche-Comt\'{e}, the Institut National Polytechnique (INP) and the Institut National de Physique nucl\'{e}aire et de Physique des Particules (IN2P3) co-funded by CNES;
\\ -- the German Aerospace Agency (Deutsches Zentrum f\"{u}r Luft- und Raumfahrt e.V., DLR) through grants 50QG0501, 50QG0601, 50QG0602, 50QG0701, 50QG0901, 50QG1001, 50QG1101, 50\-QG1401, 50QG1402, 50QG1403, 50QG1404, 50QG1904, 50QG2101, 50QG2102, and 50QG2202, and the Centre for Information Services and High Performance Computing (ZIH) at the Technische Universit\"{a}t Dresden for generous allocations of computer time;
\\ -- the Hungarian Academy of Sciences through the Lend\"{u}let Programme grants LP2014-17 and LP2018-7 and the Hungarian National Research, Development, and Innovation Office (NKFIH) through grant KKP-137523 (`SeismoLab');
\\ -- the Science Foundation Ireland (SFI) through a Royal Society - SFI University Research Fellowship (M.~Fraser);
\\ -- the Israel Ministry of Science and Technology through grant 3-18143 and the Tel Aviv University Center for Artificial Intelligence and Data Science (TAD) through a grant;
\\ -- the Agenzia Spaziale Italiana (ASI) through contracts I/037/08/0, I/058/10/0, 2014-025-R.0, 2014-025-R.1.2015, and 2018-24-HH.0 to the Italian Istituto Nazionale di Astrofisica (INAF), contract 2014-049-R.0/1/2 to INAF for the Space Science Data Centre (SSDC, formerly known as the ASI Science Data Center, ASDC), contracts I/008/10/0, 2013/030/I.0, 2013-030-I.0.1-2015, and 2016-17-I.0 to the Aerospace Logistics Technology Engineering Company (ALTEC S.p.A.), INAF, and the Italian Ministry of Education, University, and Research (Ministero dell'Istruzione, dell'Universit\`{a} e della Ricerca) through the Premiale project `MIning The Cosmos Big Data and Innovative Italian Technology for Frontier Astrophysics and Cosmology' (MITiC);
\\ -- the Netherlands Organisation for Scientific Research (NWO) through grant NWO-M-614.061.414, through a VICI grant (A.~Helmi), and through a Spinoza prize (A.~Helmi), and the Netherlands Research School for Astronomy (NOVA);
\\ -- the Polish National Science Centre through HARMONIA grant 2018/30/M/ST9/00311 and DAINA grant 2017/27/L/ST9/03221 and the Ministry of Science and Higher Education (MNiSW) through grant DIR/WK/2018/12;
\\ -- the Portuguese Funda\c{c}\~{a}o para a Ci\^{e}ncia e a Tecnologia (FCT) through national funds, grants SFRH/\-BD/128840/2017 and PTDC/FIS-AST/30389/2017, and work contract DL 57/2016/CP1364/CT0006, the Fundo Europeu de Desenvolvimento Regional (FEDER) through grant POCI-01-0145-FEDER-030389 and its Programa Operacional Competitividade e Internacionaliza\c{c}\~{a}o (COMPETE2020) through grants UIDB/04434/2020 and UIDP/04434/2020, and the Strategic Programme UIDB/\-00099/2020 for the Centro de Astrof\'{\i}sica e Gravita\c{c}\~{a}o (CENTRA);  
\\ -- the Slovenian Research Agency through grant P1-0188;
\\ -- the Spanish Ministry of Economy (MINECO/FEDER, UE), the Spanish Ministry of Science and Innovation (MICIN), the Spanish Ministry of Education, Culture, and Sports, and the Spanish Government through grants BES-2016-078499, BES-2017-083126, BES-C-2017-0085, ESP2016-80079-C2-1-R, ESP2016-80079-C2-2-R, FPU16/03827, PDC2021-121059-C22, RTI2018-095076-B-C22, and TIN2015-65316-P (`Computaci\'{o}n de Altas Prestaciones VII'), the Juan de la Cierva Incorporaci\'{o}n Programme (FJCI-2015-2671 and IJC2019-04862-I for F.~Anders), the Severo Ochoa Centre of Excellence Programme (SEV2015-0493), and MICIN/AEI/10.13039/501100011033 (and the European Union through European Regional Development Fund `A way of making Europe') through grant RTI2018-095076-B-C21, the Institute of Cosmos Sciences University of Barcelona (ICCUB, Unidad de Excelencia `Mar\'{\i}a de Maeztu’) through grant CEX2019-000918-M, the University of Barcelona's official doctoral programme for the development of an R+D+i project through an Ajuts de Personal Investigador en Formaci\'{o} (APIF) grant, the Spanish Virtual Observatory through project AyA2017-84089, the Galician Regional Government, Xunta de Galicia, through grants ED431B-2021/36, ED481A-2019/155, and ED481A-2021/296, the Centro de Investigaci\'{o}n en Tecnolog\'{\i}as de la Informaci\'{o}n y las Comunicaciones (CITIC), funded by the Xunta de Galicia and the European Union (European Regional Development Fund -- Galicia 2014-2020 Programme), through grant ED431G-2019/01, the Red Espa\~{n}ola de Supercomputaci\'{o}n (RES) computer resources at MareNostrum, the Barcelona Supercomputing Centre - Centro Nacional de Supercomputaci\'{o}n (BSC-CNS) through activities AECT-2017-2-0002, AECT-2017-3-0006, AECT-2018-1-0017, AECT-2018-2-0013, AECT-2018-3-0011, AECT-2019-1-0010, AECT-2019-2-0014, AECT-2019-3-0003, AECT-2020-1-0004, and DATA-2020-1-0010, the Departament d'Innovaci\'{o}, Universitats i Empresa de la Generalitat de Catalunya through grant 2014-SGR-1051 for project `Models de Programaci\'{o} i Entorns d'Execuci\'{o} Parallels' (MPEXPAR), and Ramon y Cajal Fellowship RYC2018-025968-I funded by MICIN/AEI/10.13039/501100011033 and the European Science Foundation (`Investing in your future');
\\ -- the Swedish National Space Agency (SNSA/Rymdstyrelsen);
\\ -- the Swiss State Secretariat for Education, Research, and Innovation through the Swiss Activit\'{e}s Nationales Compl\'{e}mentaires and the Swiss National Science Foundation through an Eccellenza Professorial Fellowship (award PCEFP2\_194638 for R.~Anderson);
\\ -- the United Kingdom Particle Physics and Astronomy Research Council (PPARC), the United Kingdom Science and Technology Facilities Council (STFC), and the United Kingdom Space Agency (UKSA) through the following grants to the University of Bristol, the University of Cambridge, the University of Edinburgh, the University of Leicester, the Mullard Space Sciences Laboratory of University College London, and the United Kingdom Rutherford Appleton Laboratory (RAL): PP/D006511/1, PP/D006546/1, PP/D006570/1, ST/I000852/1, ST/J005045/1, ST/K00056X/1, ST/\-K000209/1, ST/K000756/1, ST/L006561/1, ST/N000595/1, ST/N000641/1, ST/N000978/1, ST/\-N001117/1, ST/S000089/1, ST/S000976/1, ST/S000984/1, ST/S001123/1, ST/S001948/1, ST/\-S001980/1, ST/S002103/1, ST/V000969/1, ST/W002469/1, ST/W002493/1, ST/W002671/1, ST/W002809/1, and EP/V520342/1.

The GBOT programme  uses observations collected at (i) the European Organisation for Astronomical Research in the Southern Hemisphere (ESO) with the VLT Survey Telescope (VST), under ESO programmes
092.B-0165,
093.B-0236,
094.B-0181,
095.B-0046,
096.B-0162,
097.B-0304,
098.B-0030,
099.B-0034,
0100.B-0131,
0101.B-0156,
0102.B-0174, and
0103.B-0165;
%
%
and (ii) the Liverpool Telescope, which is operated on the island of La Palma by Liverpool John Moores University in the Spanish Observatorio del Roque de los Muchachos of the Instituto de Astrof\'{\i}sica de Canarias with financial support from the United Kingdom Science and Technology Facilities Council, and (iii) telescopes of the Las Cumbres Observatory Global Telescope Network.

\end{appendix}

\end{document}